\documentclass[%
aip,
amsmath,amssymb,
reprint,
]{revtex4-2}

\usepackage{graphicx}
\usepackage{booktabs} 
\usepackage{dcolumn}  
\usepackage{bm}       
\usepackage{array}
\usepackage{multirow}
\usepackage{xcolor}
\usepackage{gensymb}  
\usepackage{hyperref}

\makeatletter
\let\@fnsymbol\@fnsymbol@latex
\@booleanfalse\altaffilletter@sw
\makeatother

\begin{document}
	
	\title{Assessment of Reynolds-Averaged Navier-Stokes Modeling of Jet Interaction in Fan-Array Wind Generator Flows}
	
	\author{M. Hosein Niroomand}
	\author{Utku Şentürk}
	\email[Corresponding author: ]{utku.senturk@ege.edu.tr}
	\affiliation{Ege University, Faculty of Engineering Mechanical Engineering Department, 35100 Bornova, İzmir, Turkey}
	
	\date{\today}
	
	\begin{abstract}
		Fan-array wind generators (FAWGs) provide controlled turbulent inflow conditions that cannot be reproduced in conventional wind tunnels. Despite their increasing use in experimental studies, numerical modeling of FAWG-generated flows remains largely unexplored. The present study assesses the capability of Reynolds-Averaged Navier–Stokes (RANS) modeling to predict jet interaction in a $10\times10$ fan-array wind generator. Numerical predictions are compared against experimental measurements of axial velocity and turbulence intensity from a reference configuration. Individual fan units are represented using a pressure-jump boundary condition based on a reconstructed performance curve derived from manufacturer data. Grid convergence is verified, and the influence of fan representation, operating point and inflow turbulence conditions is examined. The results show that RANS modeling captures the global jet interaction topology and downstream velocity decay with reasonable accuracy. However, systematic magnitude discrepancies are observed in the near-field injection region and peripheral shear layers. Turbulence intensity predictions exhibit larger deviations, reflecting limitations of the eddy-viscosity closure in highly mixing-dominated flows. A low-aspect-ratio flat plate is included as a demonstrative application to illustrate the aerodynamic impact of FAWG-generated inflow. Overall, the study shows that RANS modeling, combined with a pressure-jump fan representation, provides a computationally efficient framework for predicting the mean-flow structure of FAWG systems, while exhibiting clear limitations in resolving localized turbulence characteristics.
	\end{abstract}
	
	\keywords{Fan-array wind generator, windshaper, jet interaction, Reynolds-Averaged Navier–Stokes.}
	
	\maketitle
	
	\section{Introduction}
	
	Conventional wind tunnels are designed to generate spatially uniform inflow conditions with low turbulence intensity. While suitable for canonical aerodynamic testing, such facilities are inherently limited in their ability to reproduce the highly turbulent, non-uniform and transient flow conditions of real atmospheric environments. This restricts their applicability for investigating realistic inflow conditions that often lead to unsteady aerodynamic loading \cite{longobardi2024aerodynamic,RajasakeraBabu2025}.
	
	Fan array wind generators (FAWGs) have recently emerged as an interesting approach for generating controlled, spatially varying turbulent inflow conditions through independently actuated fans. By modulating fan speed and activation patterns, FAWGs can generate flow fields with turbulence intensities and non-uniformities that significantly exceed those seen in conventional wind tunnels with an open test section. This capability makes FAWGs particularly attractive for applications requiring realistic inflow conditions, including atmospheric boundary-layer research, gust modeling, aerodynamic testing of wind turbines, propellers and unmanned aerial vehicles \cite{noca2021flow, li2024aerodynamic,  liu2025fanarray,arc2025WS,Velsmann2021}.
	
	Previous studies on FAWGs have primarily focused on experimental characterization, including fan arrangement strategies, duty-cycle modulation, dynamic response, and turbulence enhancement using geometric or control-based approaches \cite{wang2024coarse,arc2025WS}. Large-scale facilities with hundreds to thousands of individually controllable fans have demonstrated the capability to generate highly complex inflow conditions.
	
	Existing studies rely largely on measurements and qualitative flow observations with limited use of numerical modeling \cite{wang2024coarse}. In particular, computational fluid dynamics (CFD) investigations that consider the jet interaction of a FAWG device have not yet been reported to the authors’ knowledge. Accurate numerical prediction of such flows is challenging due to the presence of multiple interacting jets and strong turbulent shear layers. In addition, direct application of conventional turbomachinery modeling approaches is not practical for systems composed of a large number of small, independently operated fan units. These features place stringent demands on grid resolution, turbulence modeling, and fan representation strategies, requiring validation against experimental data.
	
	The present work addresses this gap by assessing the capability of Reynolds-Averaged Navier–Stokes (RANS) modeling to predict jet interaction in a fan-array wind generator. Numerical results are compared against experimental measurements of velocity and turbulence intensity from the reference configuration \cite{arc2025WS}. The effects of grid resolution, fan representation, turbulence modeling assumptions and operating conditions are examined to evaluate the applicability and limitations of a RANS-based computational framework for FAWG simulations.
	
	\section{Numerical Methodology}
	
	\subsection{FAWG Configuration}
	
	The present study considers a reference FAWG configuration recently reported in \cite{arc2025WS}, which provides detailed experimental measurements of velocity and turbulence intensity for multiple fan activation patterns. The device consists of a full-scale $10\times10$ fan array comprising 100 axial-flow fan units \cite{li2024aerodynamic, arc2025WS}. Each unit is a two-stage counter-rotating axial fan (Delta Electronics, model GFB0412EHS) housed in a $40\times40\times56$ mm enclosure. The reported flow conditions include maximum streamwise velocities of approximately 12~m/s and turbulence intensities exceeding 28\%, depending on the activation pattern \cite{arc2025WS}.
	
	Although the reference study also examines duty-cycle modulation and gust-generation capabilities, these aspects are not considered here. The analysis is restricted to steady fan activation patterns in order to isolate the ability of the CFD model to reproduce mean velocity and turbulence intensity fields. Among the configurations reported in reference \cite{arc2025WS}, two representative activation patterns (B and D) are selected for validation.
	
	\subsection{Computational Domain and Boundary Conditions}
	
	The computational geometry is constructed based on the reference fan array configuration. Each fan unit has an exit diameter of $D = 40$ mm. A clearance of 1 mm is introduced between adjacent units, resulting in a center-to-center spacing of 42 mm. The hub diameter is not specified in the reference study and is therefore assumed. For small axial fans, the hub diameter ($d$) typically ranges between 45\% and 65\% of the outer diameter. In the present model, a value of 60\% of $D$ is adopted. Based on the specified spacing, the overall dimensions of the fan array block are $420 \times 420$ mm. This length is defined as the characteristic device width, $W$. The computational domain consists of an upstream chamber and a downstream chamber divided by an interior surface. The lateral dimensions of the domain are set to $5W$ in both width and height. The upstream chamber extends $2W$ upstream of the fan array, while the downstream chamber extends $10W$ downstream. To account for ground effects, the fan array block is positioned 1200 mm above a no-slip wall representing the floor, consistent with the experimental configuration \cite{arc2025WS}. 
	
	The front, side, and top faces of the upstream chamber are defined as pressure inlet boundaries, while the corresponding faces of the downstream chamber are defined as pressure outlet boundaries. All pressure boundaries are assigned a gauge pressure of $0\,\mathrm{Pa}$. The turbulence intensity (TI) and turbulent viscosity ratio (TVR, $\mu_t/\mu$) at the inlet boundaries are set to $5\%$ and 10, respectively. These parameters are varied to assess sensitivity to inflow turbulence conditions. The overall computational domain and key geometric parameters are illustrated in Figure \ref{fig:fan_models_and_domain}. 
	
	The effect of each fan unit is modeled using a pressure-jump boundary condition, which represents the fan as a momentum source without resolving blade geometry. In this approach, the fan is treated as a surface across which a prescribed pressure rise is imposed as a function of the local normal velocity. It is consistent with the actuator-disk concept, where the detailed blade-induced flow is replaced by a lumped pressure discontinuity that reproduces the integral performance of the device \cite{ansys_fluent_fan}. The pressure jump is specified through a characteristic pressure–flow relation derived from manufacturer data, allowing the operating point to be determined implicitly from the local flow field. A fourth-degree polynomial ($c_1$=402.66, $c_2$=-15.48, $c_3$=3.2595, $c_4$=-0.2846, $c_5$=0.003) is used. To ensure numerical stability, a velocity limiter is applied with a maximum allowable normal velocity of $15.5\,\mathrm{m/s}$, corresponding to the zero-pressure-jump operating point of the reconstructed fan curve. Although the formulation permits the inclusion of tangential velocity components to represent swirl, this effect is neglected in the present study since the fan units are two-stage counter-rotating devices. This approach enables computationally efficient simulation of a large number of fan units which would be prohibitively expensive to resolve using turbomachinery models.    
	
	\subsection{Surface vs.\ Ducted Fan Representations}
	
\begin{figure*}[htbp]
		\centering
		\begin{minipage}[b]{0.55\textwidth}
			\centering
			\includegraphics[width=\textwidth]{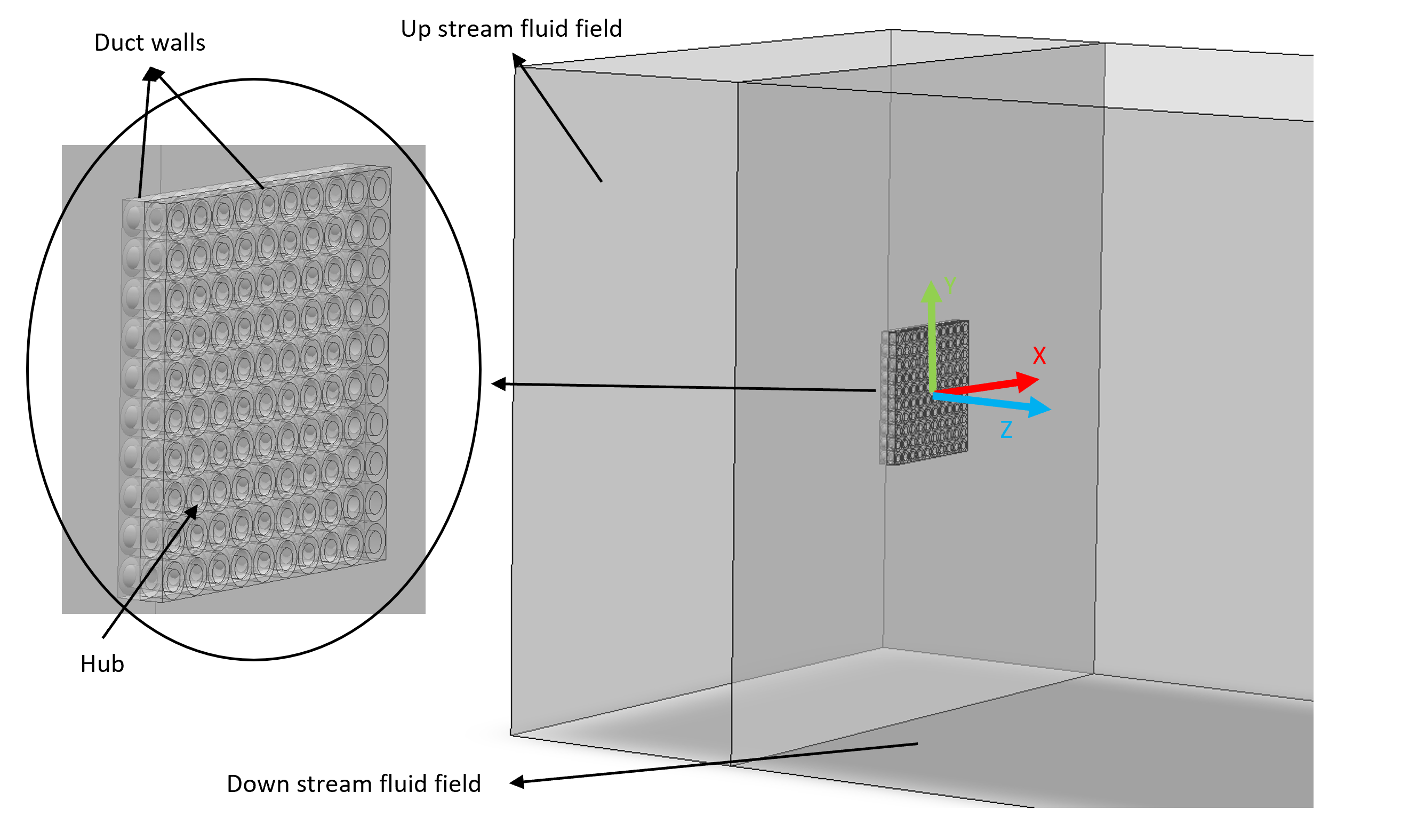}
			\vspace{5pt} \\ \small (a) 3D computational domain
		\end{minipage}
		\hfill
		\begin{minipage}[b]{0.42\textwidth}
			\centering
			\includegraphics[width=\textwidth]{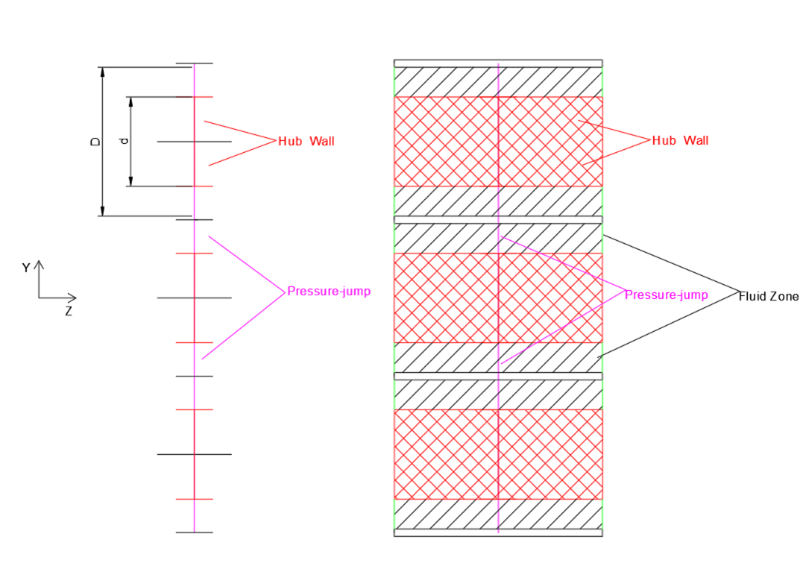}
			\vspace{5pt} \\ \small (b) 2D fan modeling schematic
		\end{minipage}
		\caption{Computational domain and fan modeling methodology. (a) 3D overview of the fluid region highlighting the upstream/downstream fluid fields, duct walls, and hub geometry. (b) 2D schematic comparing the simplified surface model (left) and the resolved ducted model (right).}
		\label{fig:fan_models_and_domain}
	\end{figure*}

	Two alternative representations are considered for implementing the pressure-jump boundary condition. In the first approach, referred to as the \textit{surface representation}, each fan is modeled as an independent, annular, zero-thickness surface extracted from the face that divides the domain. Individual fan units are activated or deactivated according to the prescribed activation patterns. This approach provides a simplified representation in which the fan effect is introduced directly at the section plane without resolving geometric confinement. A schematic of the surface model is shown in Figure~\ref{fig:fan_models_and_domain}.
	
	In the second approach, referred to as the \textit{ducted representation}, each fan is modeled as an annular fluid region bounded by the hub and outer casing. This configuration allows the influence of flow confinement and wall interactions to be explicitly represented. The pressure-jump boundary condition is applied on a ring-shaped surface within the fan zone, while the hub surface, outer duct wall, and surrounding block faces are treated as no-slip walls (Figure~\ref{fig:fan_models_and_domain}). The two approaches are considered to assess the impact of geometric fan representation on the predicted flow field, particularly in the near-field region where confinement effects are expected to be significant.
	
	\begin{figure*}[htbp]
		\centering
		\begin{minipage}[t]{0.48\textwidth}
			\centering
			\includegraphics[width=\textwidth]{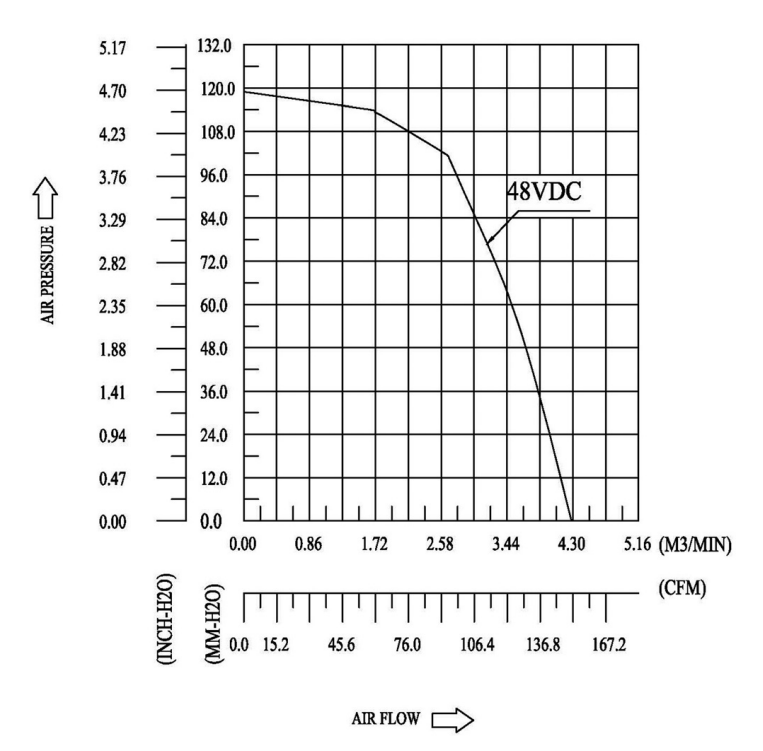}
			\vspace{5pt} \\ \small (a) Manufacturer-provided performance curve \cite{liu2025fanarray}.
		\end{minipage}
		\hfill
		\begin{minipage}[t]{0.48\textwidth}
			\centering
			\includegraphics[width=\textwidth]{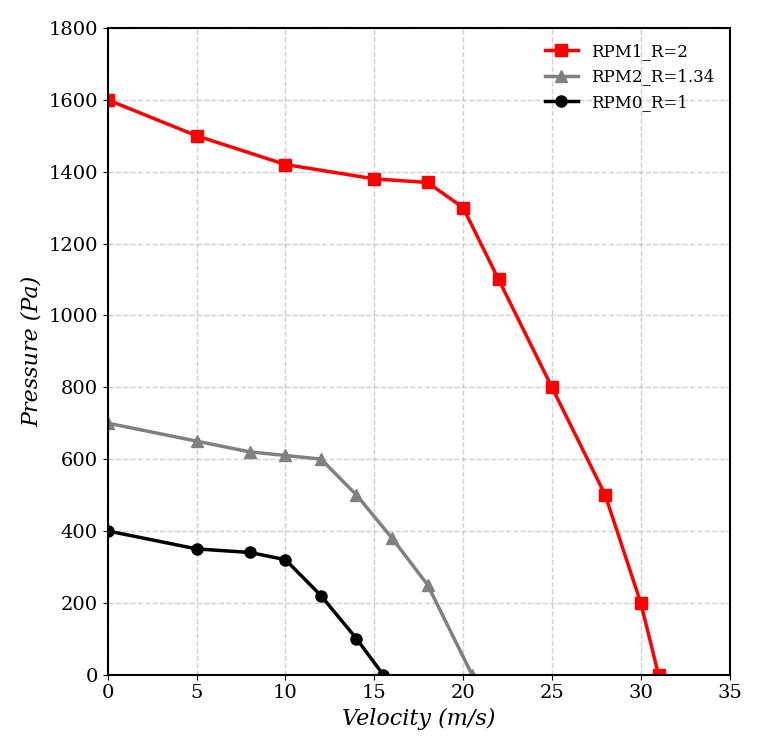}
			\vspace{5pt} \\ \small (b) Scaled performance curves for different RPM values.
		\end{minipage}
		\caption{Fan performance characteristics used in the pressure-jump modeling.}
		\label{fig:fan_curves_combined}
	\end{figure*}
	
	\subsection{Mesh Generation and Grid Convergence}
	
	A grid convergence study was conducted to assess the sensitivity of the numerical solution to spatial discretization. Three systematically refined meshes were generated, denoted as coarse (R1), medium (R2), and fine (R3), using a constant refinement ratio of $r = 1.5$.
	
	The monitored quantity for convergence assessment is the pointwise axial ($z$) velocity evaluated at the center of a plane located at a downstream distance of $z = W$ from the fan array. This location is selected as a representative point within the developing flow field.
	
	Mesh convergence is evaluated using Richardson extrapolation. The observed order of convergence is computed from the three grid solutions as
	\begin{equation}
		p = \frac{\ln \left( \frac{\Phi_3 - \Phi_2}{\Phi_2 - \Phi_1} \right)}{\ln(r)},
	\end{equation}
	where $\Phi_1$, $\Phi_2$, and $\Phi_3$ denote the solutions obtained on the fine, medium, and coarse meshes, respectively.
	
	The observed order of convergence was found to be approximately $p \approx 3$, indicating rapid reduction in the discretization error for the selected monitored quantity \cite{celik2008procedure,roache1998verification}. Since the difference between the medium and fine mesh solutions was small compared with that between the coarse and medium meshes, the medium mesh (R2) was adopted for subsequent simulations as a compromise between accuracy and computational cost. The corresponding mesh parameters are summarized in Table~\ref{tab:mesh_parameters}.
	
	\begin{table*}[htbp]
		\centering
		\caption{Summary of mesh parameters and convergence results}
		\label{tab:mesh_parameters}
		\begin{ruledtabular}
			\begin{tabular}{lccc}
				\textbf{Mesh Parameter} & \textbf{R1} & \textbf{R2} & \textbf{R3} \\
				\midrule
				Fan inlet sizing (mm) & 2 & 1.33 & 0.89 \\
				Refinement zone sizing (mm) & 10 & 6.67 & 4.44 \\
				Refinement zone -x (mm) & 215 & 215 & 215 \\
				Refinement zone +x (mm) & 215 & 215 & 215 \\
				Refinement zone -y (mm) & 215 & 215 & 215 \\
				Refinement zone +y (mm) & 215 & 215 & 215 \\
				Refinement zone -z (mm) & 40 & 40 & 40 \\
				Refinement zone +z (mm) & 400 & 400 & 400 \\
				Ratio ($r$) & - & 1.5 & 1.5 \\
				Global min. mesh size (mm) & 1 & 1 & 1 \\
				Global max. mesh size (mm) & 32 & 32 & 32 \\
				Growth rate & 1.2 & 1.2 & 1.2 \\
				Size function & curvature & curvature & curvature \\
				Curvature normal angle (deg) & 18 & 18 & 18 \\
				Volume mesh type & Poly-hexcore & Poly-hexcore & Poly-hexcore \\
				Min. orthogonal quality & 0.2076 & 0.2034 & 0.2019 \\
				\midrule
				Mesh size & 1,481,518 & 2,060,707 & 5,087,454 \\
				Centerpoint velocity (m/s) & 10.607 & 12.186 & 11.721 \\
				\midrule
				$p$ & \multicolumn{3}{c}{3.016} \\
			\end{tabular}
		\end{ruledtabular}
	\end{table*}
	
	\begin{figure*}[htbp]
		\centering
		\begin{minipage}[c]{0.58\textwidth}
			\centering
			\includegraphics[width=\linewidth]{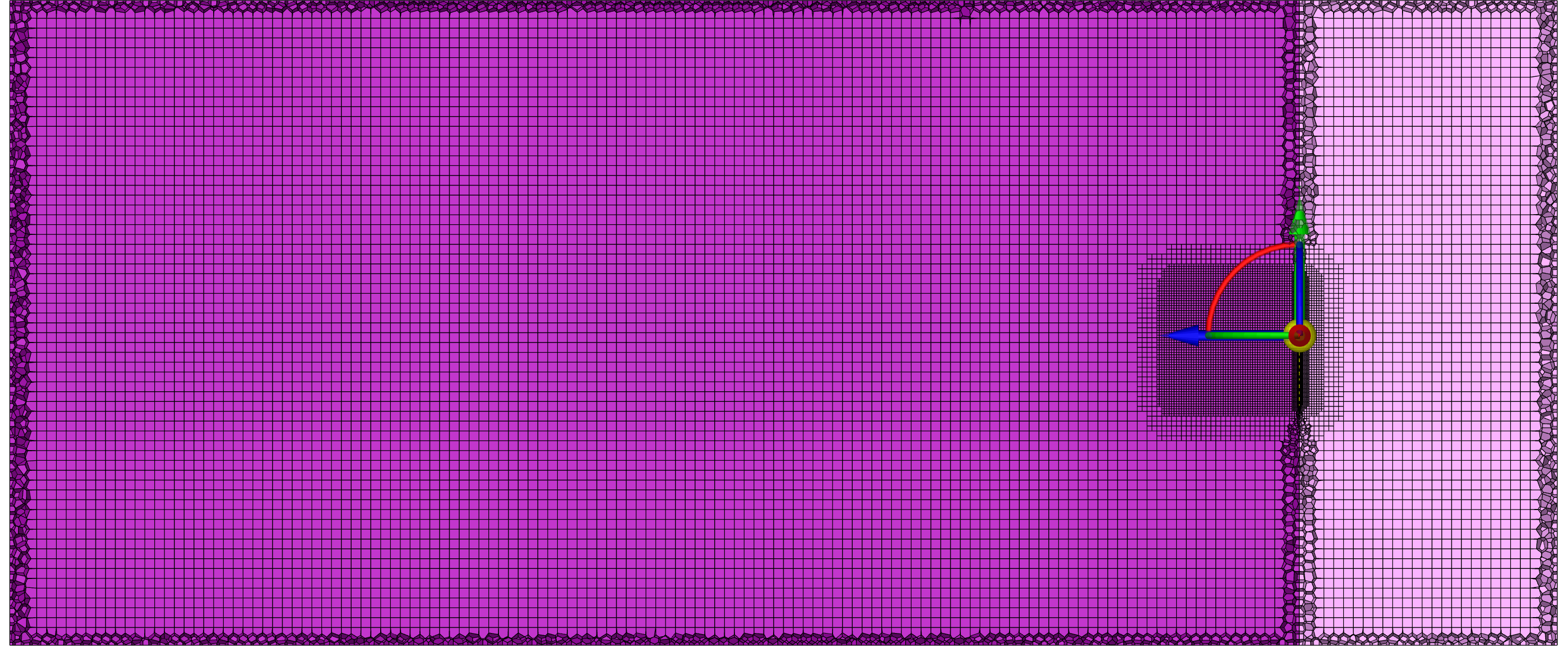}
			\vspace{5pt} \\ \small (a) Global mesh view
		\end{minipage}
		\hfill
		\begin{minipage}[c]{0.38\textwidth}
			\centering
			\includegraphics[width=\linewidth]{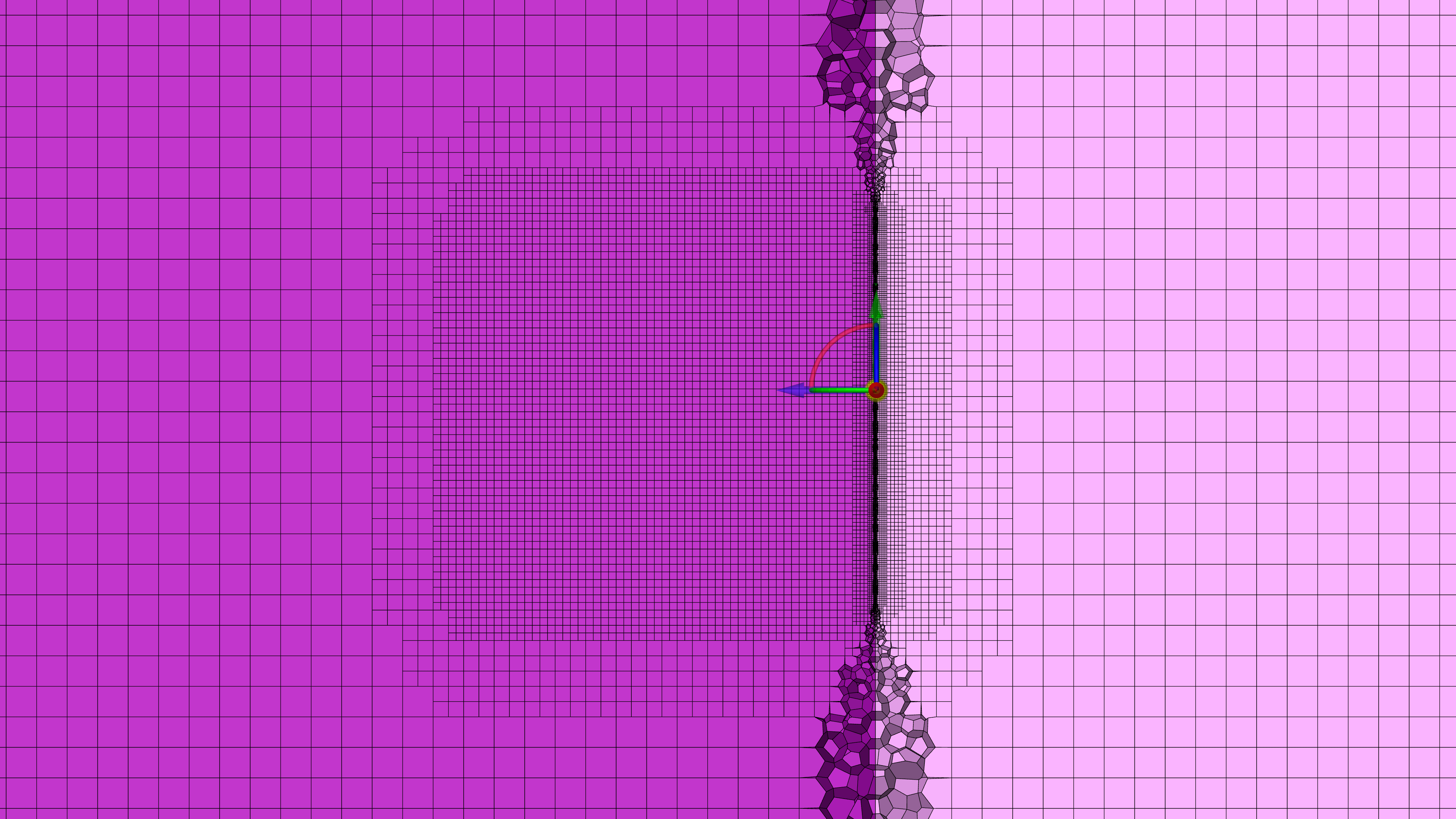}
			\vspace{5pt} \\ \small (b) Cross-sectional detail
		\end{minipage}
		\caption{Computational grid of the fluid domain. (a) Global view of the unstructured mesh at the main duct intersections, and (b) detailed cross-sectional view highlighting the structured inflation layers applied at the wall boundaries to resolve near-wall velocity gradients.}
		\label{fig:domain_mesh}
	\end{figure*}
	
	\subsection{Numerical Solution Procedure}
	
	All simulations are performed using a pressure-based, steady-state solver with an absolute velocity formulation. Air is modeled as an incompressible, single-phase fluid with constant properties. The density and dynamic viscosity are taken as $\rho = 1.225~\mathrm{kg/m^3}$ and $\mu = 1.7894\times10^{-5}~\mathrm{Pa\cdot s}$, respectively.
	
	Turbulence is modeled using the $k$--$\omega$ Shear Stress Transport (SST) model with default constants \cite{menter1994two}. This model is selected to provide a computationally feasible framework for predicting the mean flow structure in a configuration involving multiple interacting jets. Although scale-resolving approaches such as Large Eddy Simulation (LES) are more suitable for capturing transient shear-layer dynamics and anisotropic turbulence, their computational cost is prohibitive for the present study, considering the large number of jet cores.
	
	The primary flow quantities considered are the axial ($z$) velocity component and the local turbulence intensity. The axial velocity is obtained directly from the steady-state solution. The local turbulence intensity is defined as
	\begin{equation}
		TI = \frac{\sqrt{2k/3}}{|U|},
	\end{equation}
	where $k$ is the modeled turbulent kinetic energy and $U$ is the local mean velocity magnitude. This definition is consistent with the isotropic turbulence assumption inherent in eddy-viscosity models and enables direct comparison with hot-wire anemometry data reported in the reference study.
	
\begin{figure*}[htbp]
		\centering
		\begin{minipage}[b]{0.48\textwidth}
			\centering
			\includegraphics[width=\linewidth]{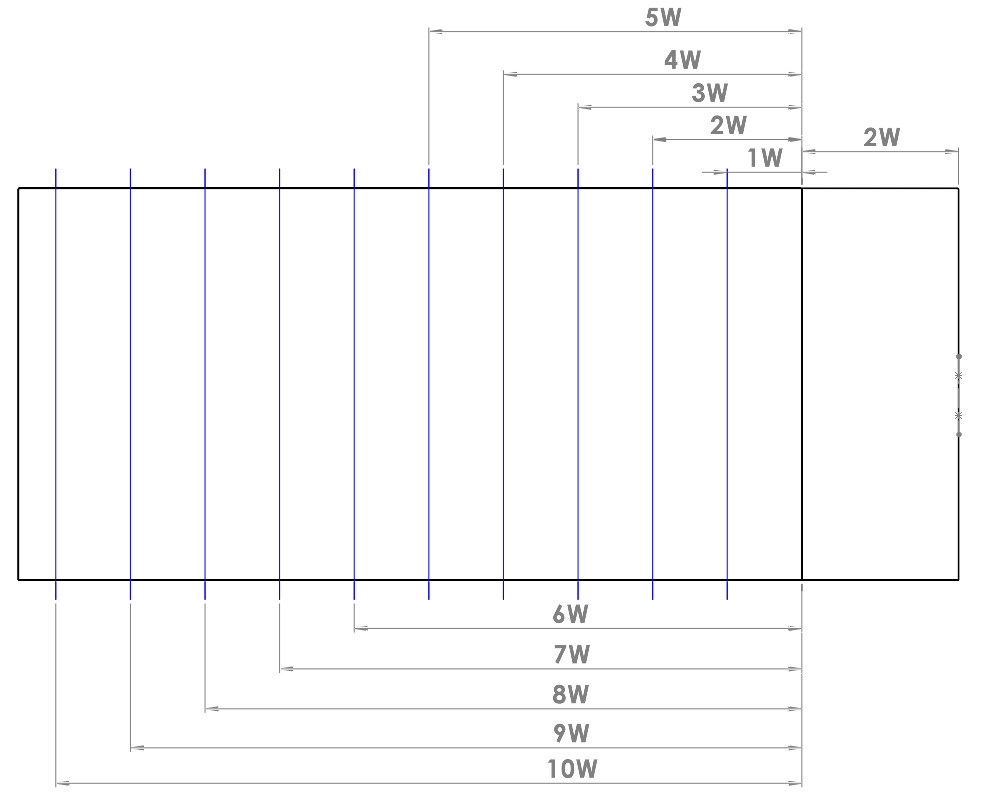}
			\vspace{5pt} \\ \small (a) Side view of the domain
		\end{minipage}
		\hfill
		\begin{minipage}[b]{0.48\textwidth}
			\centering
			\includegraphics[width=\linewidth]{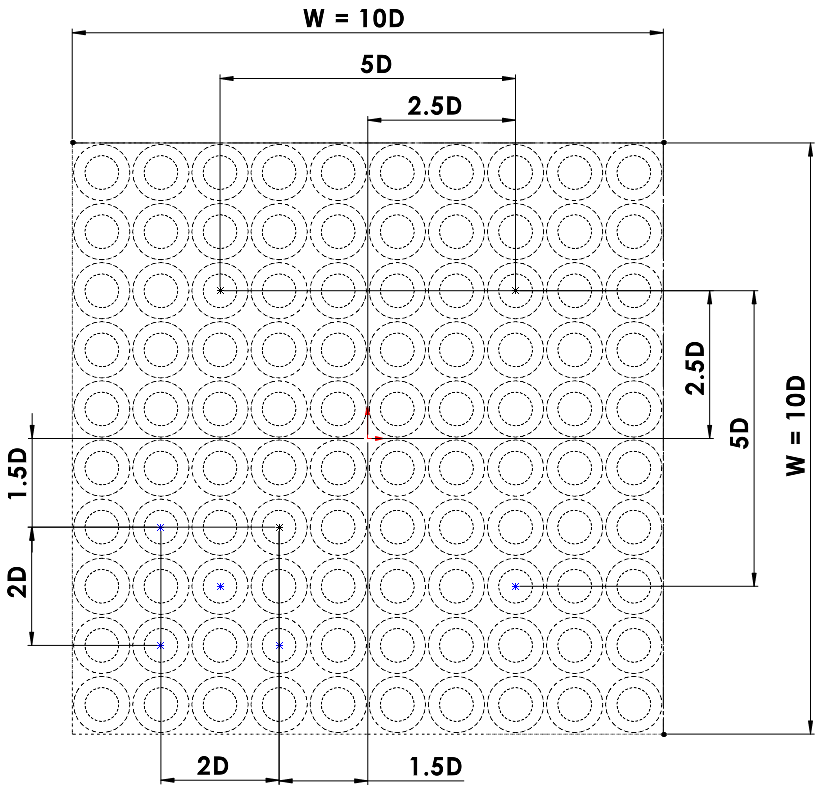}
			\vspace{5pt} \\ \small (b) Point Locations in the XY Plane
		\end{minipage}
		\caption{Geometric locations used for spatial data extraction.}
		\label{fig:measurement_locations}
	\end{figure*}

	To ensure direct comparability with the experimental data, all flow quantities are evaluated at the same measurement locations and planes. The coordinates of these points are listed in Table~\ref{tab:measurement_points} and illustrated in Figure~\ref{fig:measurement_locations}. For the coarse chessboard arrangement (Pattern B), results are presented as pointwise distributions of axial velocity and turbulence intensity. For Pattern D, contour plots of the same quantities are reported on selected planes. Since the focus of the present work is on the near- and mid-field jet interaction region, the analysis is restricted to the first 10 planes, corresponding to a downstream extent of $10W$. 
	
	\begin{table}[htbp]
		\centering
		\caption{Point locations used for data extraction.}
		\label{tab:measurement_points}
		\begin{ruledtabular}
			\begin{tabular}{cccc}
				\textbf{Set} & \textbf{Name} & \textbf{X} & \textbf{Y} \\
				\midrule
				Set 1 & s1p1 & $-2.5D$ & $ 2.5D$ \\
				& s1p2 & $ 2.5D$ & $ 2.5D$ \\
				& s1p3 & $-2.5D$ & $-2.5D$ \\
				& s1p4 & $ 2.5D$ & $-2.5D$ \\
				\midrule
				Set 2 & s2p1 & $-3.5D$ & $-1.5D$ \\
				& s2p2 & $-1.5D$ & $-1.5D$ \\
				& s2p3 & $-3.5D$ & $-3.5D$ \\
				& s2p4 & $-1.5D$ & $-3.5D$ \\
			\end{tabular}
		\end{ruledtabular}
	\end{table}
	
	\subsection{Model Sensitivity to Fan Speed and Upstream Turbulence}
	
	Two additional fan operating conditions are considered, $1.34N$ and $2N$ where $N$ corresponds to the speed of the manufacturer test data. The associated fan curves are scaled using affinity laws to obtain the pressure jump characteristics at each speed (Figure~\ref{fig:fan_curves_combined}). This approach allows for a consistent assessment of the effect of jet speed on the generated FAWG flow.
	
	In addition, a set of additional simulations was conducted by varying the turbulence intensity and turbulent viscosity ratio independently to assess the sensitivity of the flow to inlet turbulence specifications. A reference case was defined with $\mathrm{TI} = 5\%$ and $\mathrm{TVR} = 10$. The effect of turbulence intensity was investigated by considering $\mathrm{TI} = 1\%$ and $25\%$ while keeping $\mathrm{TVR} = 10$ fixed. Conversely, the influence of turbulent viscosity ratio was examined by varying $\mathrm{TVR} = 1$ and $50$ at a constant $\mathrm{TI} = 5\%$. In two-equation RANS models, the inlet turbulence state is defined by two independent parameters. Turbulence intensity is selected due to its direct experimental relevance, with values reaching up to $25\%$ upstream of the FAWG, while the turbulent viscosity ratio is varied to control the initial eddy viscosity level, thereby influencing the relative strength of modeled turbulent transport and diffusion.
	
	\subsection{Demonstrative Application: Flat Plate in FAWG Flow}
	
	To illustrate the aerodynamic implications of FAWG-generated inflow conditions, a canonical flat plate configuration is considered as a demonstrative application. The objective is to examine how the non-uniform, turbulent flow produced by the fan array influences the response of a simple lifting surface.
	
	The plate geometry is based on the low-Reynolds-number experimental study of Torres and Mueller \cite{torres2004}. The model is a low-aspect ratio ($AR=1.0$) square plate with a chord length of 300 mm, a thickness of 5.88 mm, and a 5:1 elliptical leading edge, followed by a blunt trailing edge. A single case with an angle of attack $\alpha=2\degree$ is considered. 
	
	The flow conditions correspond to a Reynolds number of approximately $Re \approx 1.1 \times 10^5$, consistent with the mean velocity levels generated by the FAWG. At this Reynolds number, transitional effects are expected to play a significant role. To account for this behavior, the $k$--$\omega$ SST model with the $\gamma$-transition formulation is activated in the code.
	
	To isolate the effect of the inflow conditions, two configurations are considered. In the first, the plate is placed within the FAWG-generated flow field. In the second, the same plate is simulated in a uniform freestream domain under equivalent Reynolds number conditions. Same problem domain is used, with the FAWG zone removed and the upstream boundary set as prescribed velocity boundary.
	
	The aerodynamic response is evaluated in terms of lift and drag coefficients, and further examined through comparisons of surface flow features, including velocity and shear distributions. By contrasting the FAWG and uniform inflow cases, this example highlights the influence of jet interaction and elevated turbulence levels on the predicted aerodynamic behavior.
	
	\section{Results and Discussion}   
	
	\subsection{Baseline Validation}
	
	\subsubsection{Spatial Distributions}
	
	A qualitative comparison was first conducted using axial ($z$-direction) velocity and turbulence intensity contours. The predicted flow structures capture the general macro-scale topology reported in the reference paper, particularly for patterns B and D \cite{arc2025WS}. While exact spatial matching is not achieved due to the inherent nature of the RANS formulation, the fundamental physical mechanisms -specifically spatial jet core development- are structurally consistent with the experimental observations. The axial velocity contours reproduce the expected core region and shear layer development, while the turbulence intensity contours capture the signature of turbulent mixing downstream of the fan exit. Other arrangements were not further analyzed at this stage, as their flow topology did not exhibit additional distinguishing characteristics relevant to the objectives of this study. Hence, the global flow topology and overall jet merging behavior are qualitatively reproduced, capturing the relevant aspects of the FAWG array flow.
	
	\begin{table*}[htbp]
		\centering
		\caption{Comparison of reference and CFD results for two different fan activation patterns at $z/D=10$. The scales are same for both figures. Reproduced from \cite{arc2025WS} under the Creative Commons Attribution (CC BY 4.0) license.}
		\label{tab:activation_patterns}
		\begin{ruledtabular}
			\begin{tabular}{ccc}
				Pattern & Reference & This study \\
				\midrule
				
				\multirow{2}{*}{\includegraphics[width=2.5cm]{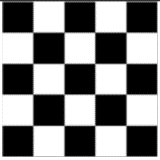}} & 
				\includegraphics[width=6cm]{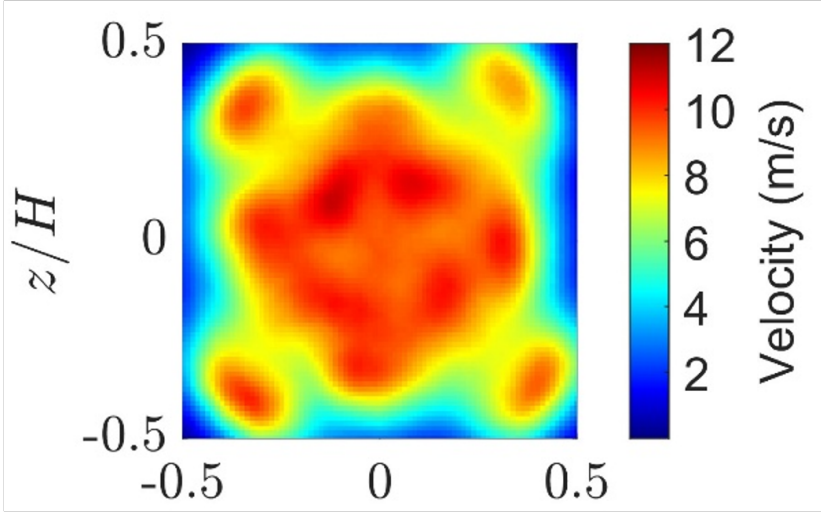} & 
				\includegraphics[width=6cm]{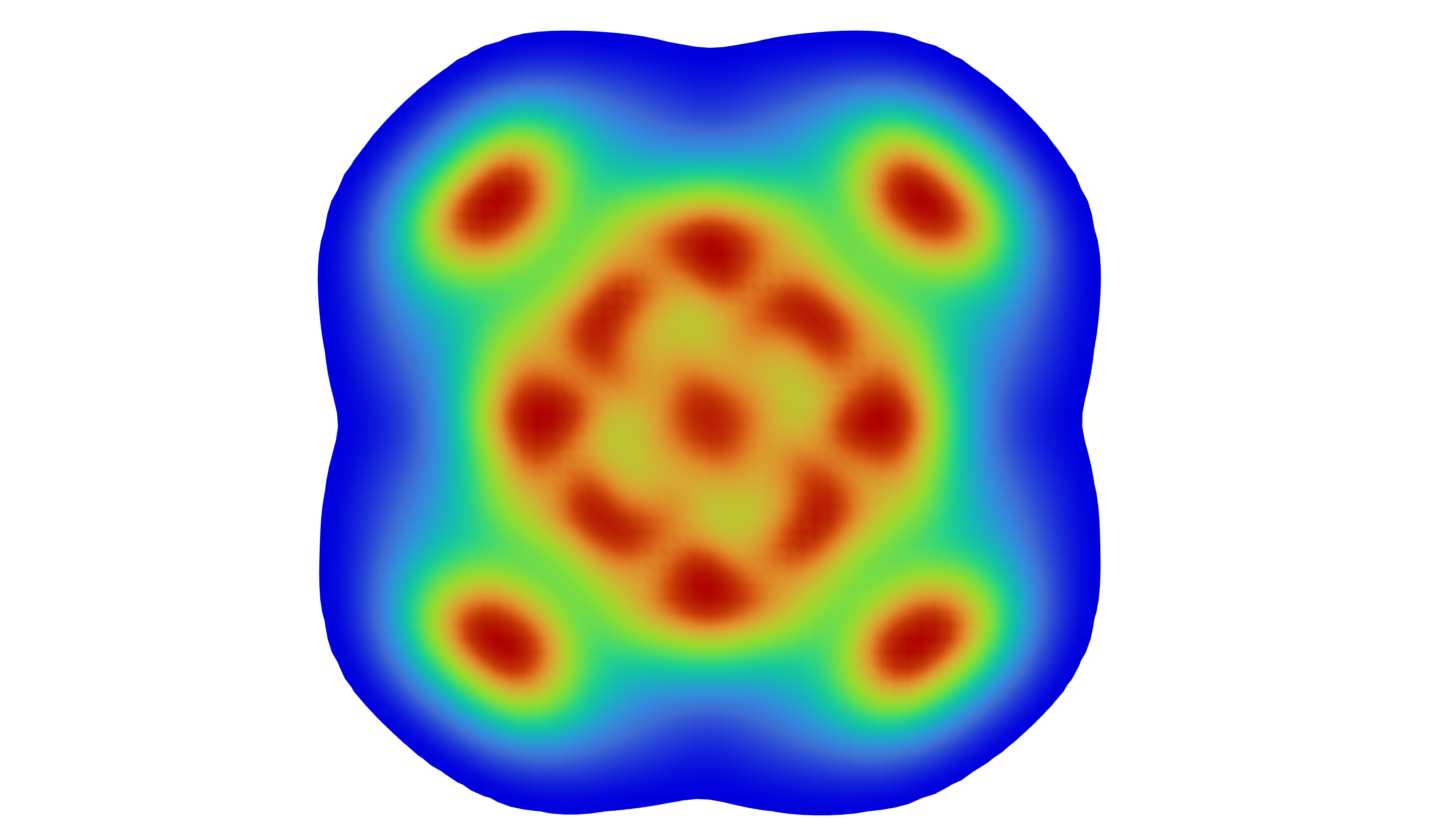} \\
				& 
				\includegraphics[width=6cm]{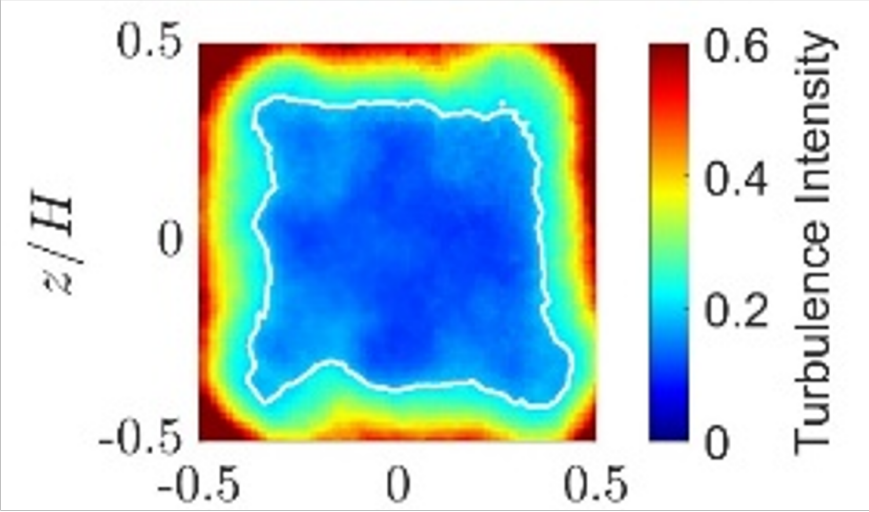} & 
				\includegraphics[width=6cm]{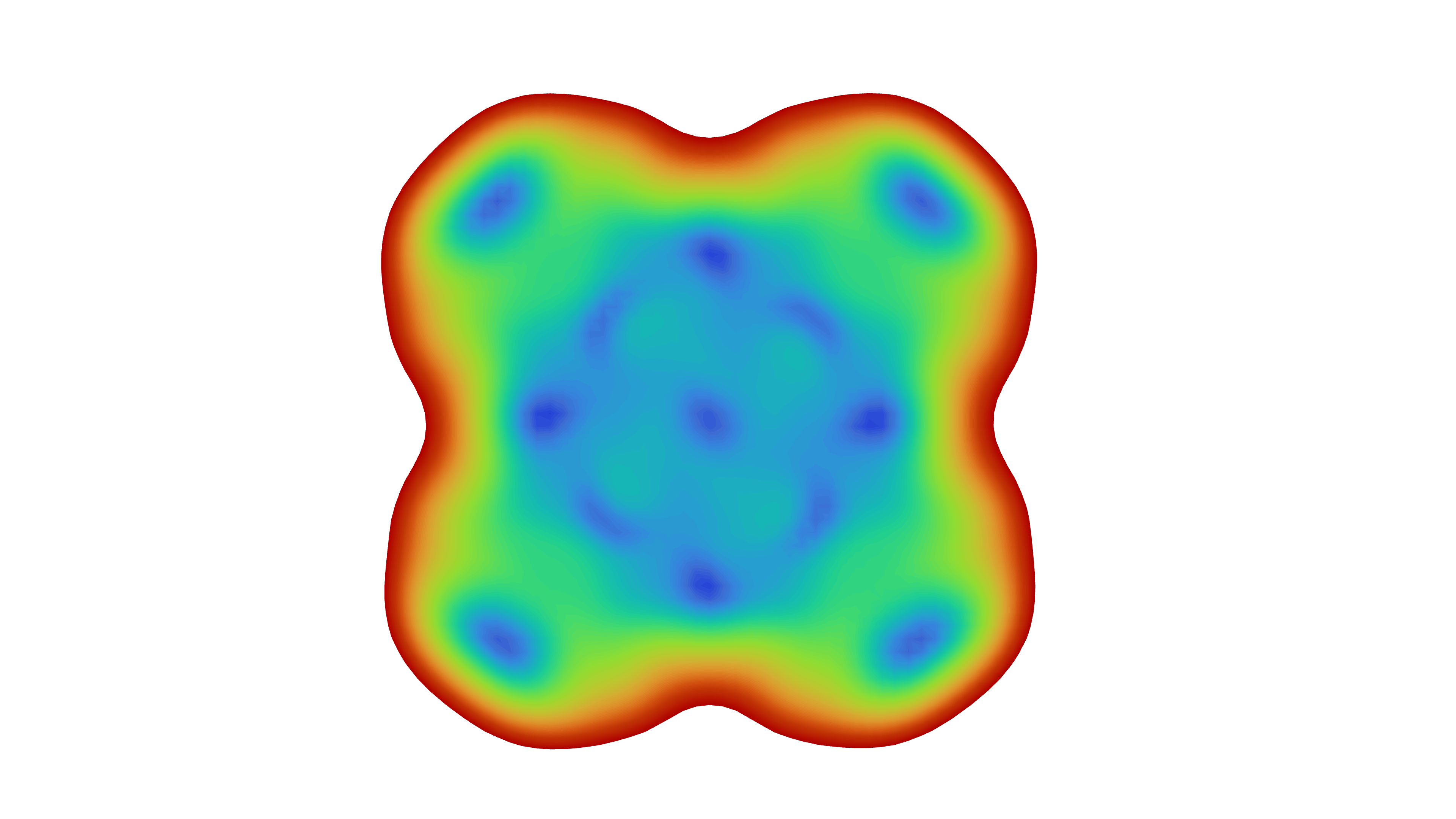} \\
				\midrule
				
				\multirow{2}{*}{\includegraphics[width=2.5cm]{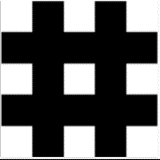}} & 
				\includegraphics[width=6cm]{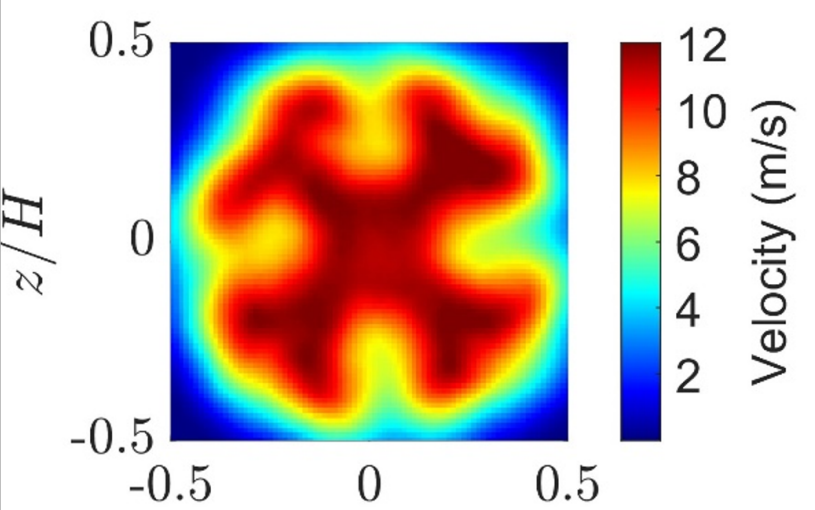} & 
				\includegraphics[width=6cm]{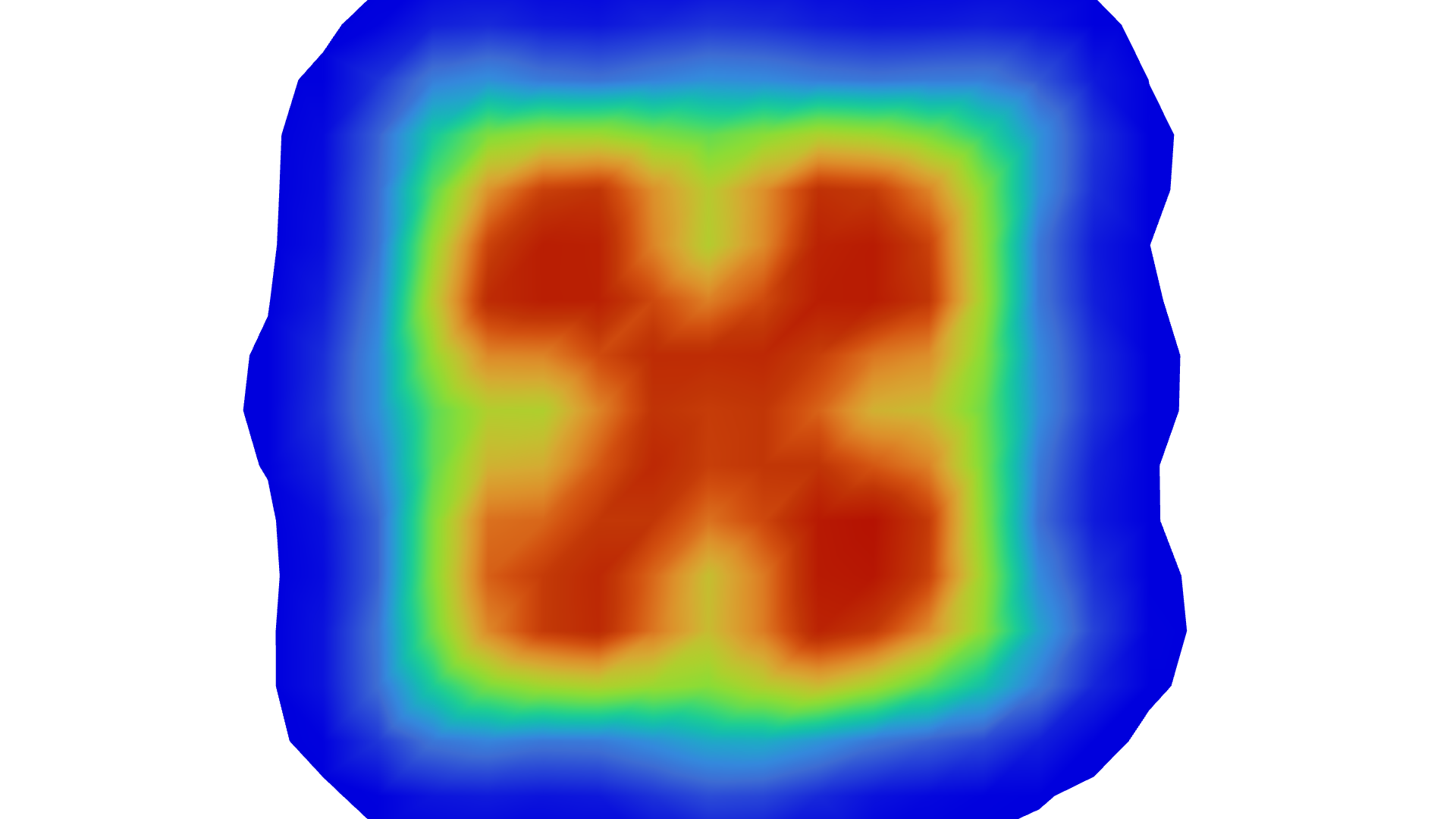} \\
				& 
				\includegraphics[width=6cm]{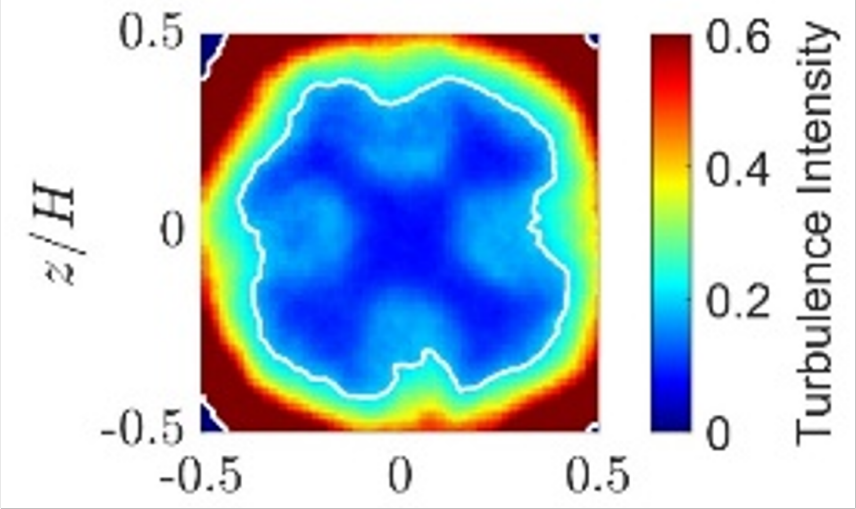} & 
				\includegraphics[width=6cm]{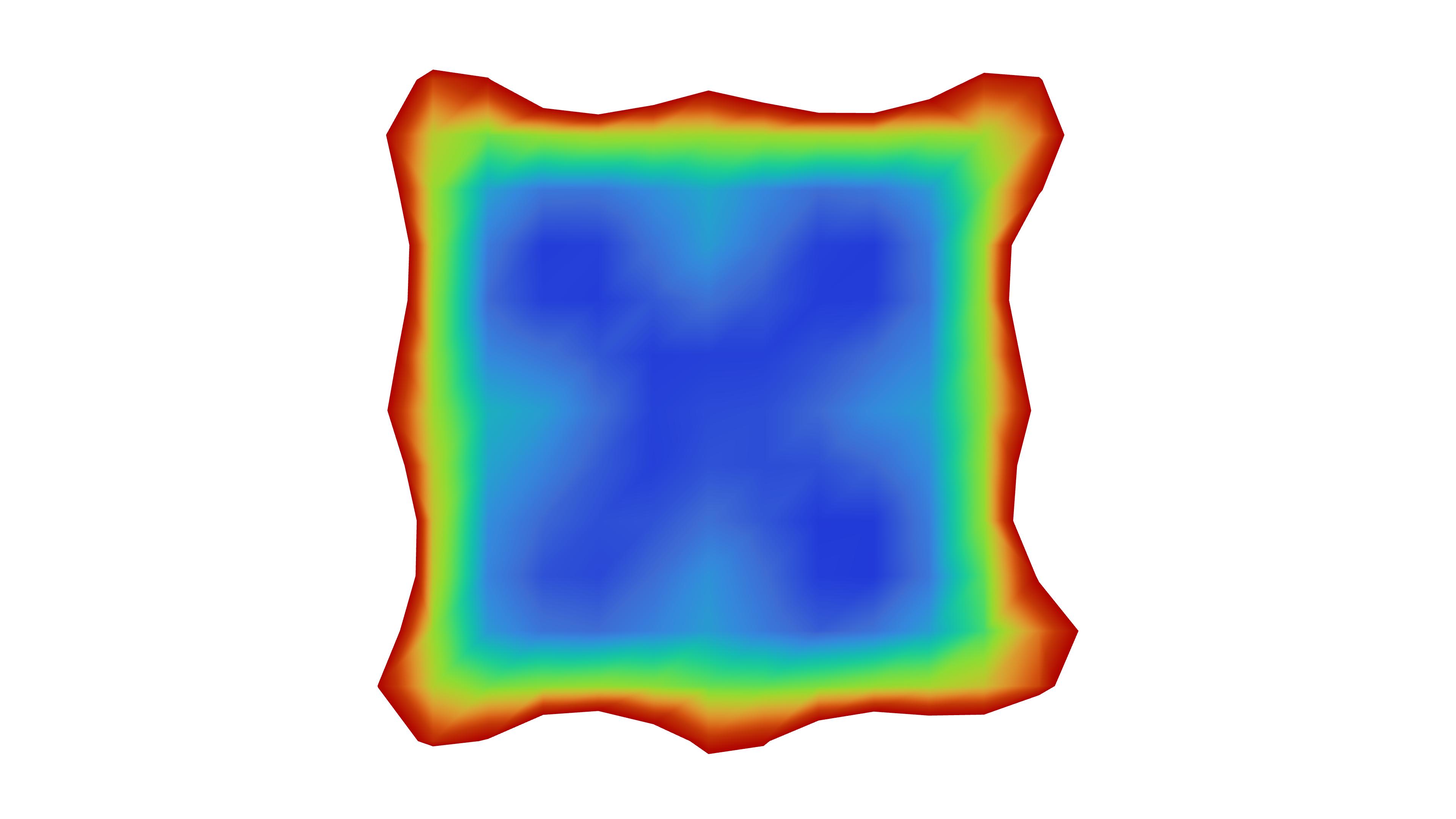} \\
			\end{tabular}
		\end{ruledtabular}
	\end{table*}
	
	\subsubsection{Pointwise Comparison}
	
\begin{figure*}[htbp]
		\centering
		\begin{minipage}[b]{\textwidth}
			\centering
			\includegraphics[width=0.9\textwidth, height=0.42\textheight, keepaspectratio]{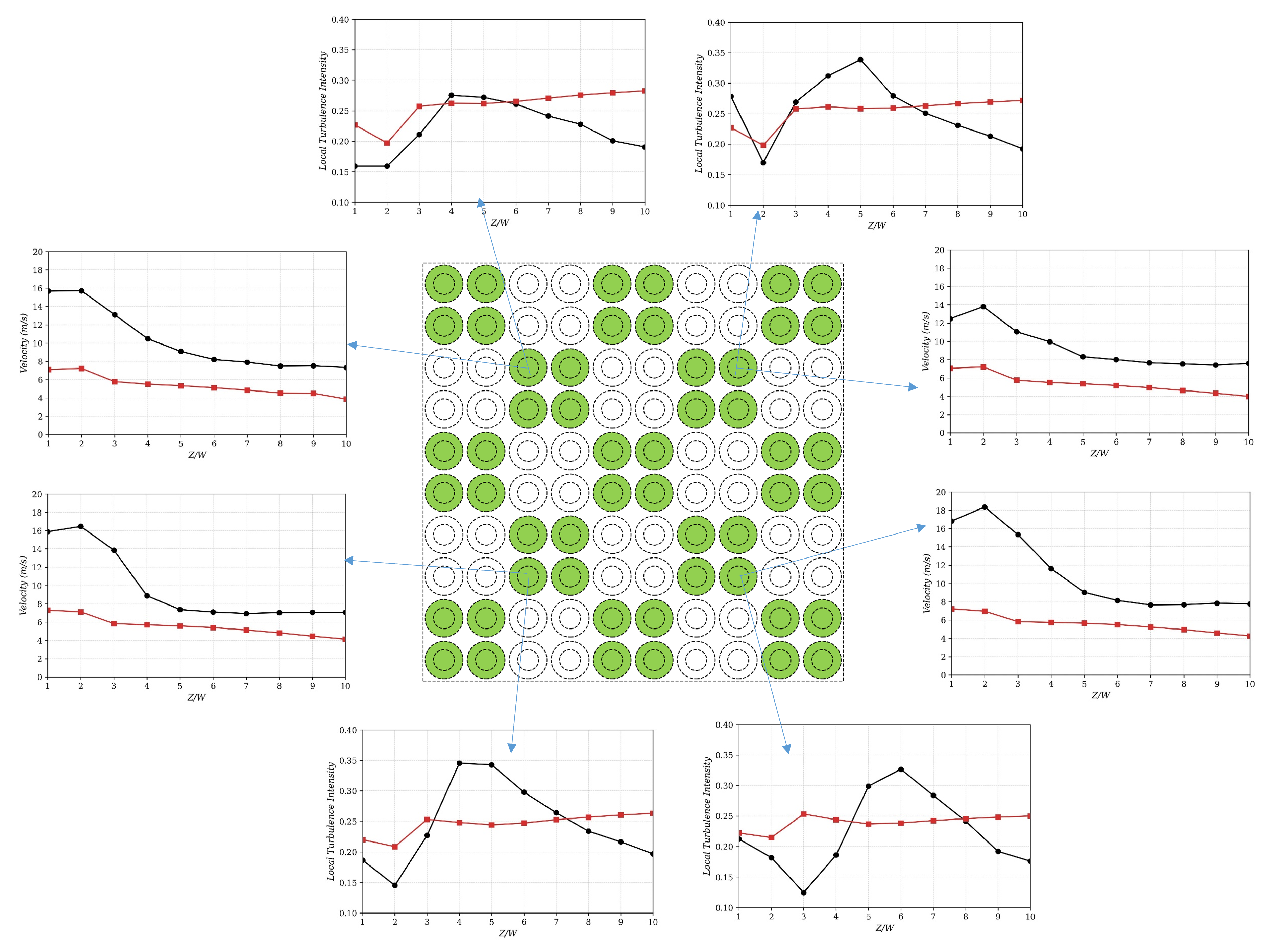}
			\vspace{5pt} \\ \small (a) Central core region (Set 1)
		\end{minipage}
		
		\vspace{10pt}
		
		\begin{minipage}[b]{\textwidth}
			\centering
			\includegraphics[width=0.9\textwidth, height=0.42\textheight, keepaspectratio]{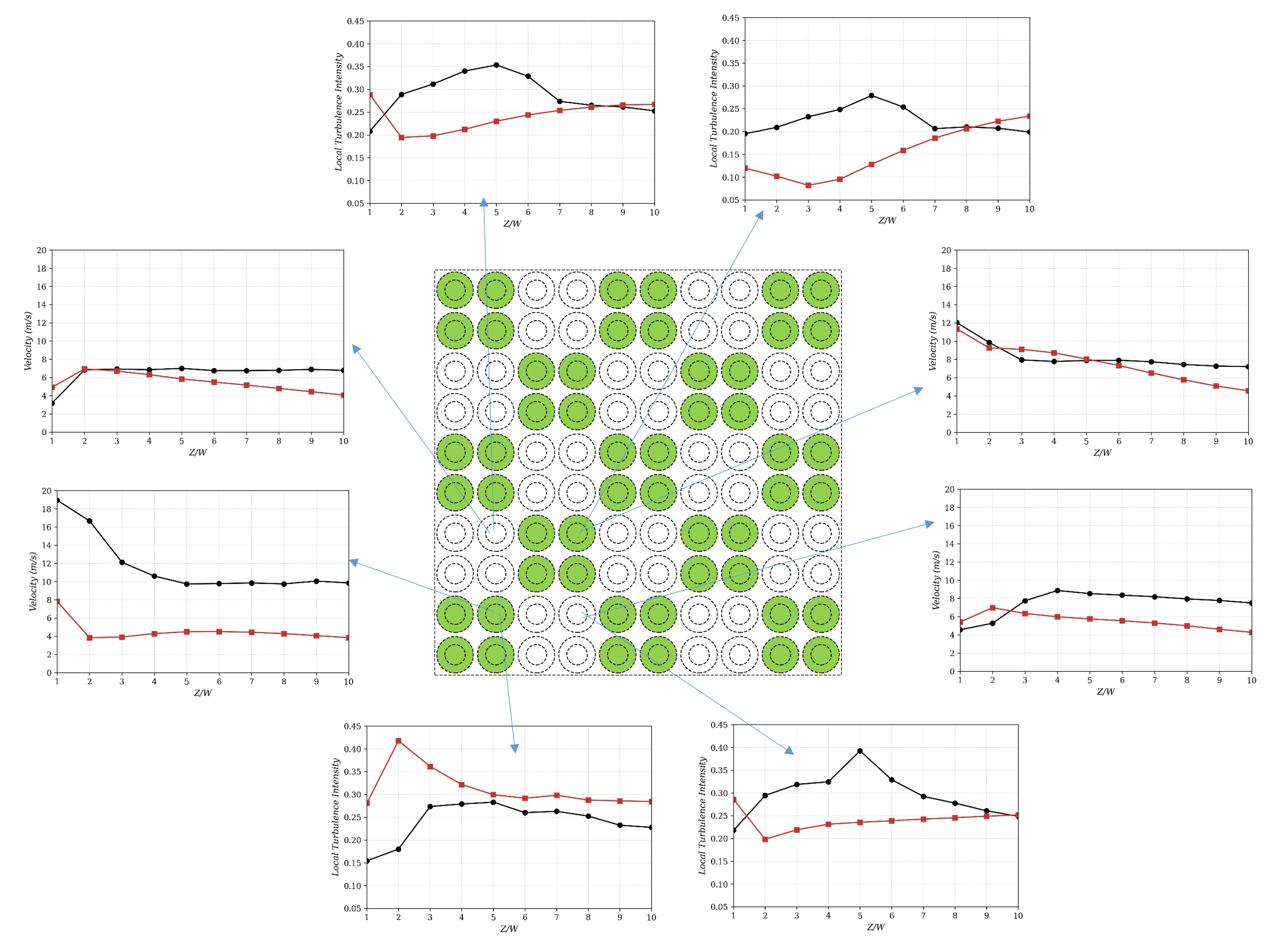}
			\vspace{5pt} \\ \small (b) Peripheral shear-layer region (Set 2)
		\end{minipage}
		
		\caption{Axial velocity and turbulence intensity distributions for Pattern B. CFD (Red) vs experiment (Black). (a) Central region. (b) Peripheral region.}
		\label{fig:master_spatial_validation}
	\end{figure*}

	Following the qualitative assessment, pointwise data were extracted at the measurement locations and compared with the experimental results that were digitized from reference \cite{arc2025WS} (Figure~\ref{fig:master_spatial_validation}).   
	
	In the central jet core (Set 1), the RANS model captures the overall downstream decay of axial velocity. The trend is monotonic and consistent with the experimental data, indicating that the dominant momentum transport mechanisms are reasonably represented. However, discrepancies are observed in the near-field region (Planes 1--3), where the numerical solution underpredicts the peak velocity magnitude and produces a smoother profile compared to the sharper experimental peak.
	
	In the peripheral region (Set 2), the flow is expected to be more dominated by shear-layer interactions. The numerical results show a continuous downstream decay as before, whereas the experimental data exhibits an apparent plateau in the far field. This difference may be associated with measurement limitations at low velocity levels, as well as differences in how shear-layer mixing is represented by RANS simulations. Additional localized deviations are observed at certain points (e.g., Set 2, Point 3), where the experimental data shows non-monotonic behavior that is not captured by the numerical model. Overall, axial velocity predictions are satisfactory in the core region and exhibit moderate deviations in shear-dominated regions.
	
	Turbulence intensity levels in similar to the experiments, but they show larger local discrepancies compared to axial velocity. In the central region, the numerical model captures the initial level in TI; however, the experimental data exhibits a pronounced peak in the mid-field (Planes 4--6) followed by decay, while the numerical results produce a smoother distribution without a clear peak. This indicates that the model diffuses turbulence more strongly than observed experimentally.
	
	In the peripheral region, the discrepancies are more pronounced than the axial velocity. The experimental data shows localized peaks  whereas the numerical results underpredict these peaks and display a more gradual variation. These differences are consistent with the limitations of eddy-viscosity models in representing highly anisotropic and mixing-dominated flows. The observed discrepancies in turbulence intensity may also be attributed to a combination of factors, including the simplified model of the fan jet flow, sensitivity to boundary conditions and grid resolution in regions of strong gradients. Compared to axial velocity, TI is more sensitive to modeling assumptions, resulting in larger deviations in both magnitude and spatial distribution.    The mean relative error (MRE) is approximately 17.6\% in the central region and 15.8\% in the periphery. Overall, the results indicate that the RANS framework provides a reasonable prediction of mean flow behavior, while exhibiting expected limitations in representing detailed turbulence structure in shear-layer regions.    
	
	\subsection{Influence of Fan Representation (Surface vs.\ Ducted)}
	
	The influence of fan geometry representation is examined through comparison of longitudinal and cross-sectional velocity fields. 
	
\begin{figure*}[htbp]
		\centering
		\setlength{\tabcolsep}{1pt} 
		\begin{tabular}{c c c}
			~ & \textbf{Surface Model} & \textbf{Ducted Model} \\ [5pt]
			
			\multirow{2}{*}{\begin{minipage}{0.20\textwidth}
					\centering
					\includegraphics[width=0.9\linewidth]{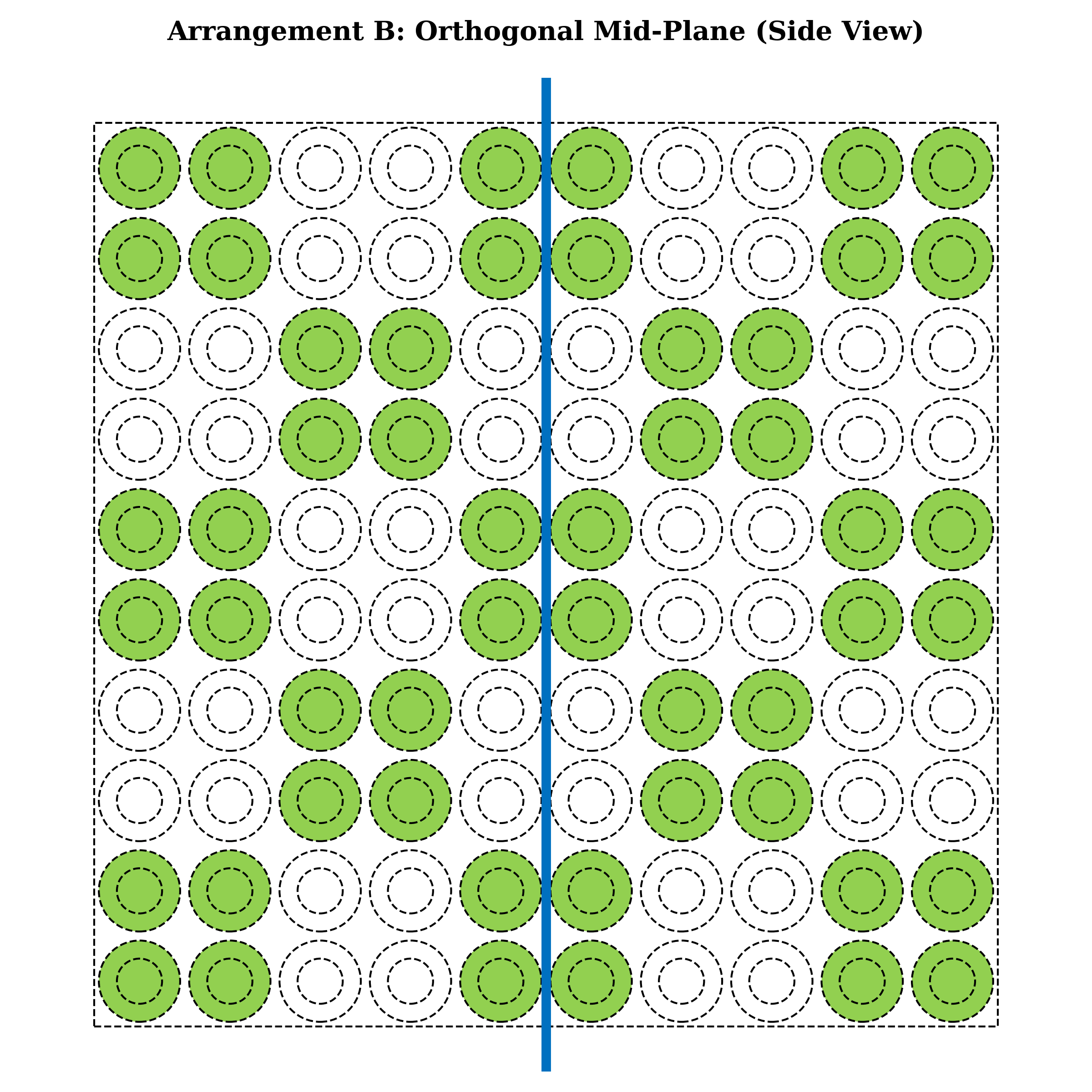} \\ 
					\small Vertical Mid-Plane \\[15pt] 
					\includegraphics[height=3cm]{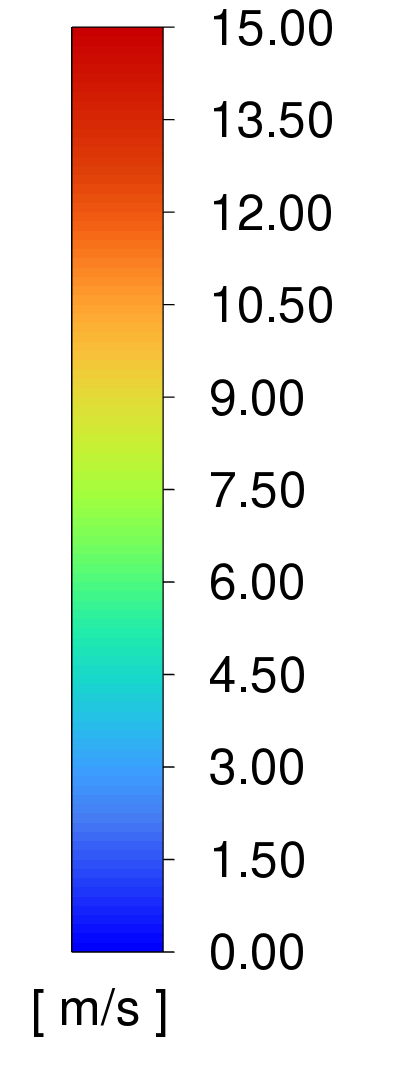}
			\end{minipage}} &
			\includegraphics[width=0.38\textwidth]{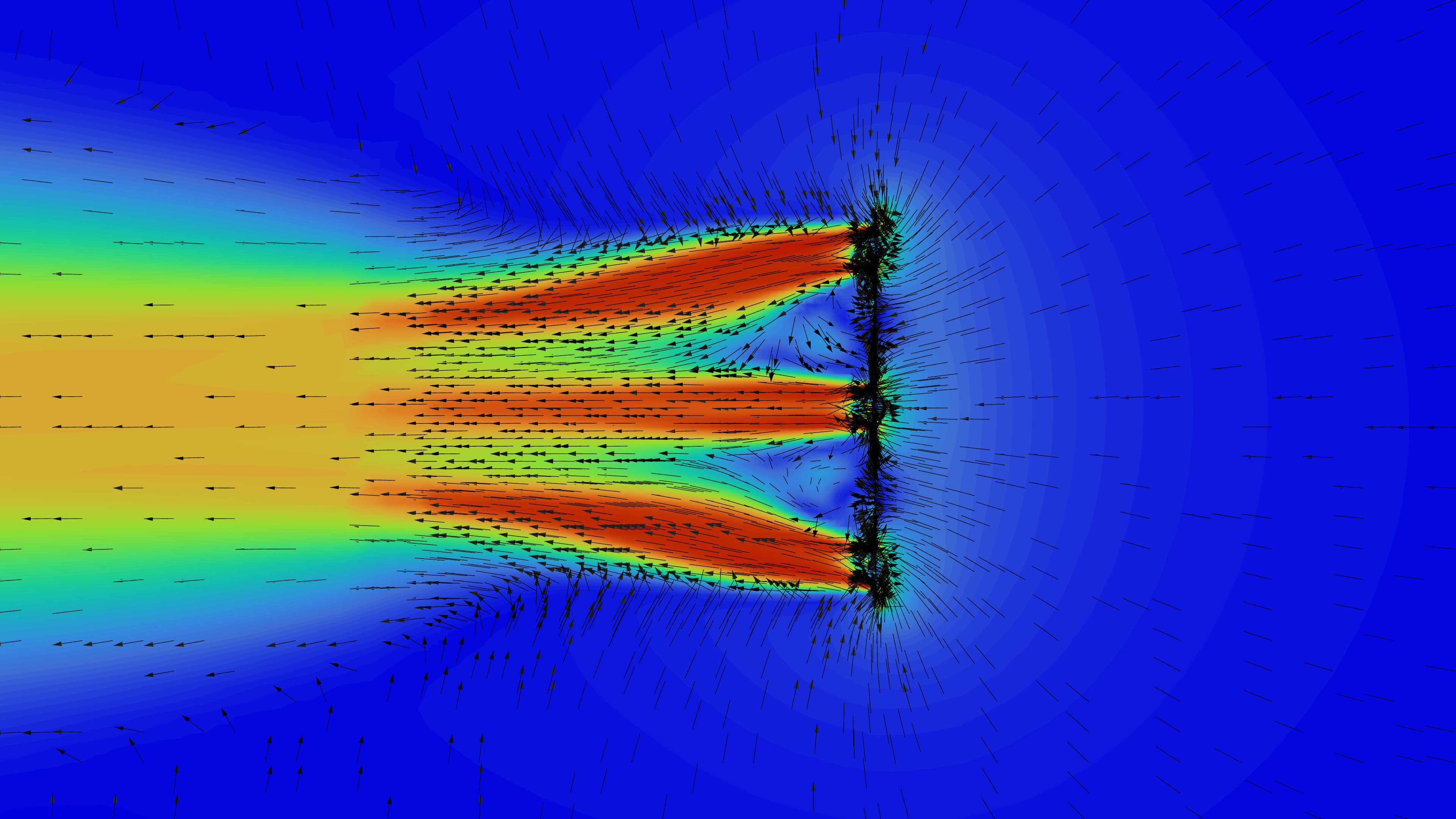} &
			\includegraphics[width=0.38\textwidth]{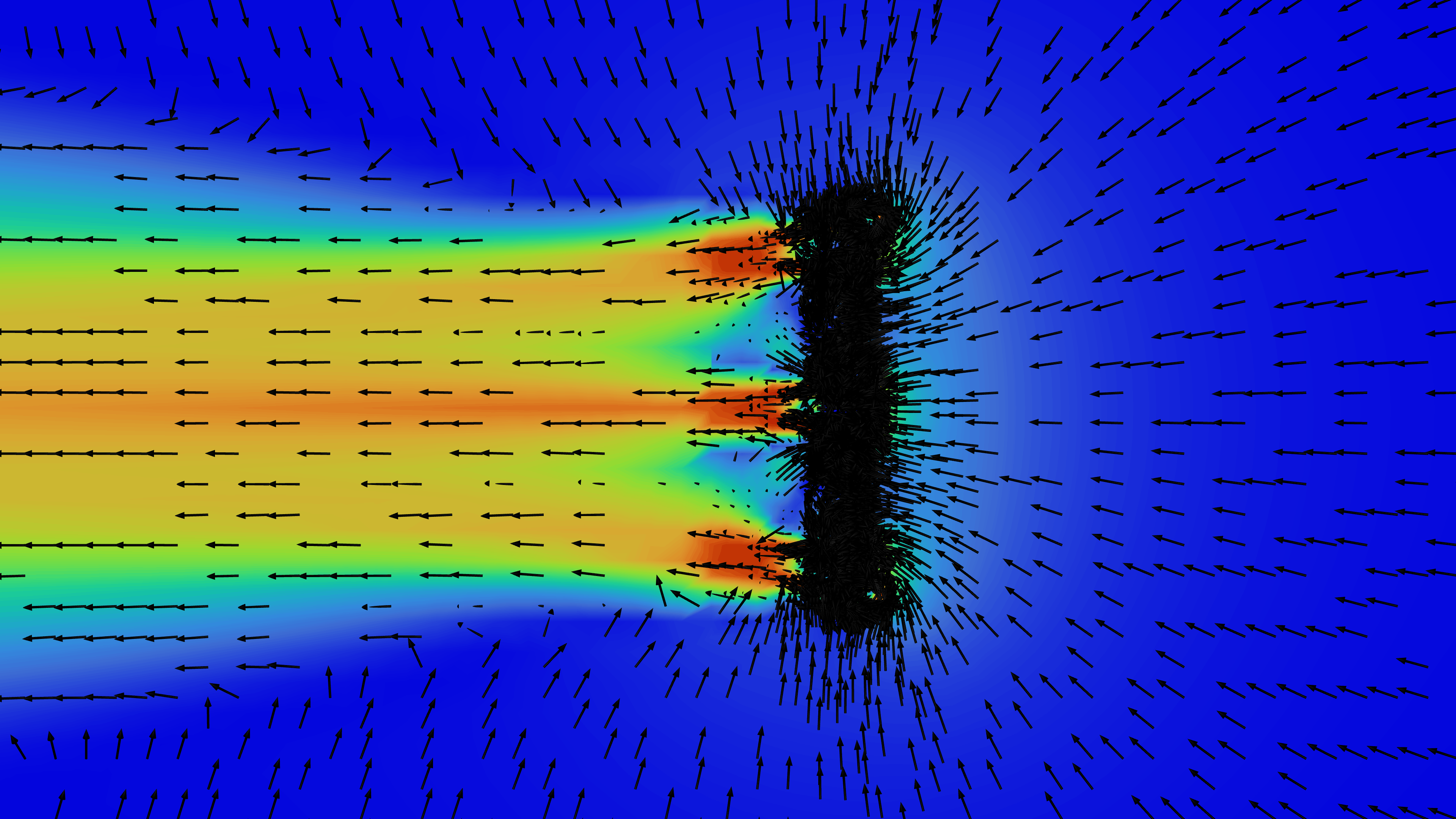} \\
			~ & \small (a) Near-field side view & 
			\small (b) Near-field side view \\[10pt]
			
			~ & \includegraphics[width=0.38\textwidth]{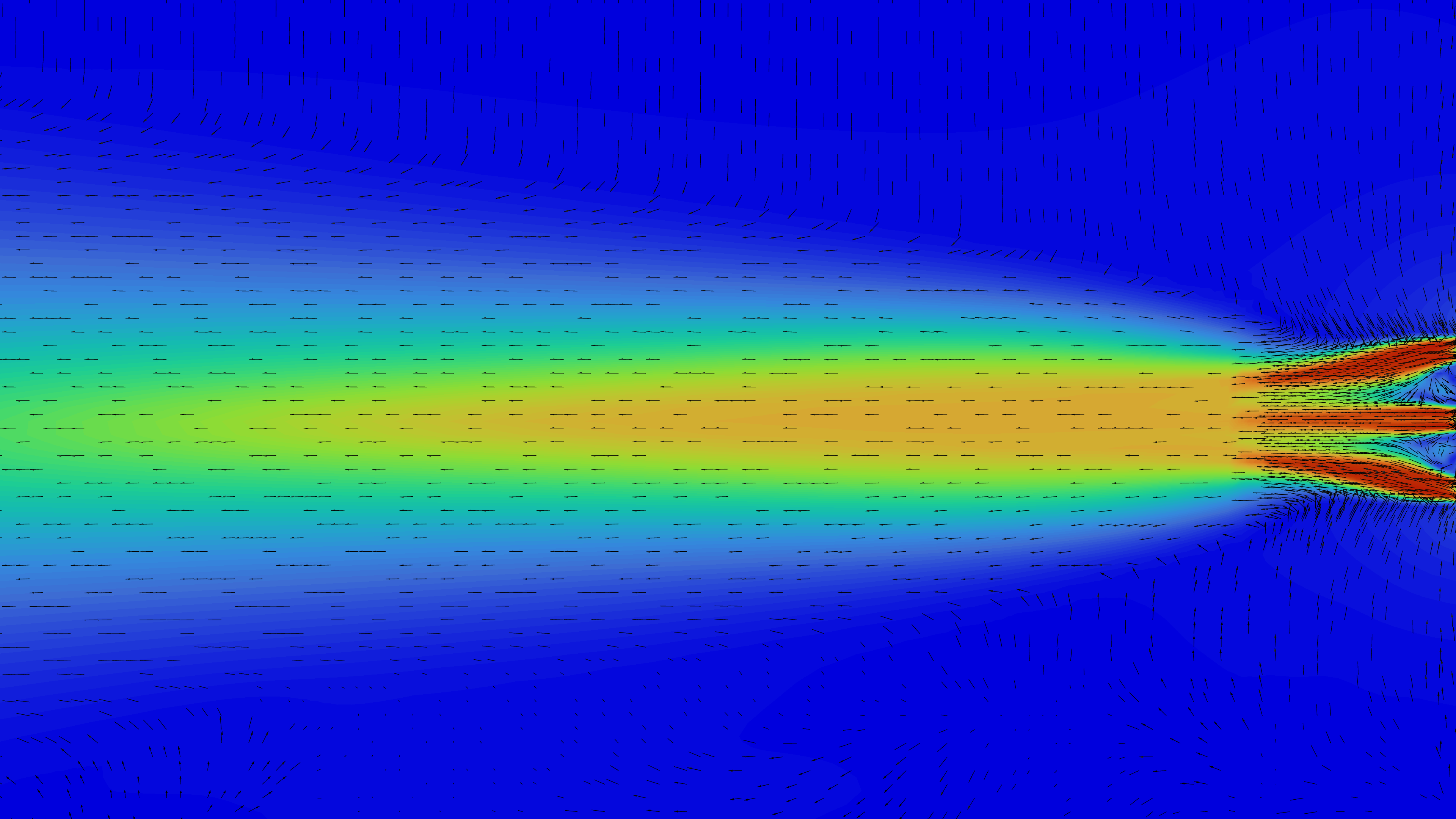} &
			\includegraphics[width=0.38\textwidth]{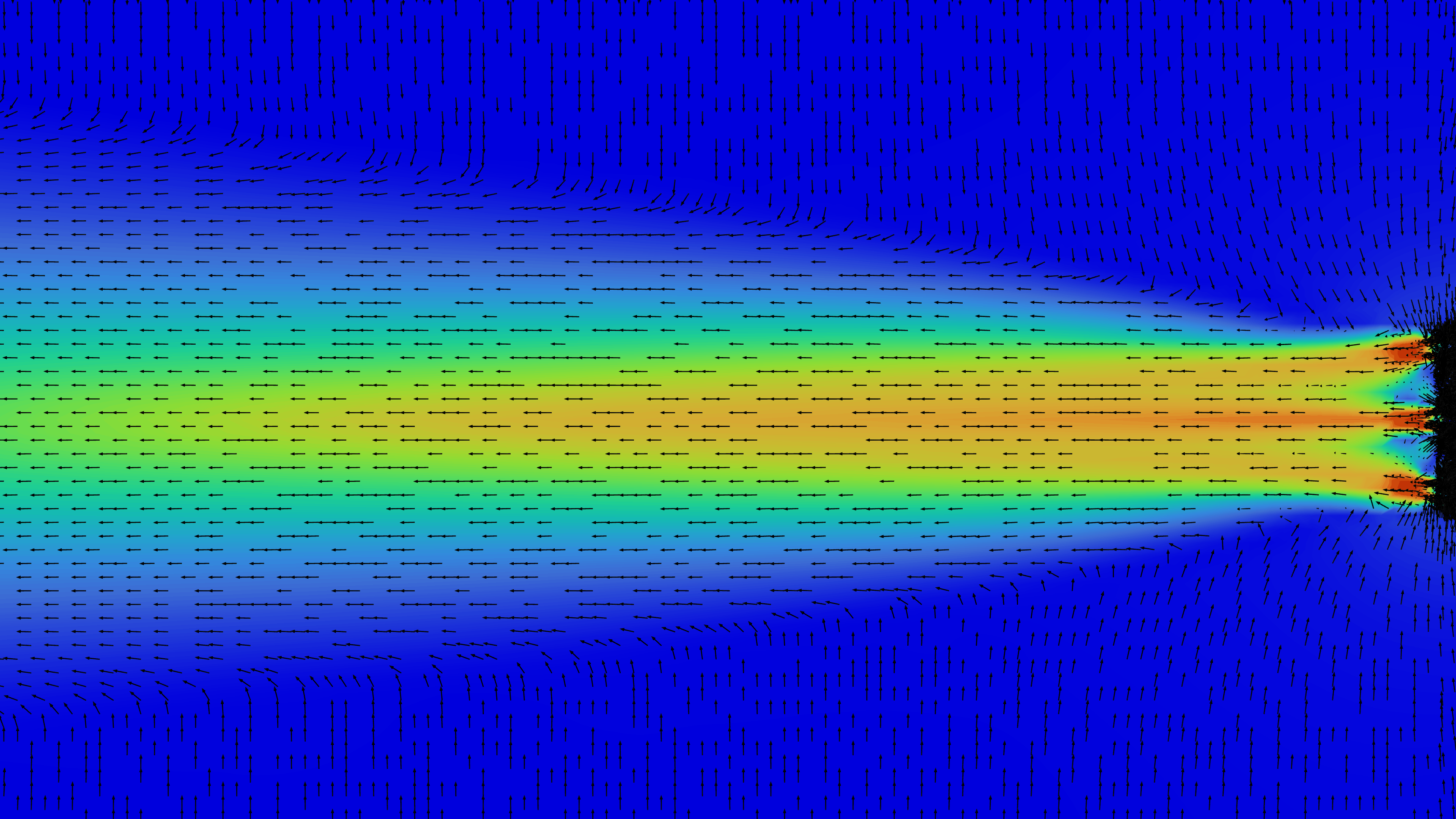} \\
			~ & \small (c) Far-field side view & 
			\small (d) Far-field side view \\[15pt]
			
			\\
			~ & \textbf{Surface Model} & \textbf{Ducted Model} \\ [5pt]
			
			\multirow{2}{*}{\begin{minipage}{0.20\textwidth}
					\centering
					\includegraphics[width=0.9\linewidth]{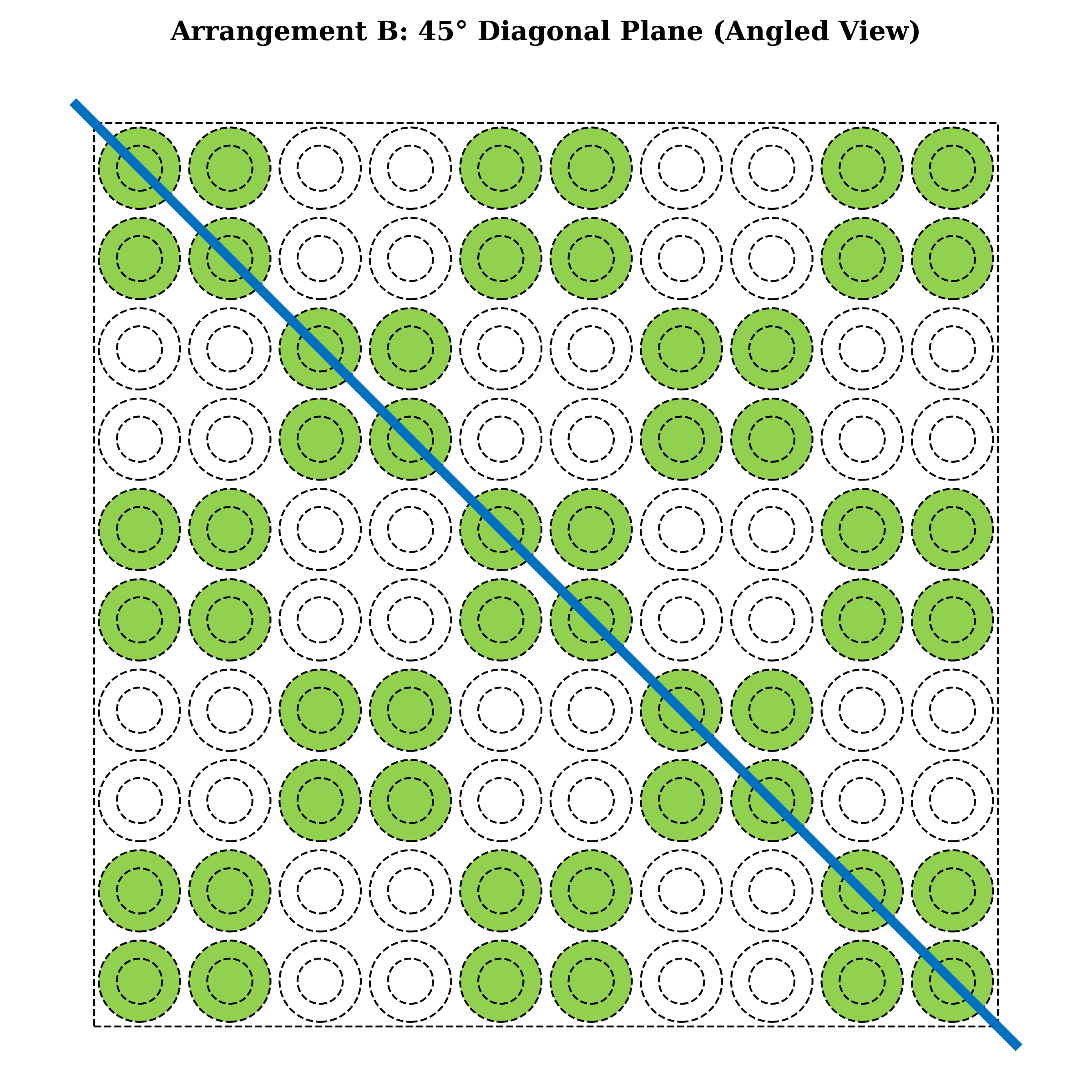} \\ 
					\small $45^\circ$ Diagonal Plane \\[15pt] 
					\includegraphics[height=3cm]{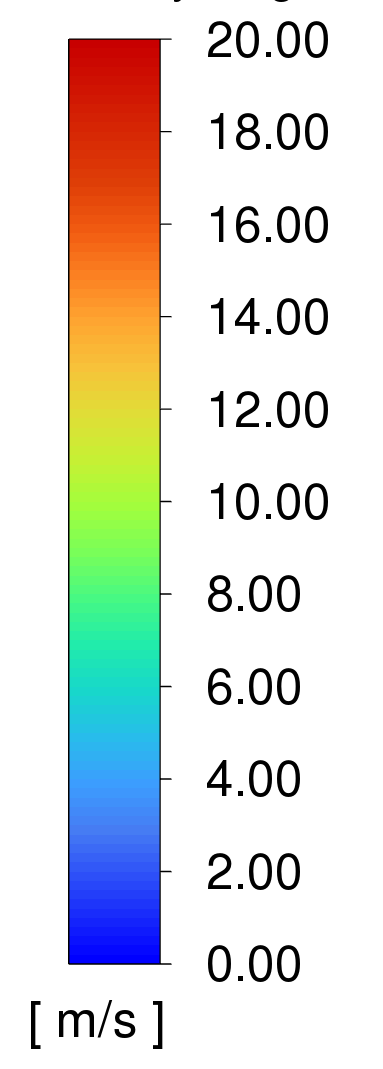}
			\end{minipage}} &
			\includegraphics[width=0.38\textwidth]{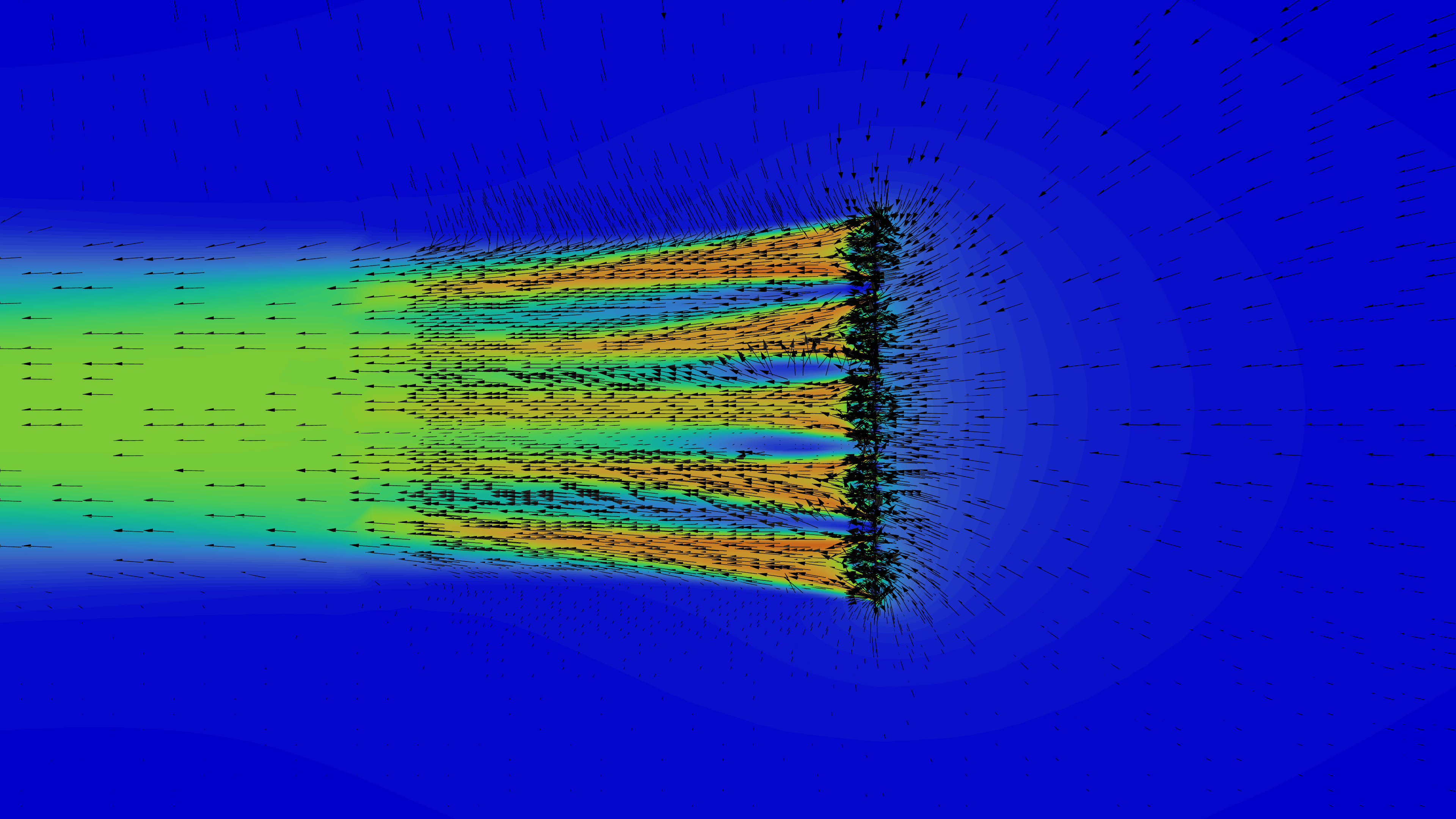} &
			\includegraphics[width=0.38\textwidth]{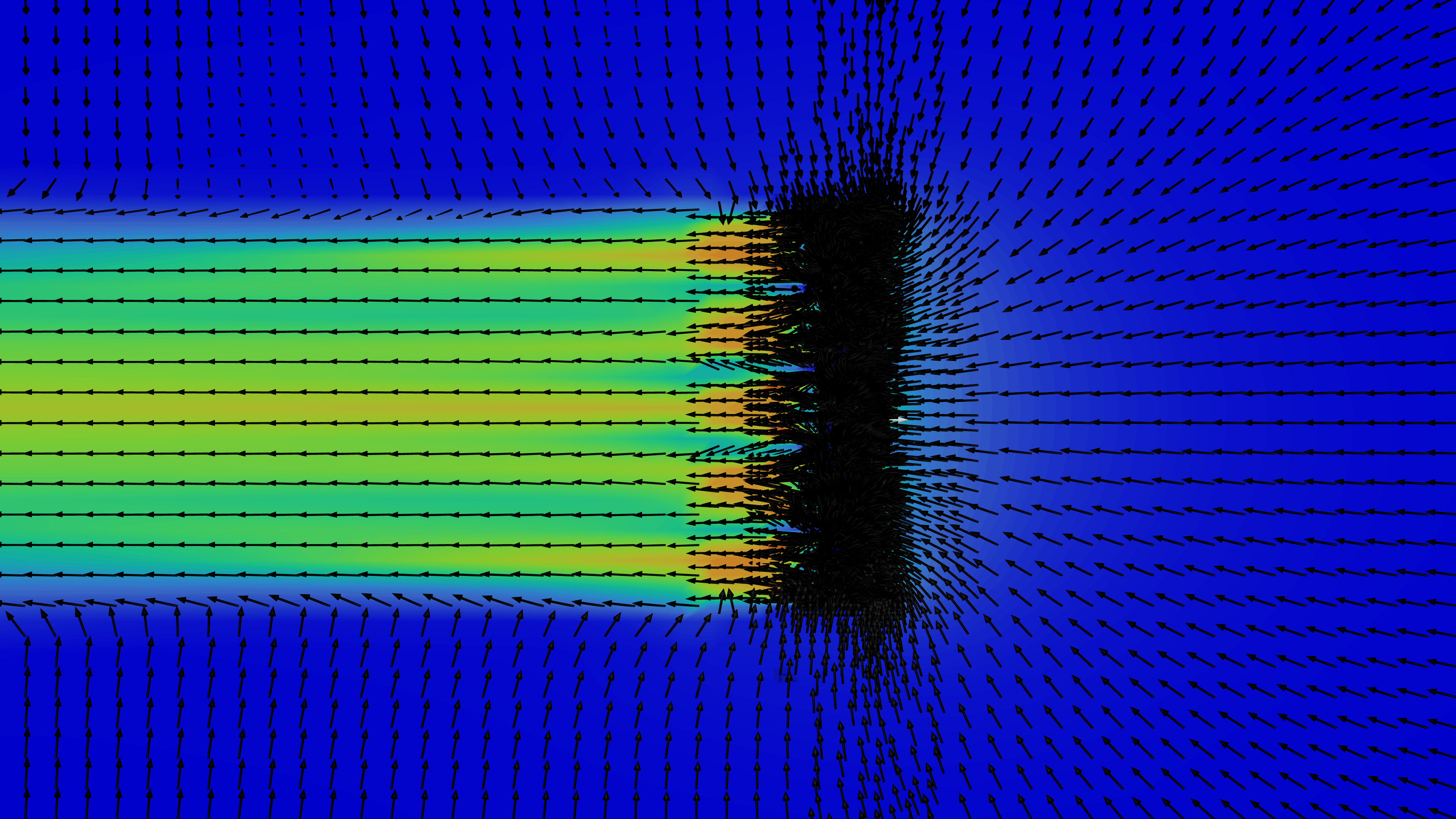} \\
			~ & \small (e) Near-field angled view & 
			\small (f) Near-field angled view \\[10pt]
			
			~ & \includegraphics[width=0.38\textwidth]{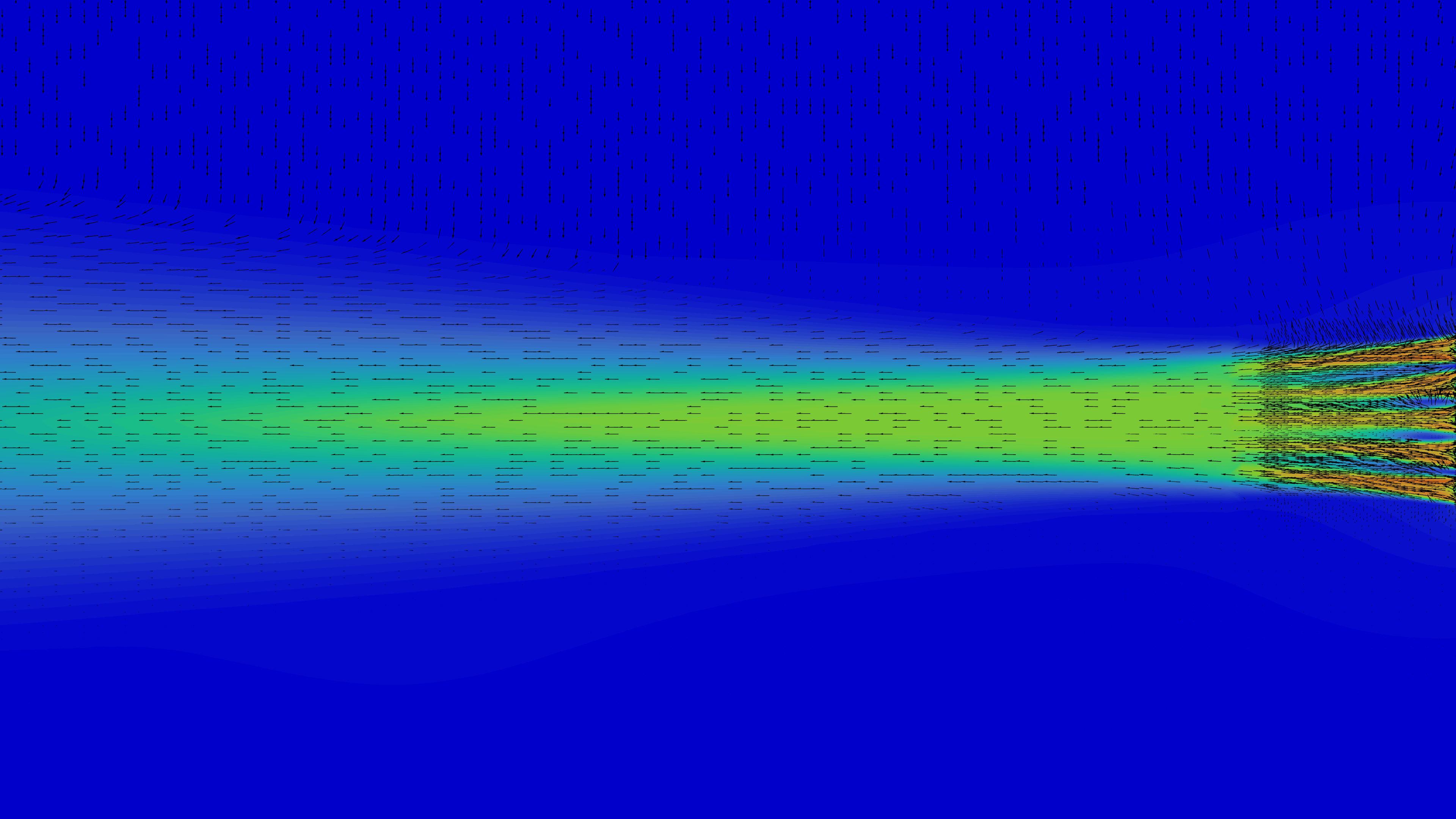} &
			\includegraphics[width=0.38\textwidth]{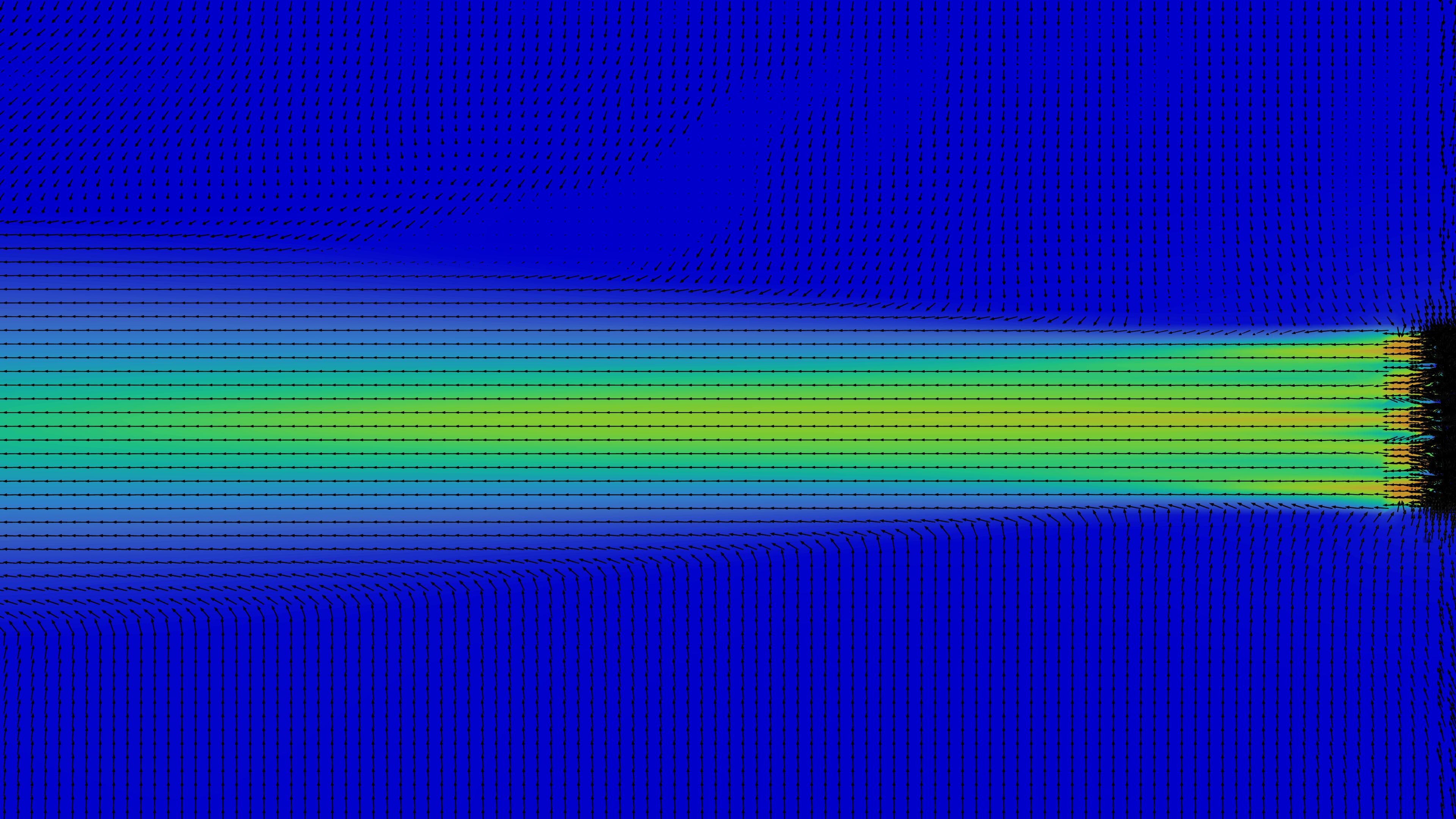} \\
			~ & \small (g) Far-field angled view & 
			\small (h) Far-field angled view \\
			\vspace{35pt}
		\end{tabular}
		\caption{Comparison of longitudinal flow topology. The leftmost schematics indicate the respective geometric extraction planes.}
		\label{fig:longitudinal_comparison}
	\end{figure*}

	In the surface-based model, the pressure-jump boundary condition introduces axial momentum directly at the inlet plane. The velocity field shows an initial inward deflection of adjacent jet cores, leading to a locally contracted flow region near the inlet. This contraction is followed by rapid interaction between neighboring jets and subsequent spreading downstream, indicating strong near-field mixing (Figure~\ref{fig:longitudinal_comparison}a, c).
	
	In contrast, the ducted representation includes the hub and casing walls, constraining the flow within an annular region and introducing additional losses. The velocity field shows a similar inward deflection of the jet cores as the surface model, but accompanied by lower jet velocities due to confinement and the presence of the hub wake. Mixing between adjacent jets initiates relatively early; however, the interaction is less energetic, resulting in a more gradual spreading of the flow. The overall jet structure remains less sharply defined, with reduced core strength and smoother velocity gradients in the near field (Figure~\ref{fig:longitudinal_comparison}b, d).
	
\begin{figure*}[htbp]
		\centering
		\setlength{\tabcolsep}{1pt}
		\begin{tabular}{c @{\hspace{15pt}} c c c c}
			~ & \textbf{$z/D = 4$} & \textbf{$z/D = 6$} & \textbf{$z/D = 8$} & \textbf{$z/D = 10$} \\[5pt]
			
			\multirow{4}{*}{\begin{minipage}{0.08\textwidth}
					\centering
					\includegraphics[height=3cm]{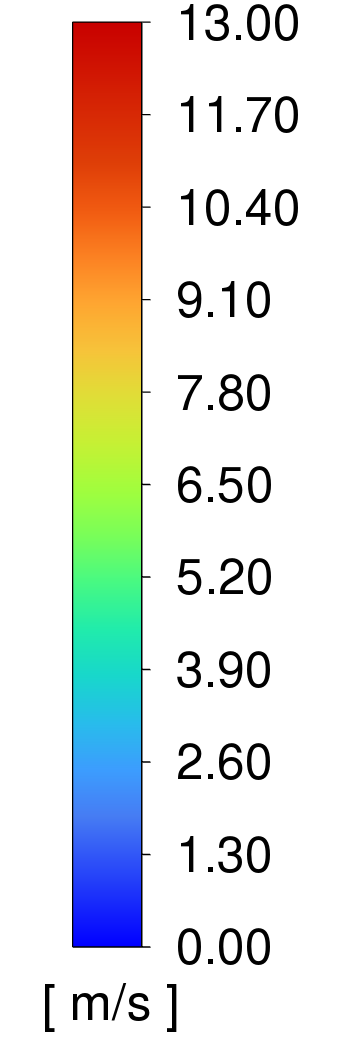}
			\end{minipage}} &
			\includegraphics[width=0.22\textwidth]{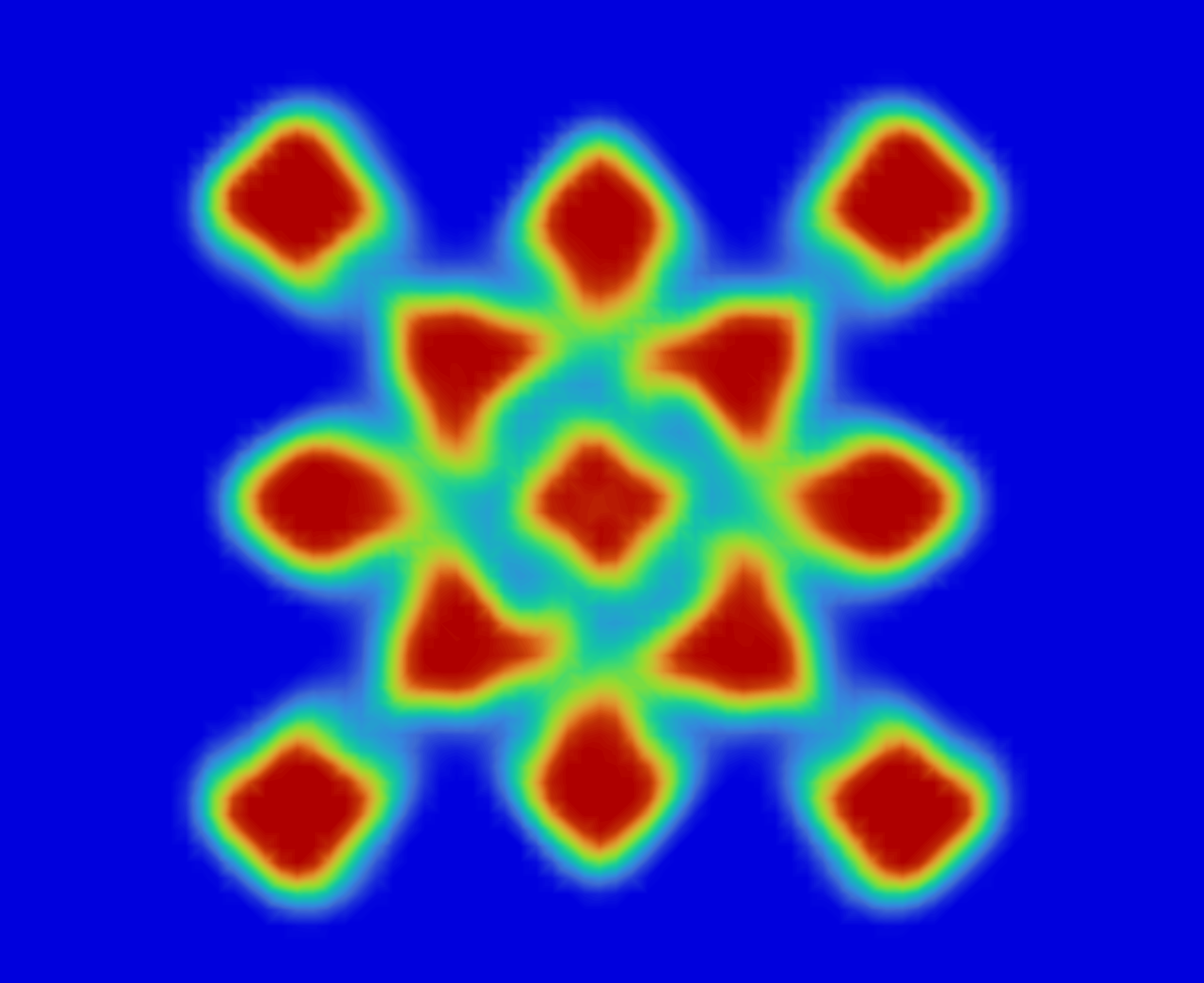} &
			\includegraphics[width=0.22\textwidth]{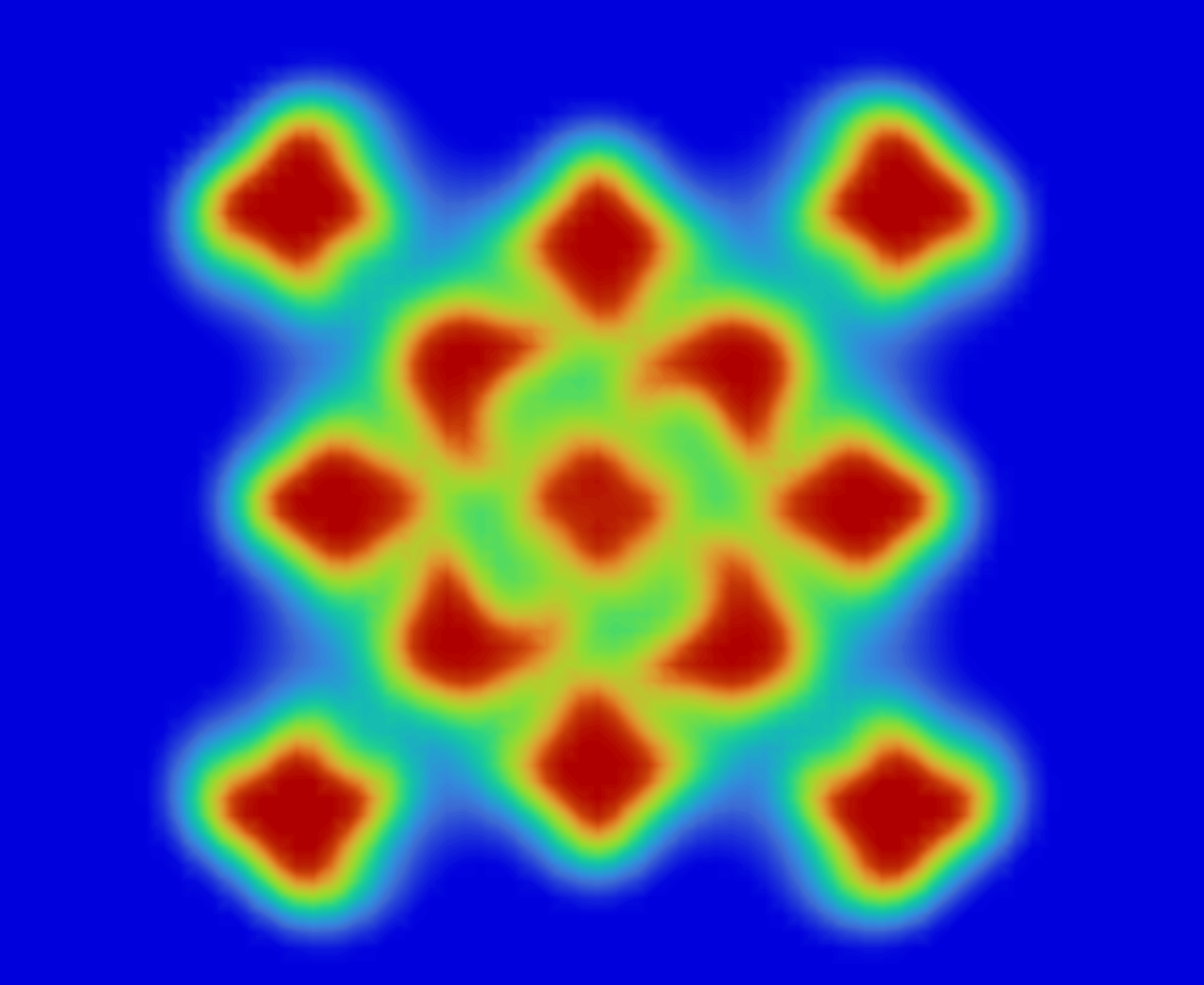} &
			\includegraphics[width=0.22\textwidth]{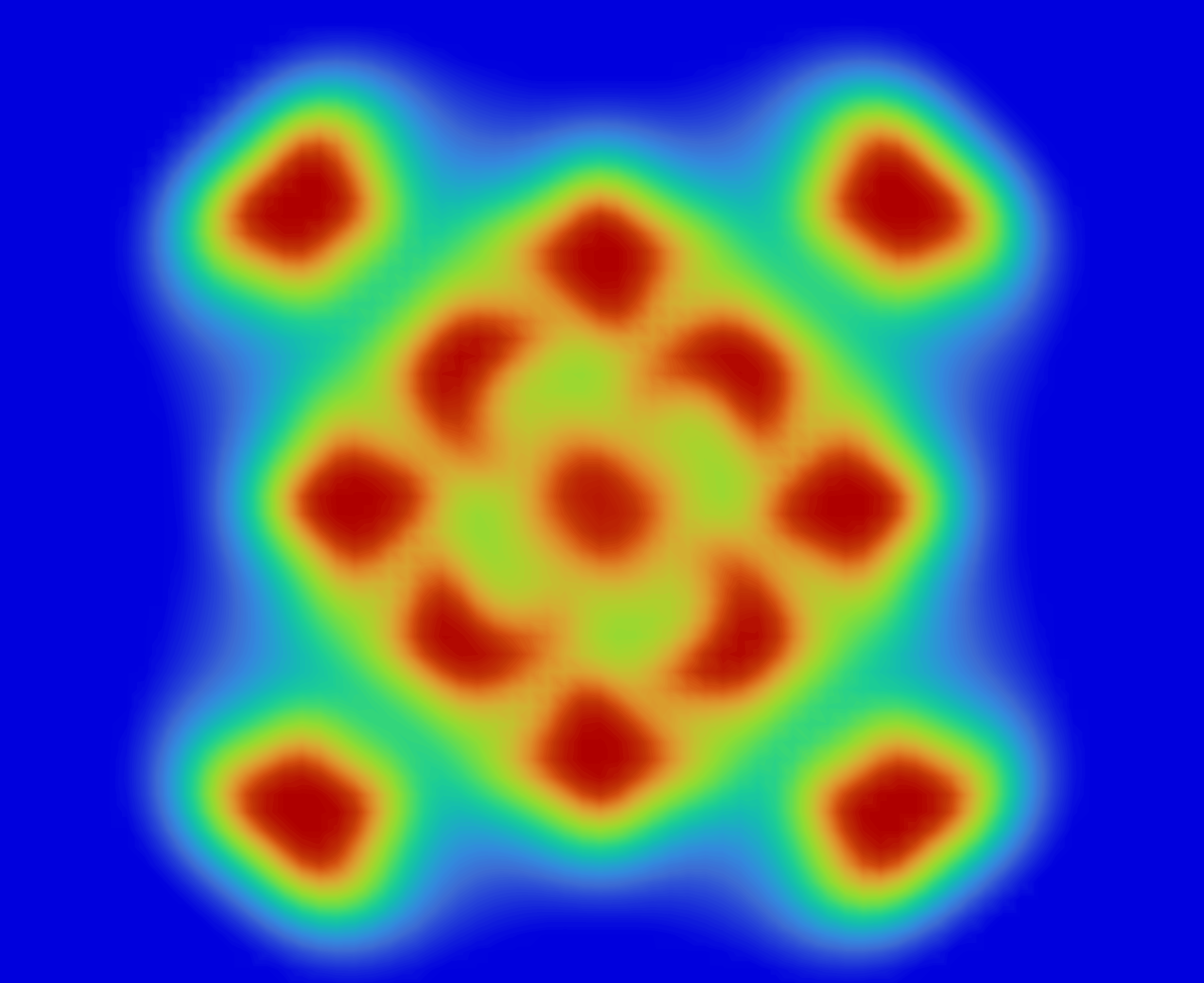} &
			\includegraphics[width=0.22\textwidth]{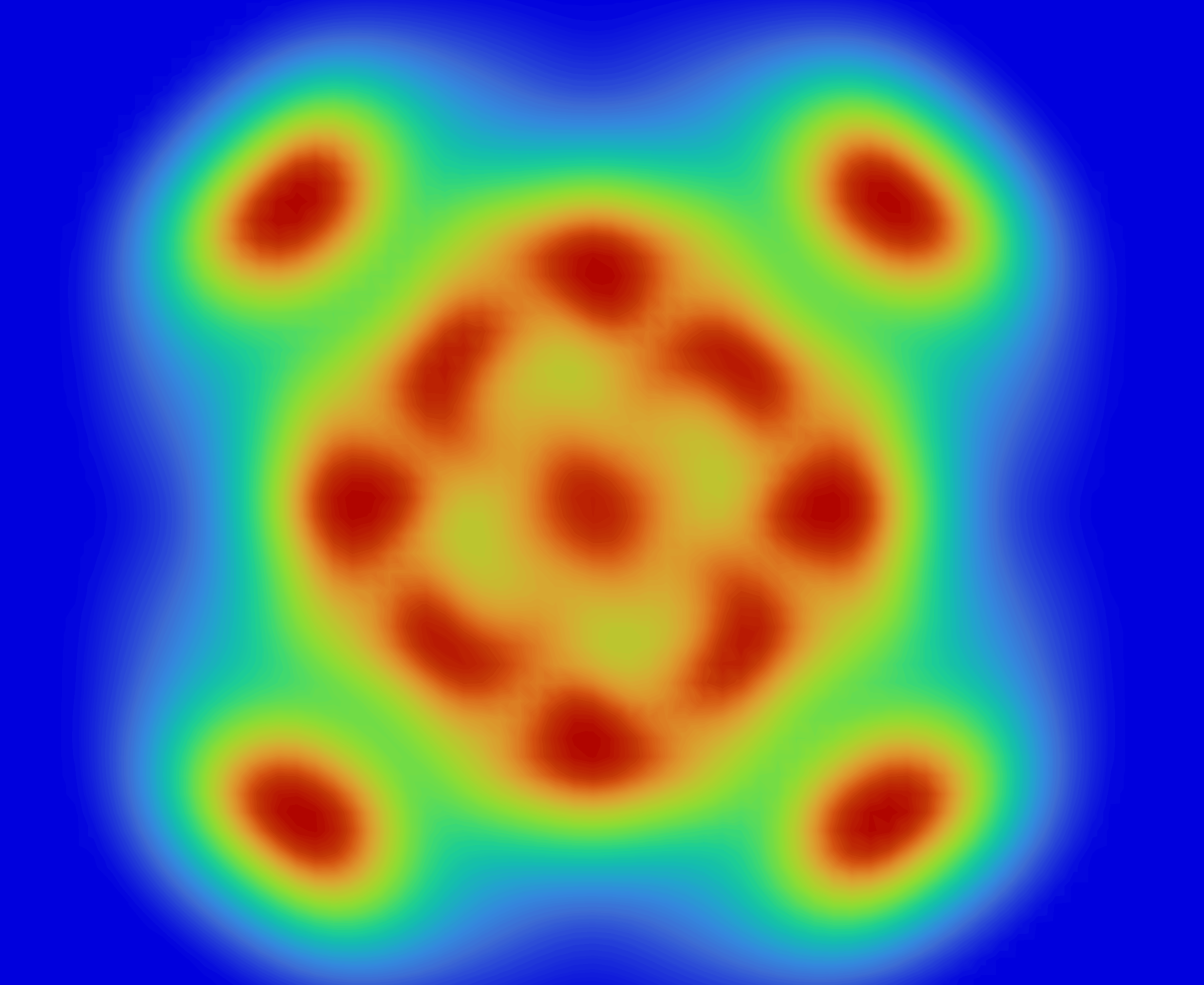} \\
			~ & \multicolumn{4}{c}{\small (a) Surface Model: Axial Velocity} \\[8pt]
			
			~ & \includegraphics[width=0.22\textwidth]{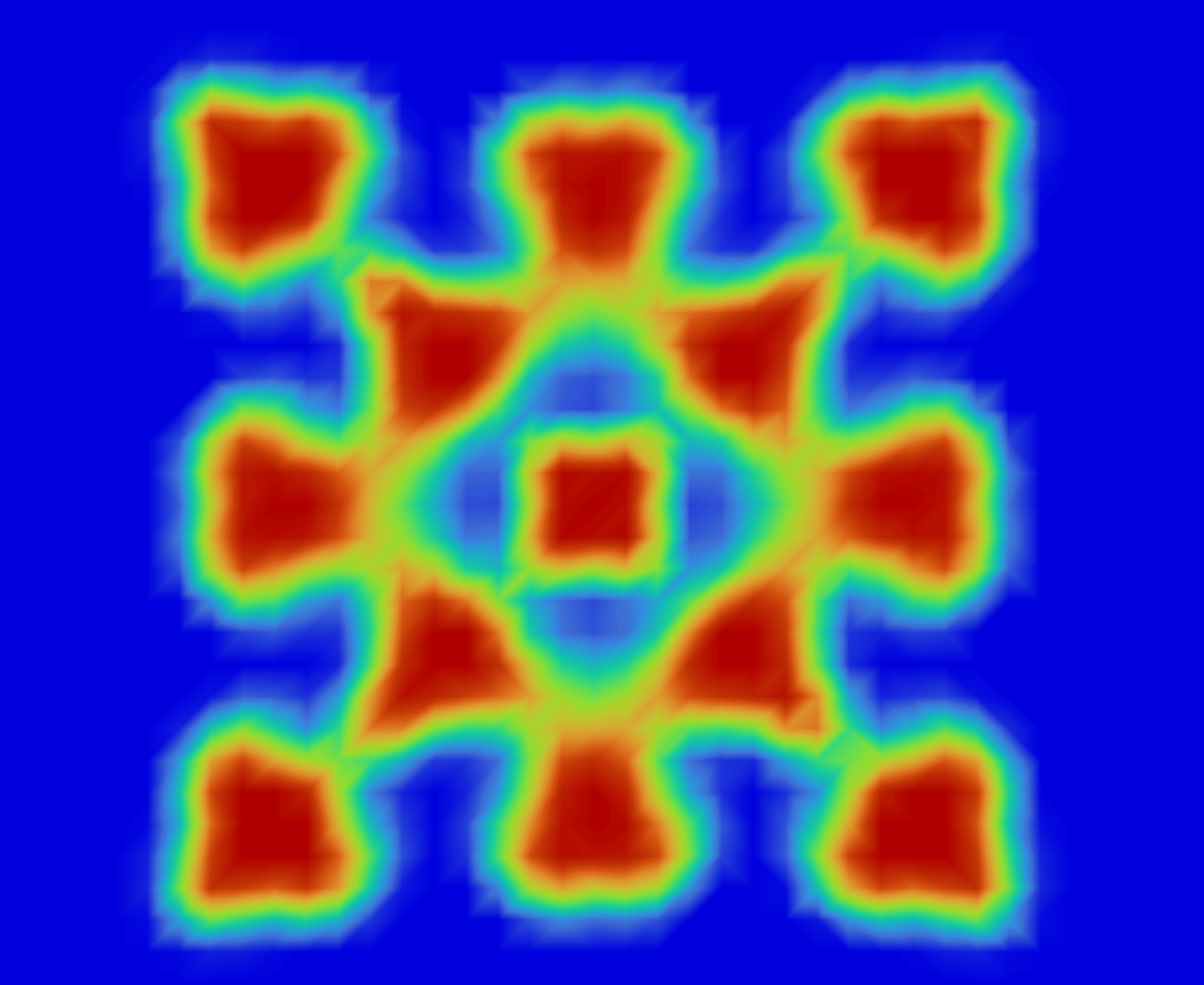} &
			\includegraphics[width=0.22\textwidth]{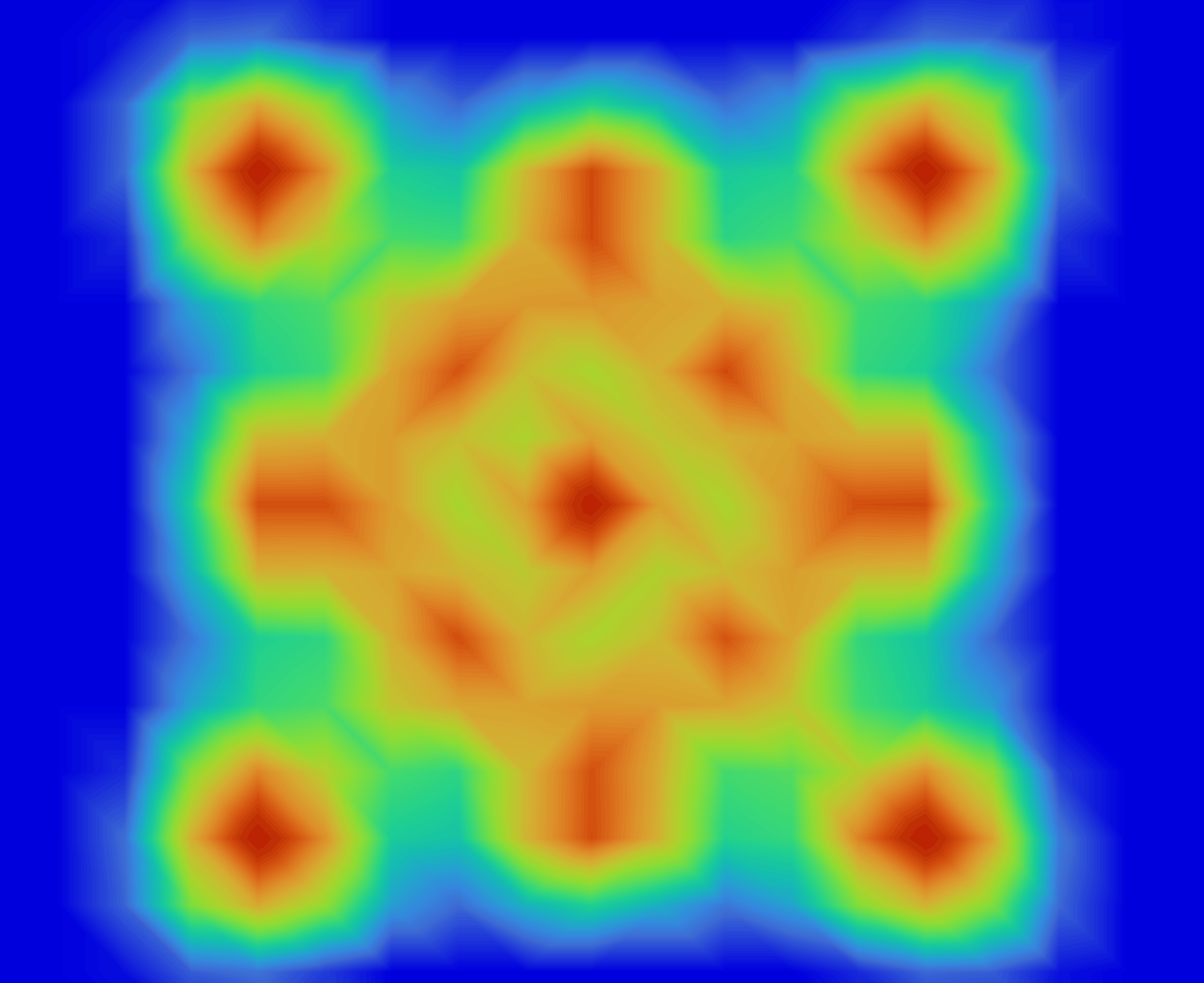} &
			\includegraphics[width=0.22\textwidth]{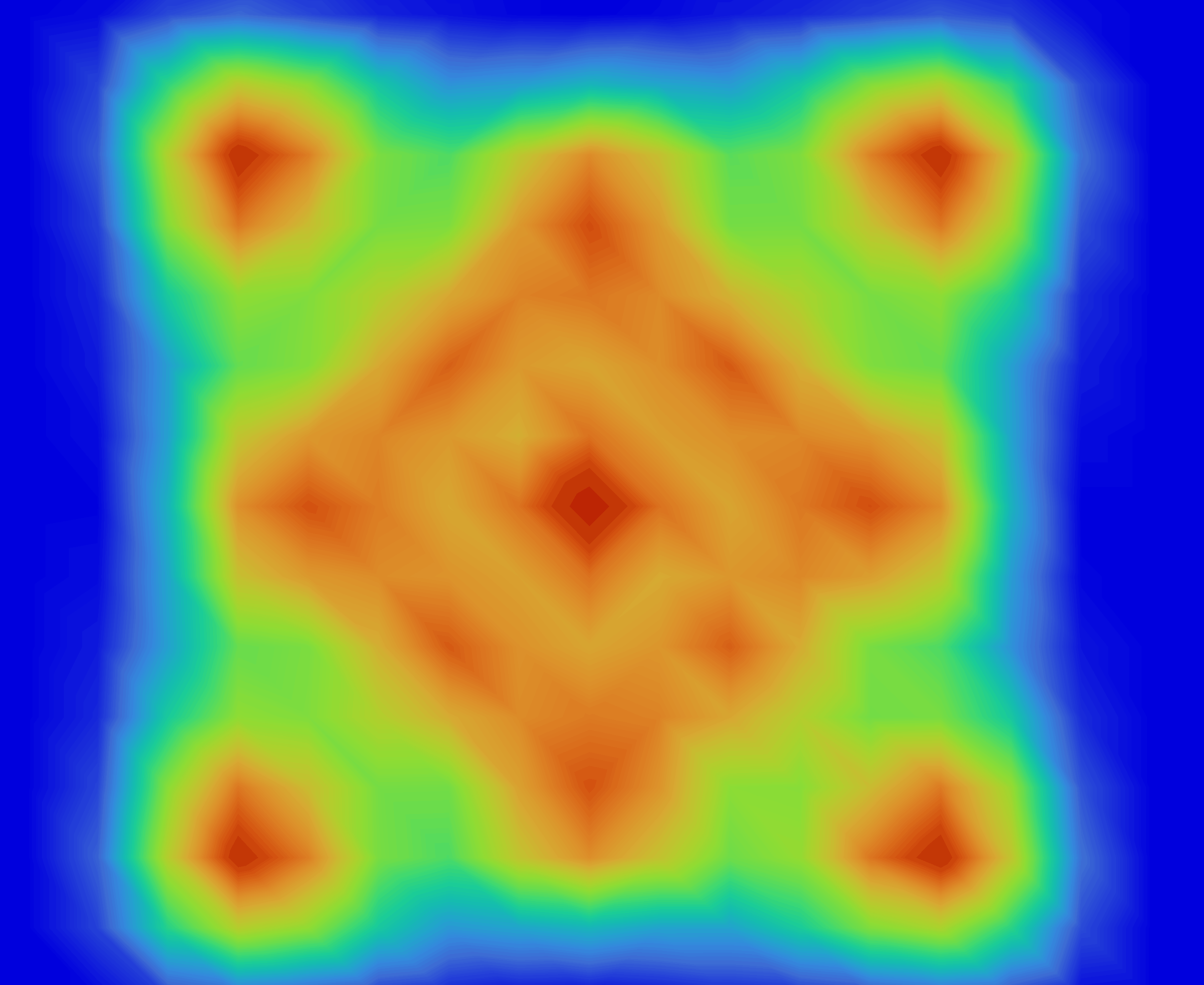} &
			\includegraphics[width=0.22\textwidth]{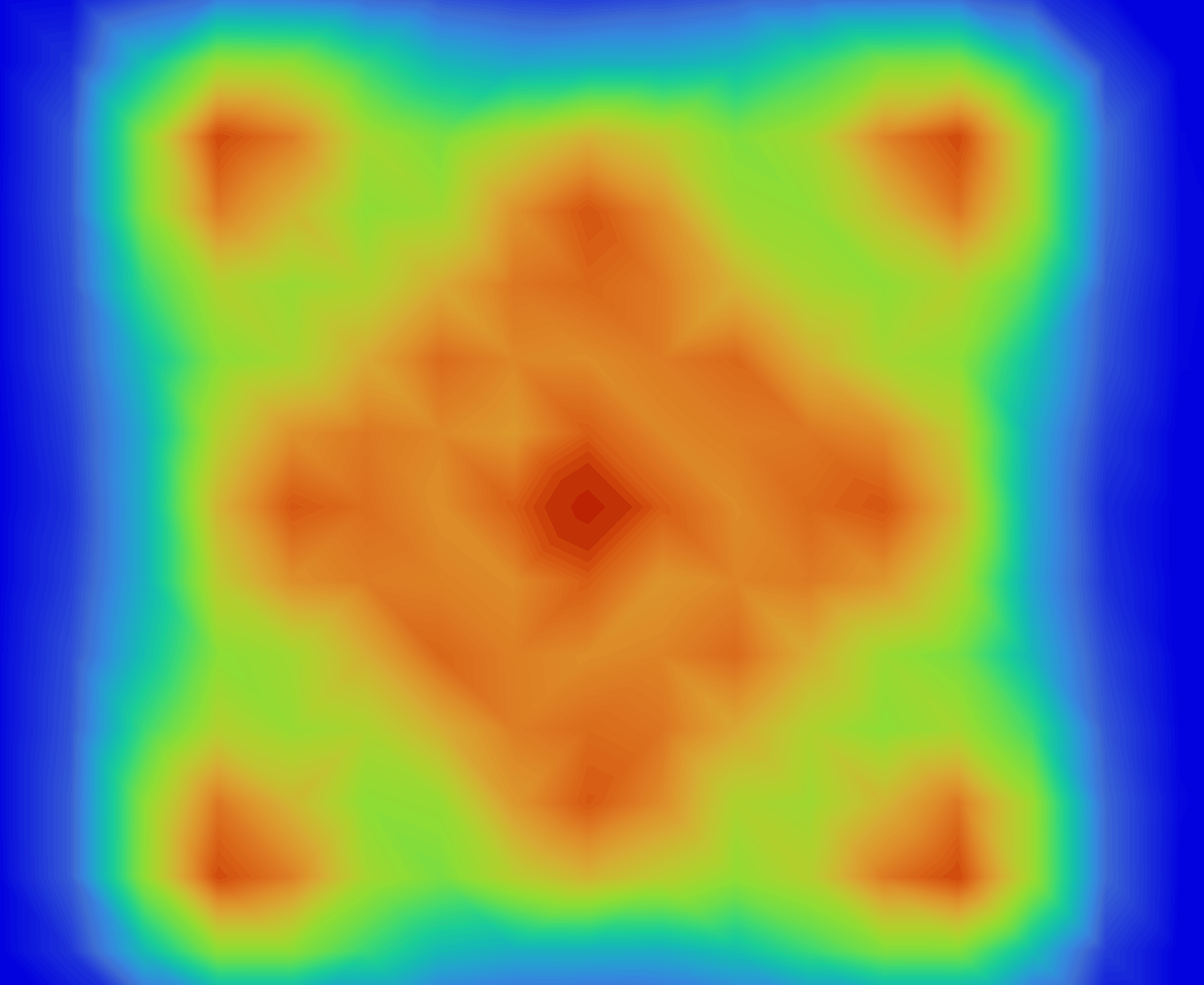} \\
			~ & \multicolumn{4}{c}{\small (b) Ducted Model: Axial Velocity} \\[15pt]
			
			\multirow{4}{*}{\begin{minipage}{0.08\textwidth}
					\centering
					\includegraphics[height=3cm]{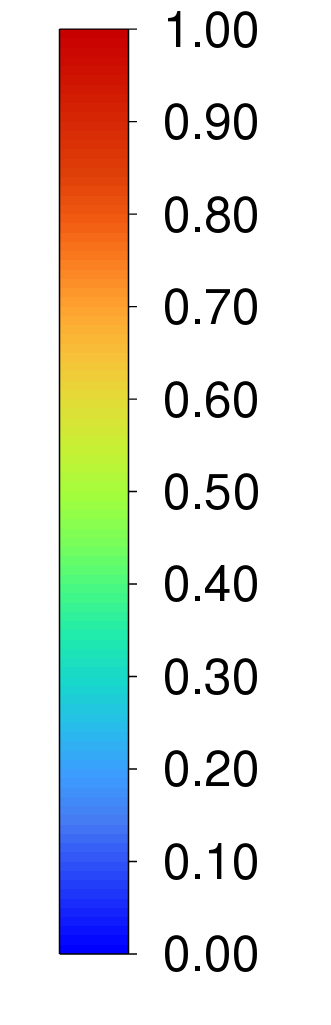} 
			\end{minipage}} & 
			\includegraphics[width=0.22\textwidth]{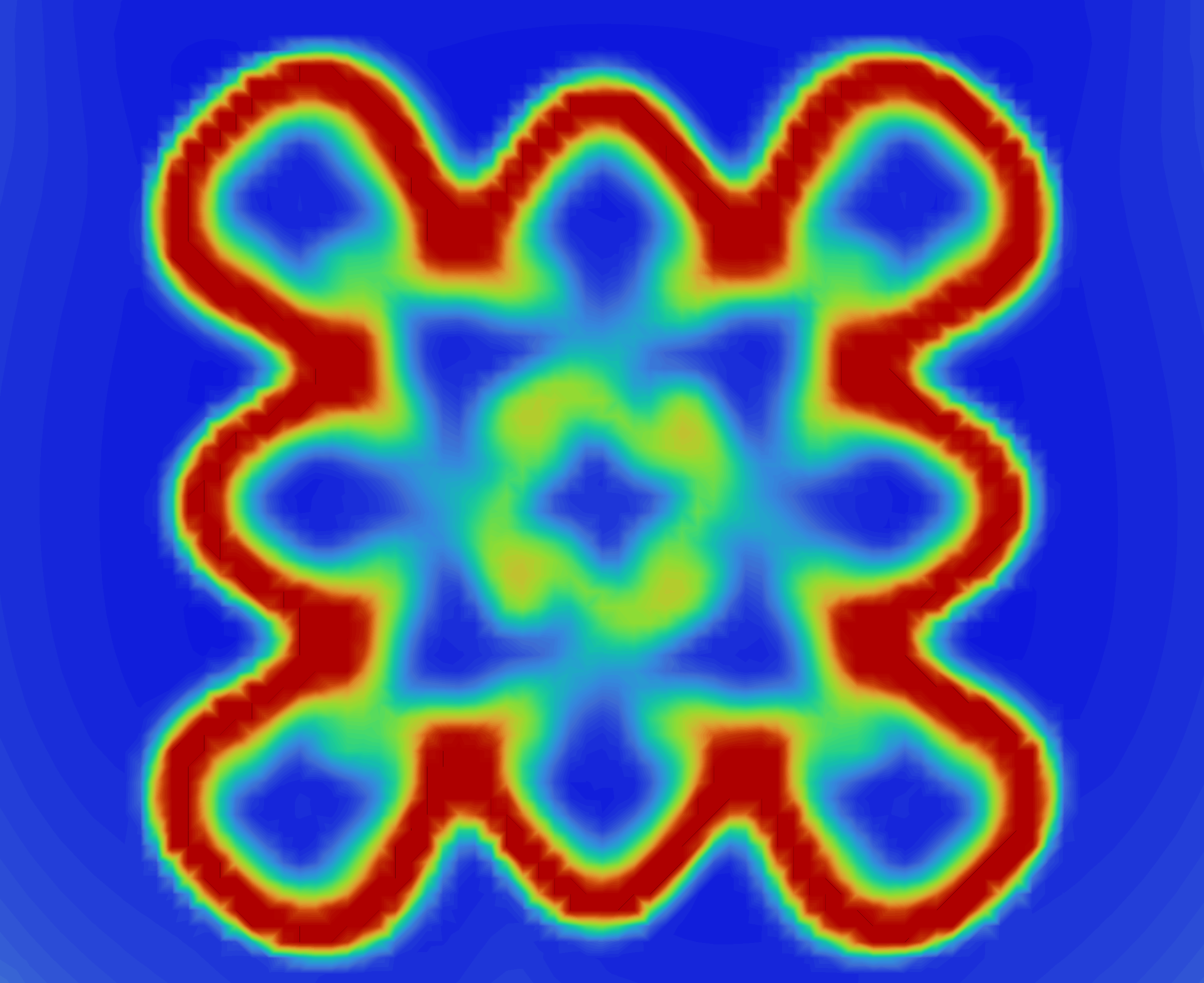} &
			\includegraphics[width=0.22\textwidth]{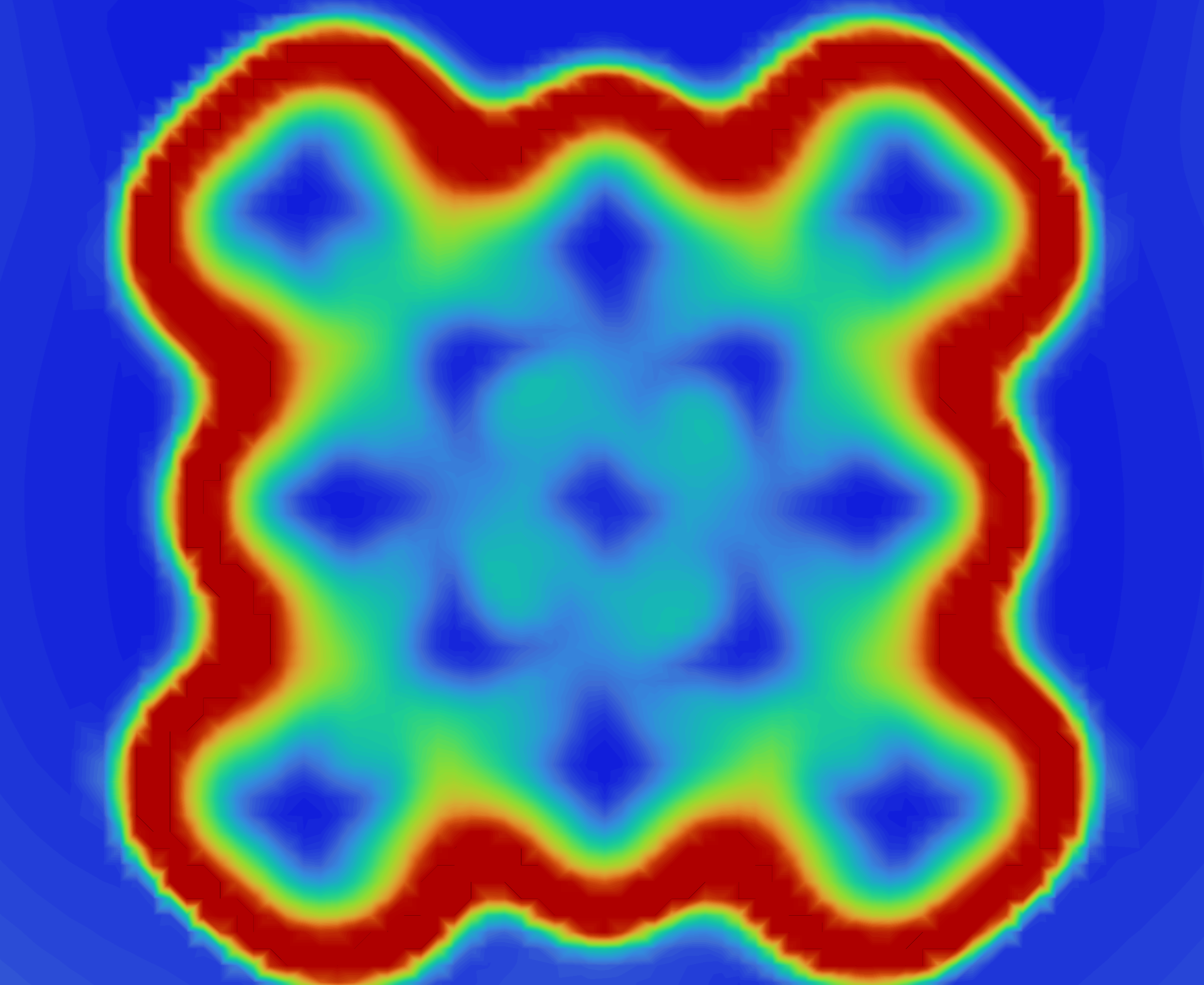} &
			\includegraphics[width=0.22\textwidth]{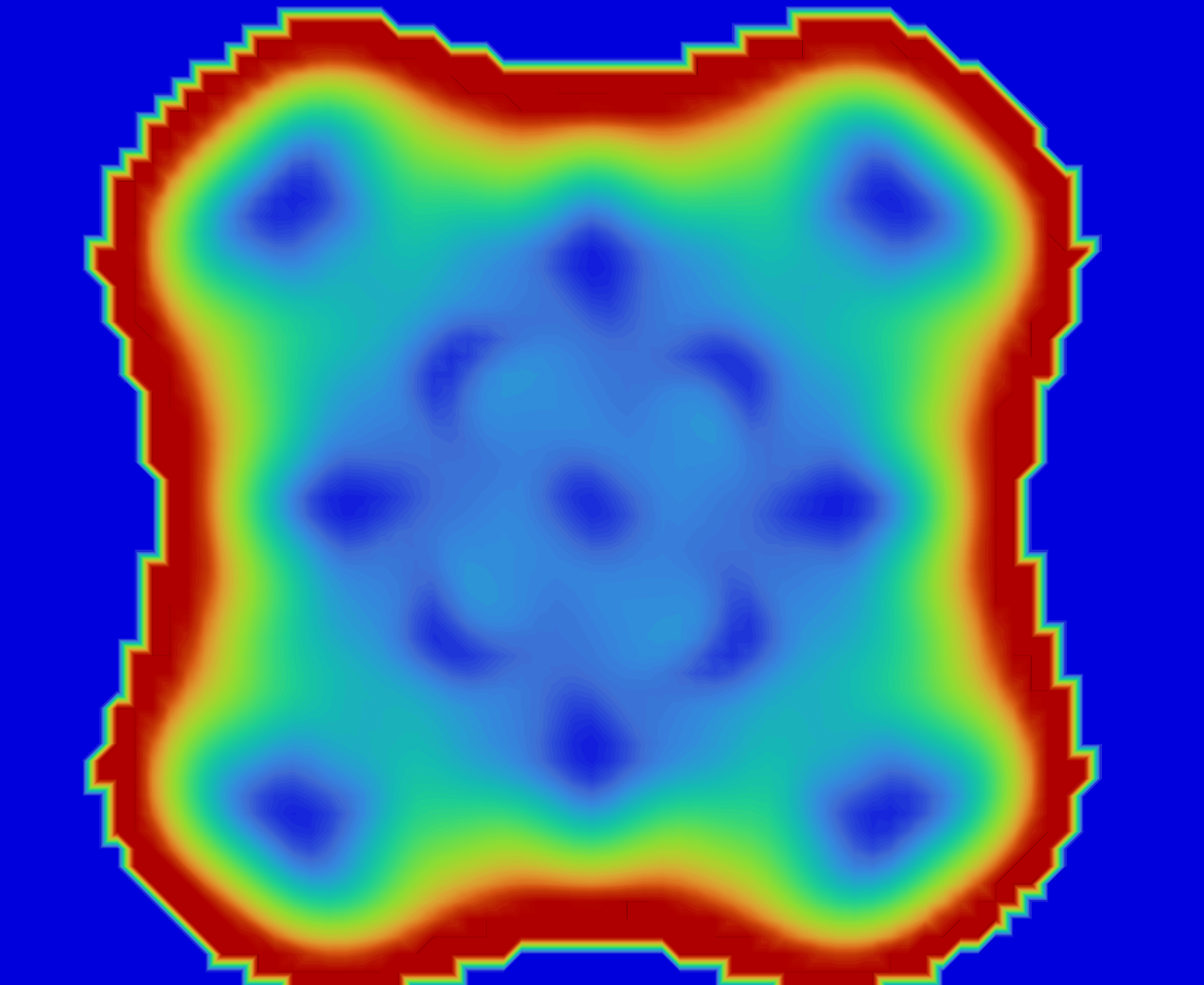} &
			\includegraphics[width=0.22\textwidth]{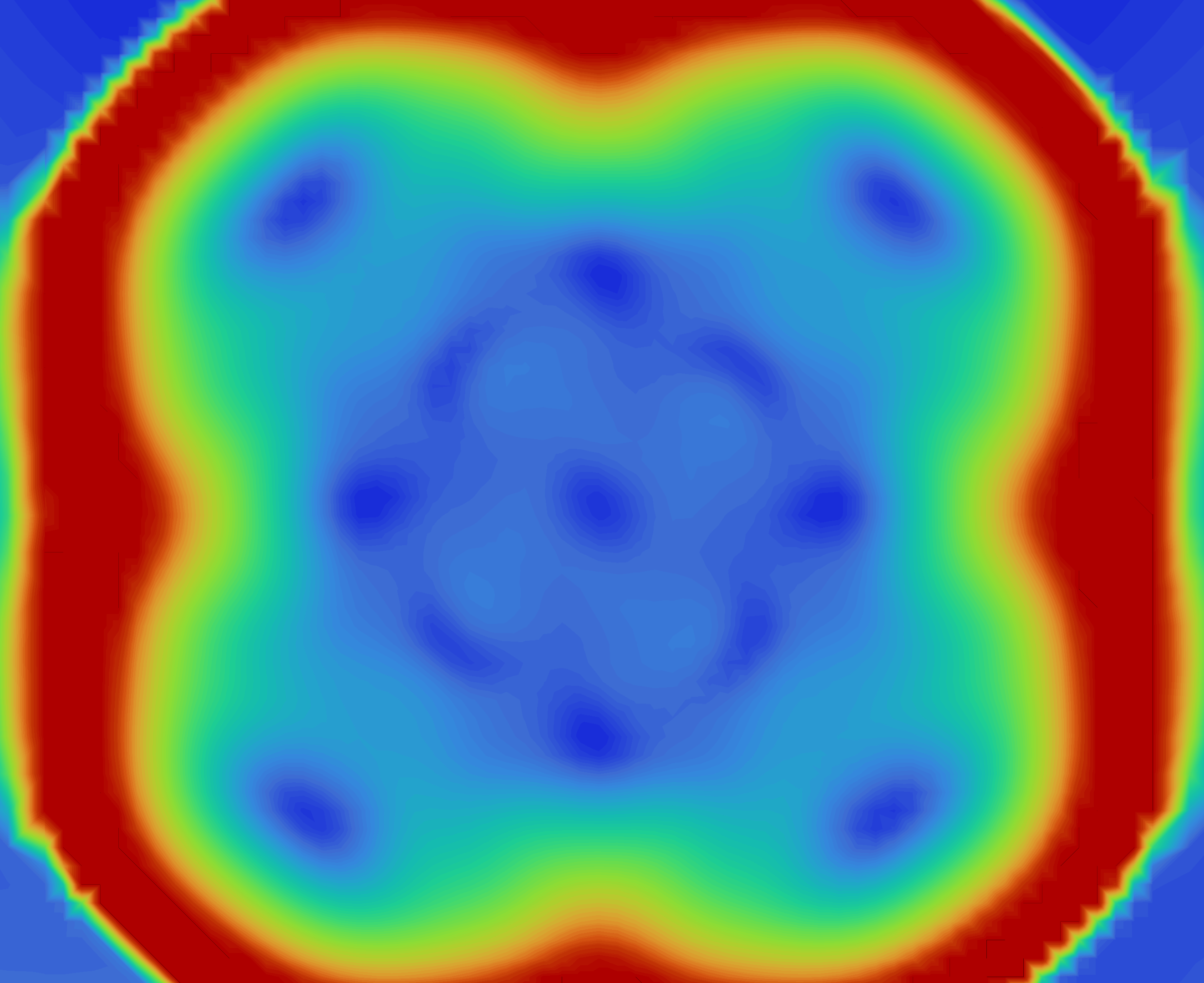} \\
			~ & \multicolumn{4}{c}{\small (c) Surface Model: Turbulence Intensity} \\[8pt]
			
			~ & \includegraphics[width=0.22\textwidth]{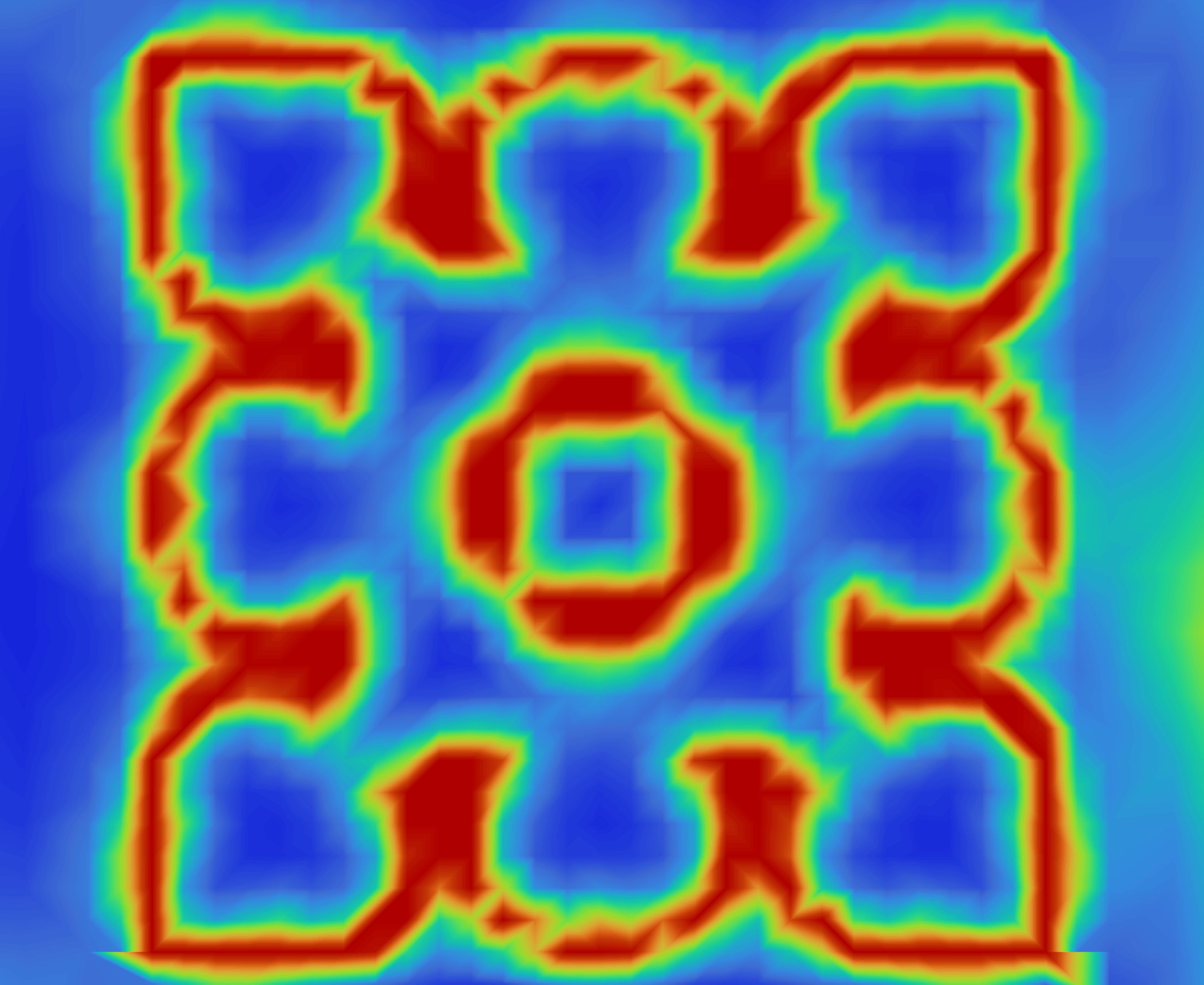} &
			\includegraphics[width=0.22\textwidth]{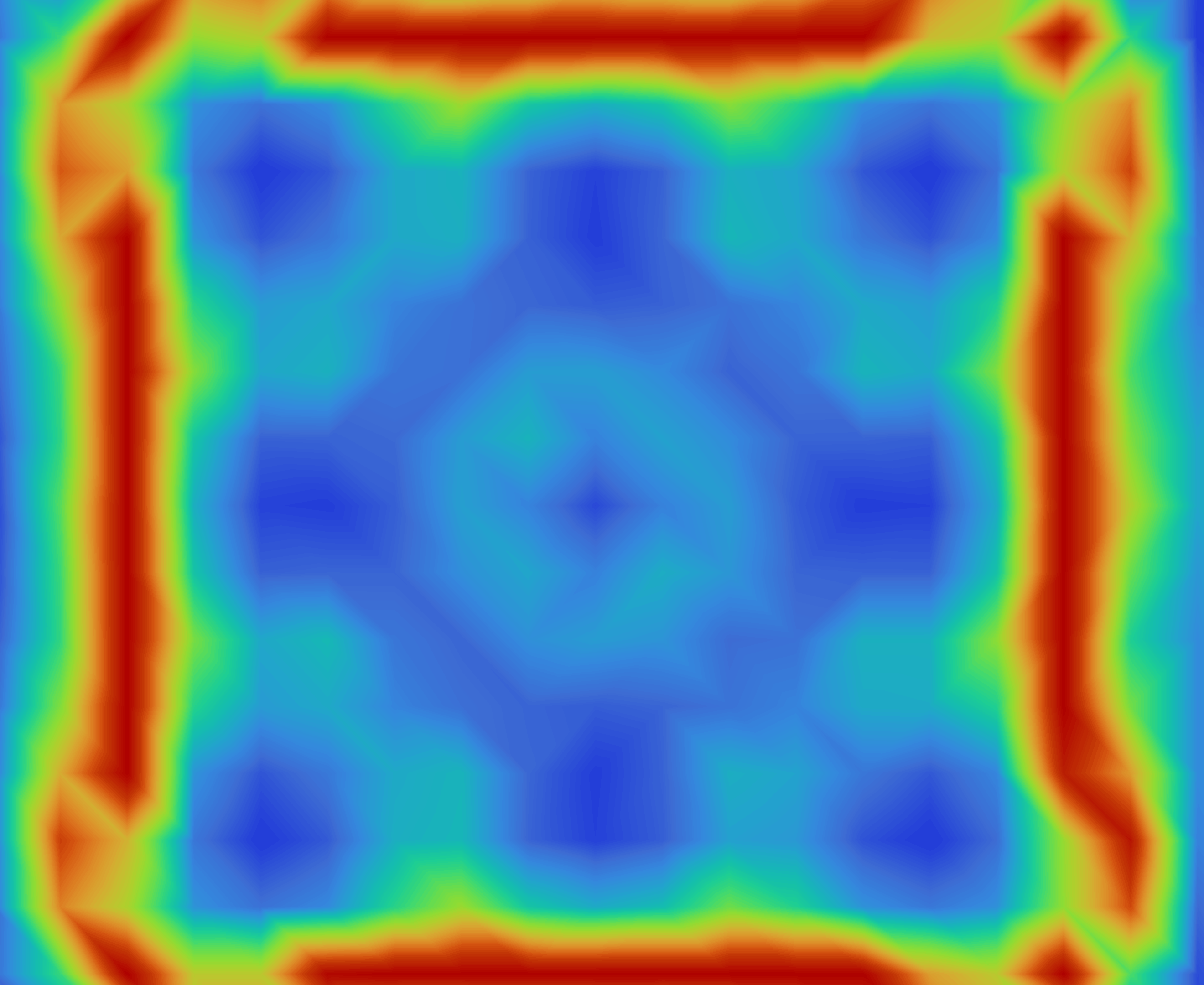} &
			\includegraphics[width=0.22\textwidth]{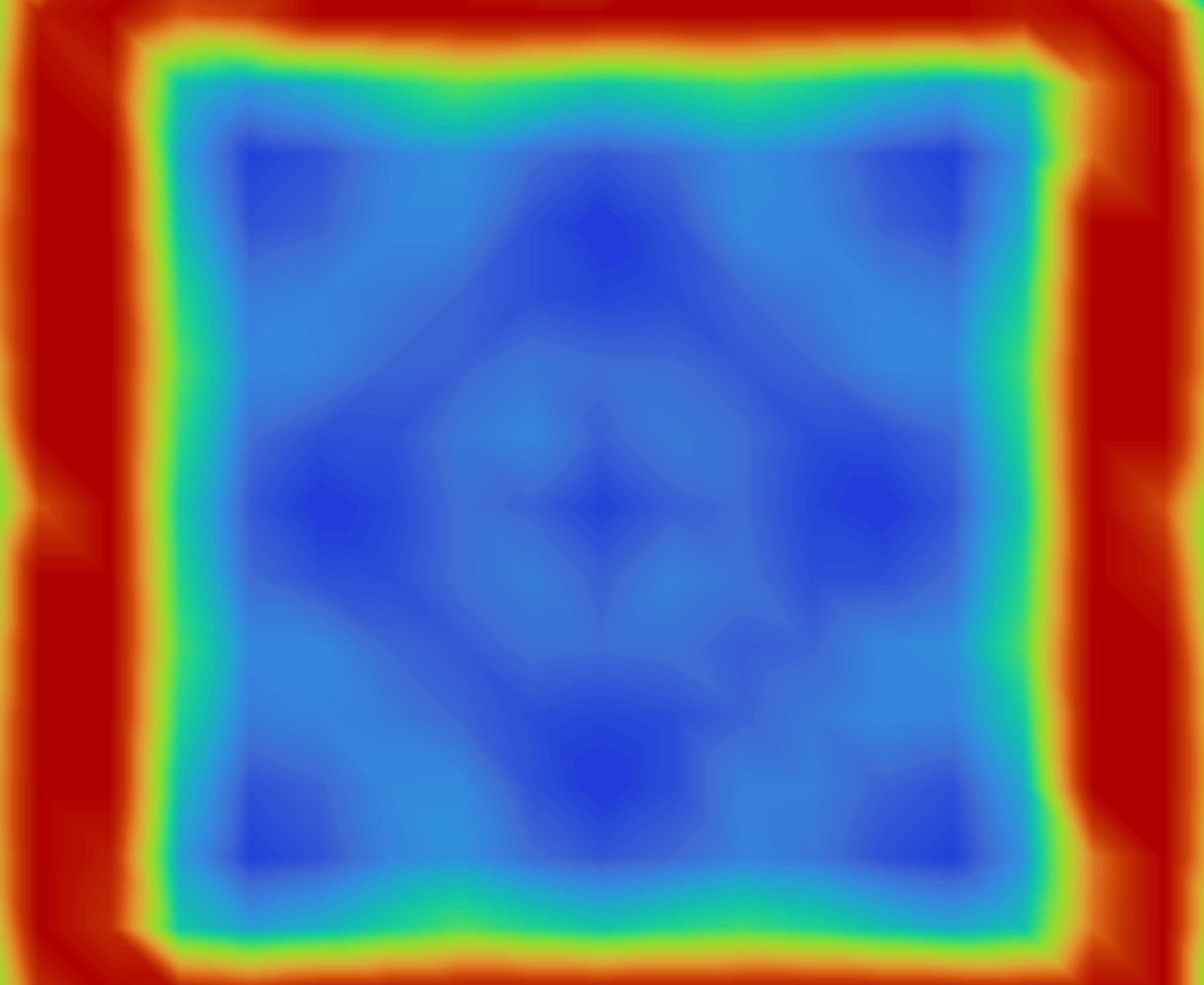} &
			\includegraphics[width=0.22\textwidth]{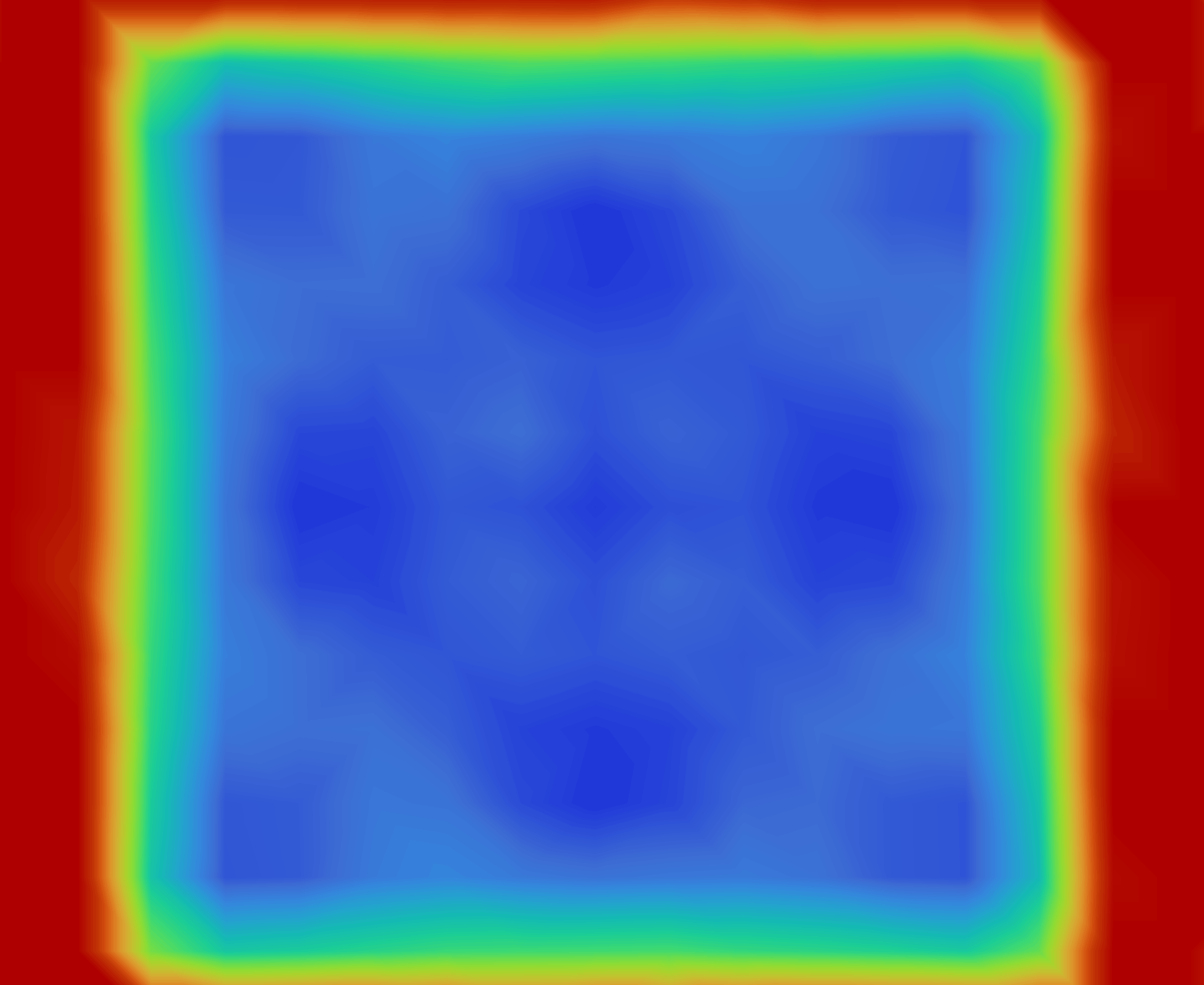} \\
			~ & \multicolumn{4}{c}{\small (d) Ducted Model: Turbulence Intensity} \\
		\end{tabular}
		\caption{Downstream spatial evolution highlighting progressive jet coalescence (top) and shear-layer structure (bottom).}
		\label{fig:spatial_evolution_master}
	\end{figure*}

	Differences between the two representations are also evident in the diagonal planes (Figure~\ref{fig:longitudinal_comparison}e--h). In the surface model, adjacent jets contract, leading to rapid deformation of the jet cores and pronounced mixing in the near field. This results in sharper velocity gradients which transition into a relatively uniform flow further downstream. In the ducted model, the jet cores are weaker and less sharply defined due to confinement and hub-induced losses. Interaction between neighboring jets is still present in the near field; however, it is less energetic, resulting in smoother velocity gradients and a more gradual merging process. Consequently, the flow appears more diffused, with reduced contrast between jet cores and surrounding regions even at relatively short downstream distances (Figure~\ref{fig:longitudinal_comparison}f, h). Downstream sectional contours in Figure~\ref{fig:spatial_evolution_master} also show that the surface model maintains fast jet cores that merge progressively with downstream distance, whereas the ducted model produces a more diffuse and uniform velocity field at earlier planes due to reduced core momentum. Similarly, the ducted model yields a smoother and more uniformly distributed turbulence field with weaker peak intensities.
	
	\begin{figure*}[htbp]
		\centering
		\begin{minipage}[b]{0.48\textwidth}
			\centering
			\includegraphics[width=\textwidth]{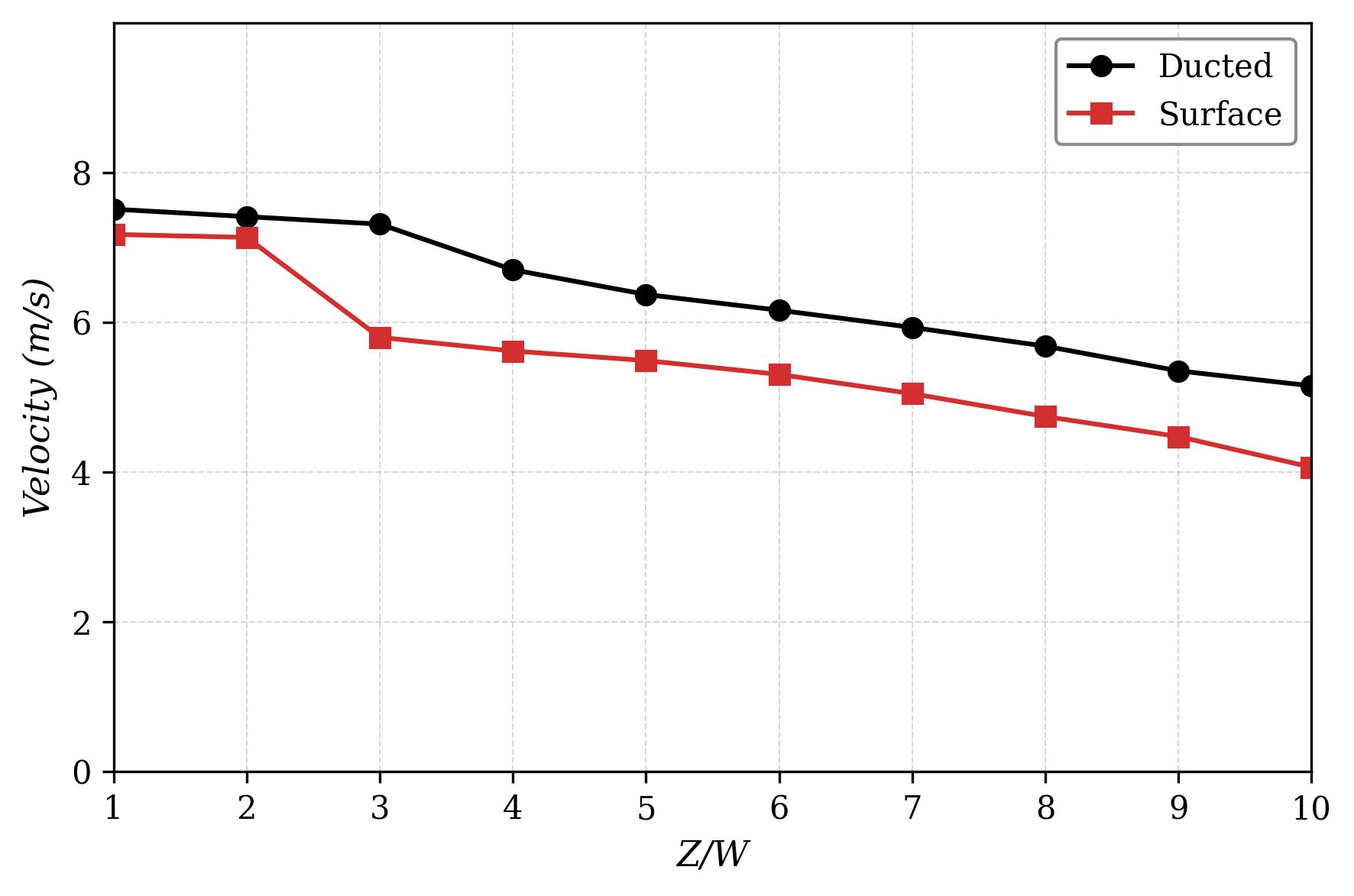}
			\vspace{5pt} \\ \small (a) Average Velocity - Set 1 points
		\end{minipage}
		\hfill
		\begin{minipage}[b]{0.48\textwidth}
			\centering
			\includegraphics[width=\textwidth]{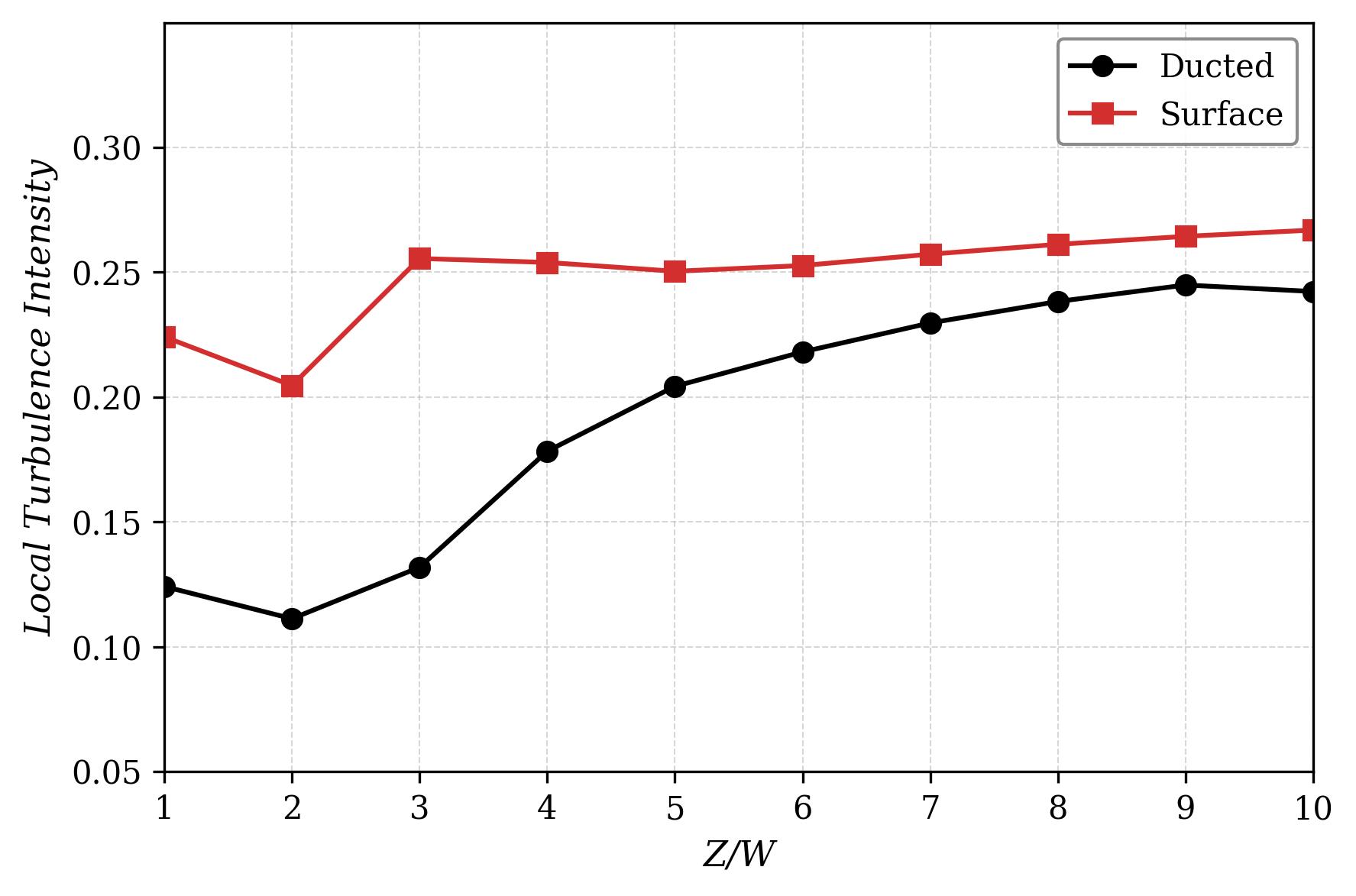}
			\vspace{5pt} \\ \small (b) Average Turbulence Intensity - Set 1 points
		\end{minipage}
		
		\vspace{10pt}
		
		\begin{minipage}[b]{0.48\textwidth}
			\centering
			\includegraphics[width=\textwidth]{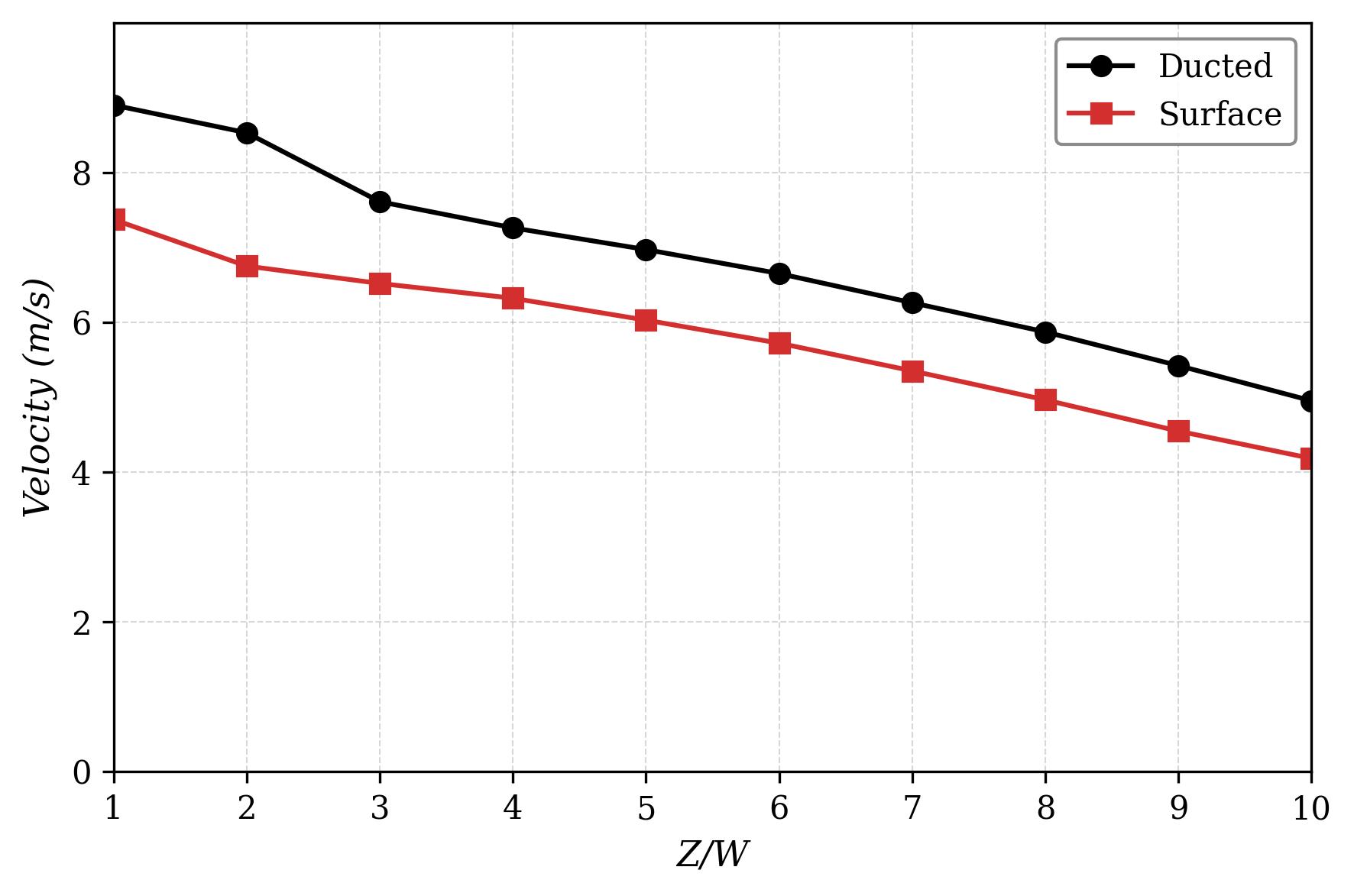}
			\vspace{5pt} \\ \small (c) Average Velocity - Set 2 points
		\end{minipage}
		\hfill
		\begin{minipage}[b]{0.48\textwidth}
			\centering
			\includegraphics[width=\textwidth]{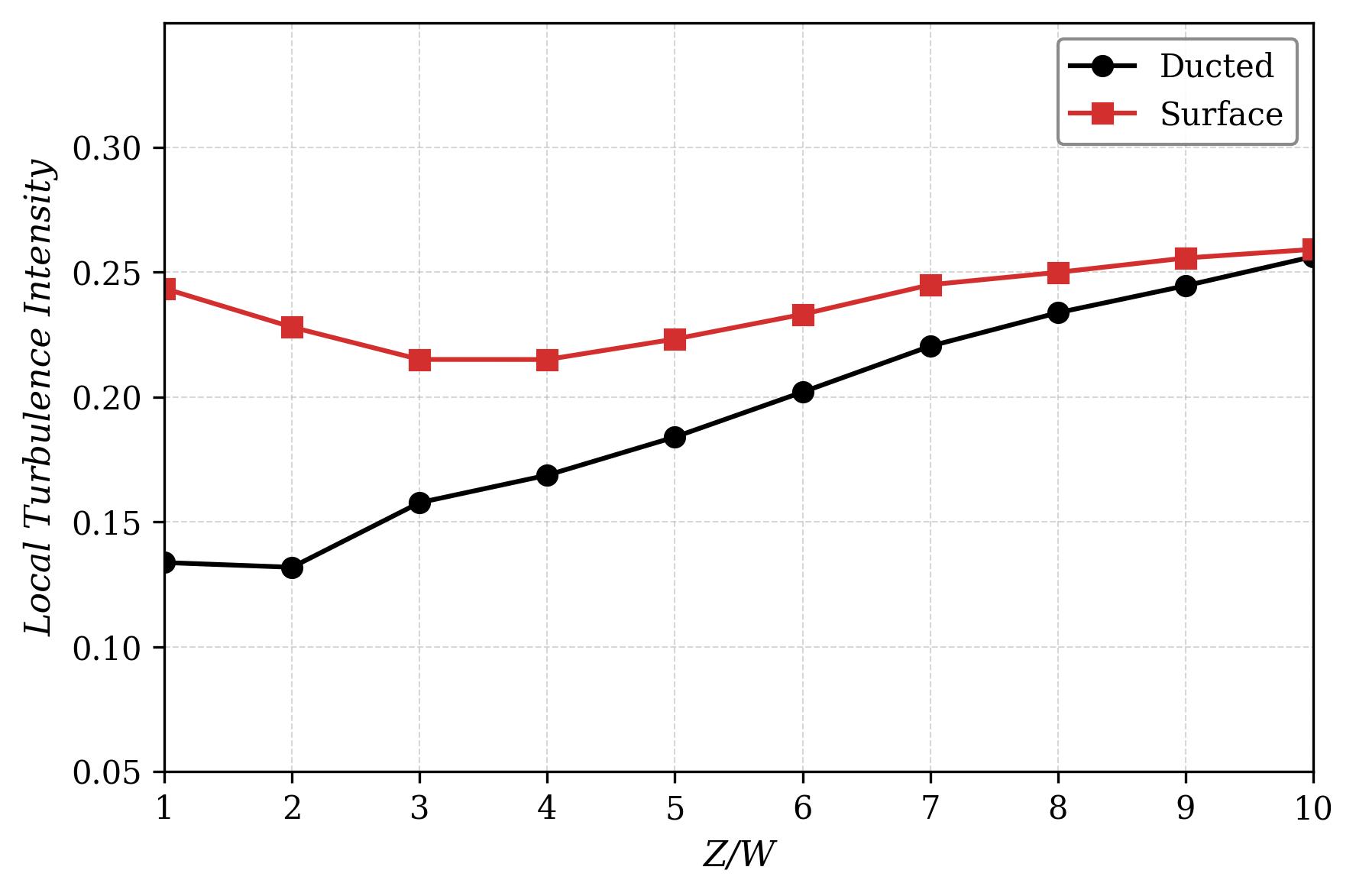}
			\vspace{5pt} \\ \small (d) Average Turbulence Intensity - Set 2 points
		\end{minipage}
		
		\caption{Comparison of averaged fields between the surface fan model and the ducted fan model.}
		\label{fig:matched_flow_comparisons}
	\end{figure*}
	
	\subsection{Modeled Effects of Jet Speed}
	
	The effect of jet speed is assessed by uniformly scaling the fan rotational speed using affinity laws applied to the fan curve (Figure~\ref{fig:fan_curves_combined}). The resulting momentum source is imposed via the fan boundary condition.
	
\begin{figure*}[htbp]
		\centering
		\begin{minipage}[b]{0.48\textwidth}
			\centering
			\includegraphics[width=\textwidth]{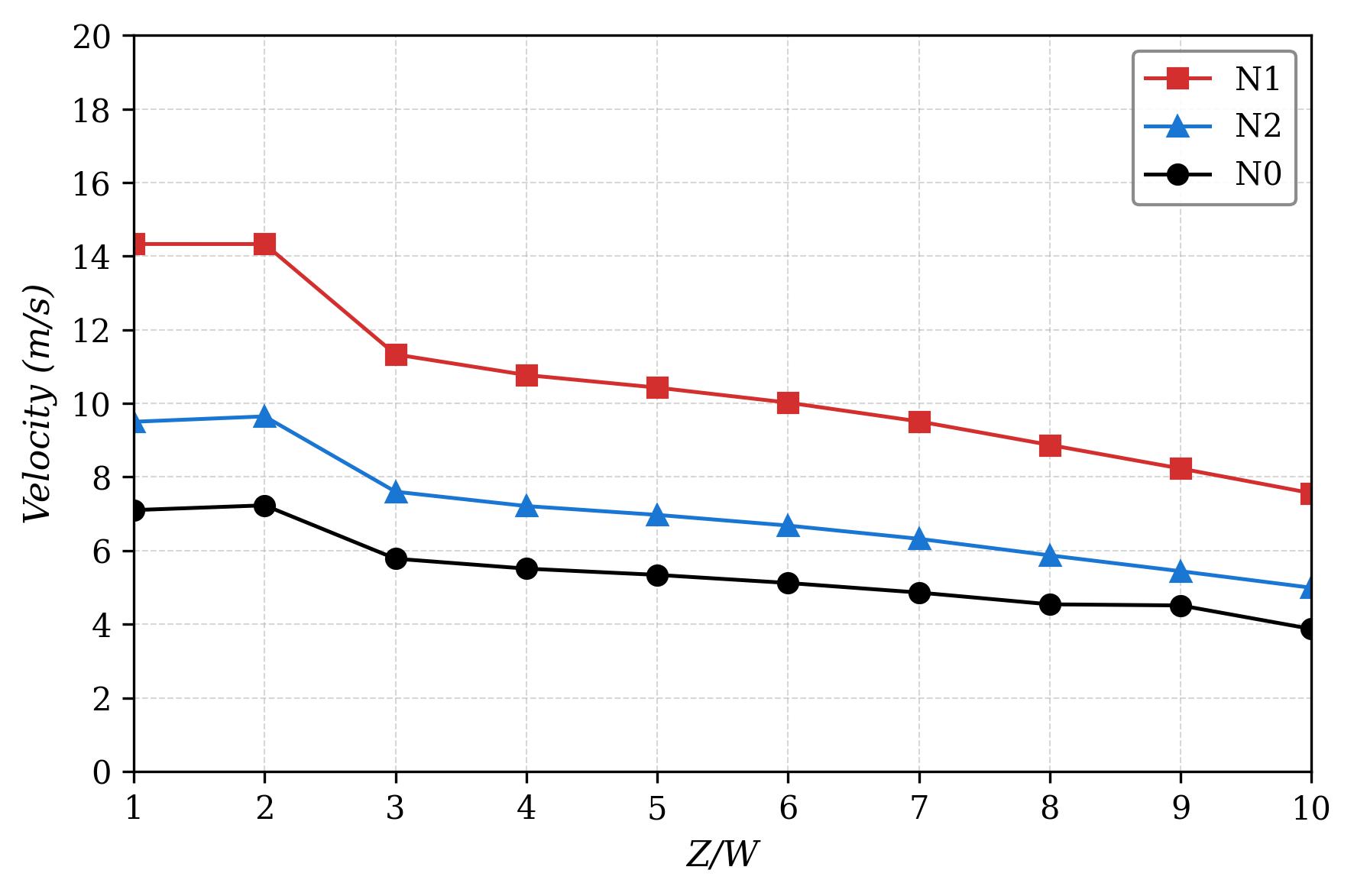}
			\vspace{5pt} \\ \small (a) Axial velocity at s1p1
		\end{minipage}
		\hfill
		\begin{minipage}[b]{0.48\textwidth}
			\centering
			\includegraphics[width=\textwidth]{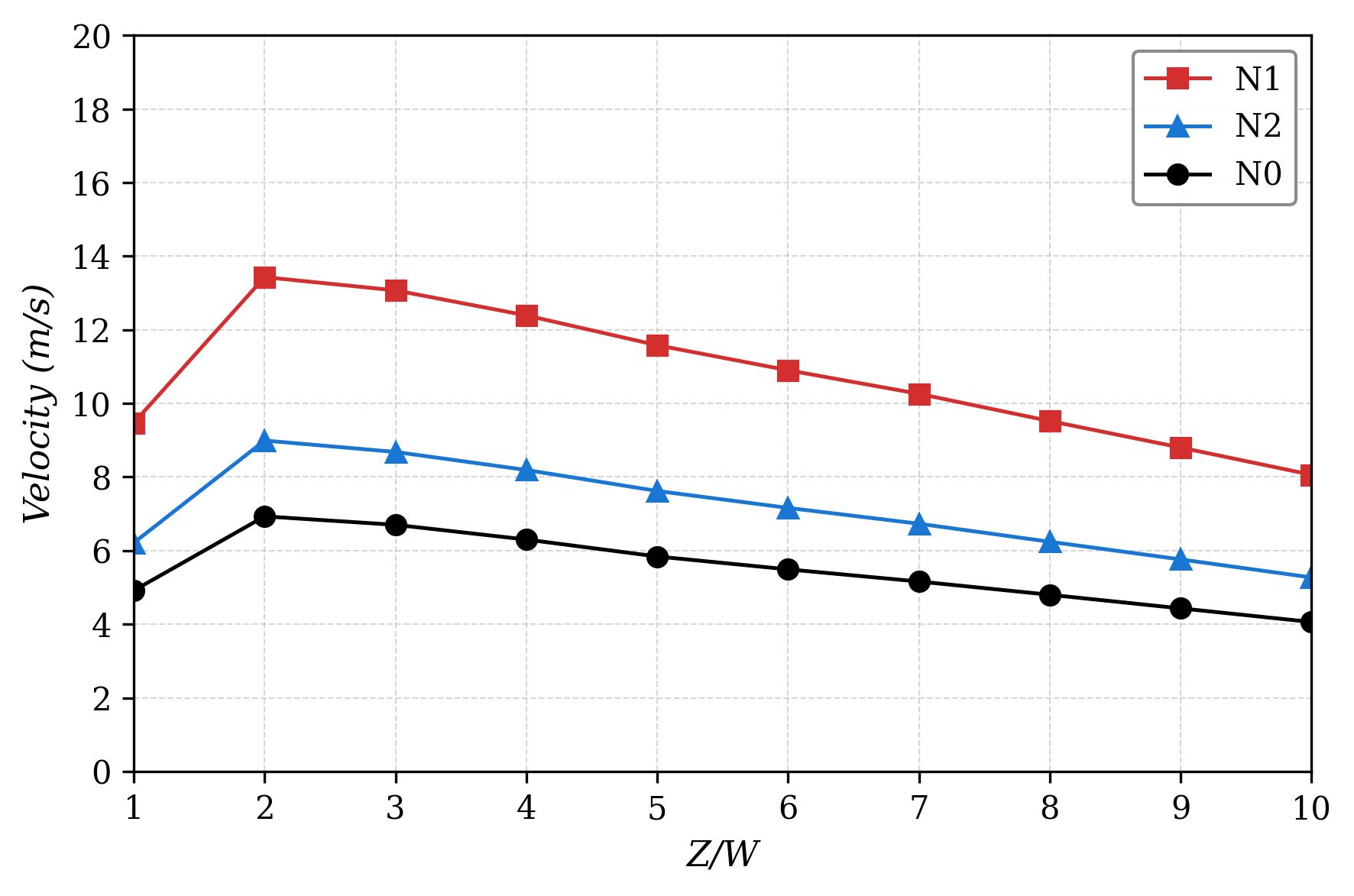}
			\vspace{5pt} \\ \small (b) Axial velocity at s2p1
		\end{minipage}
		
		\vspace{10pt} 
		
		\begin{minipage}[b]{0.48\textwidth}
			\centering
			\includegraphics[width=\textwidth]{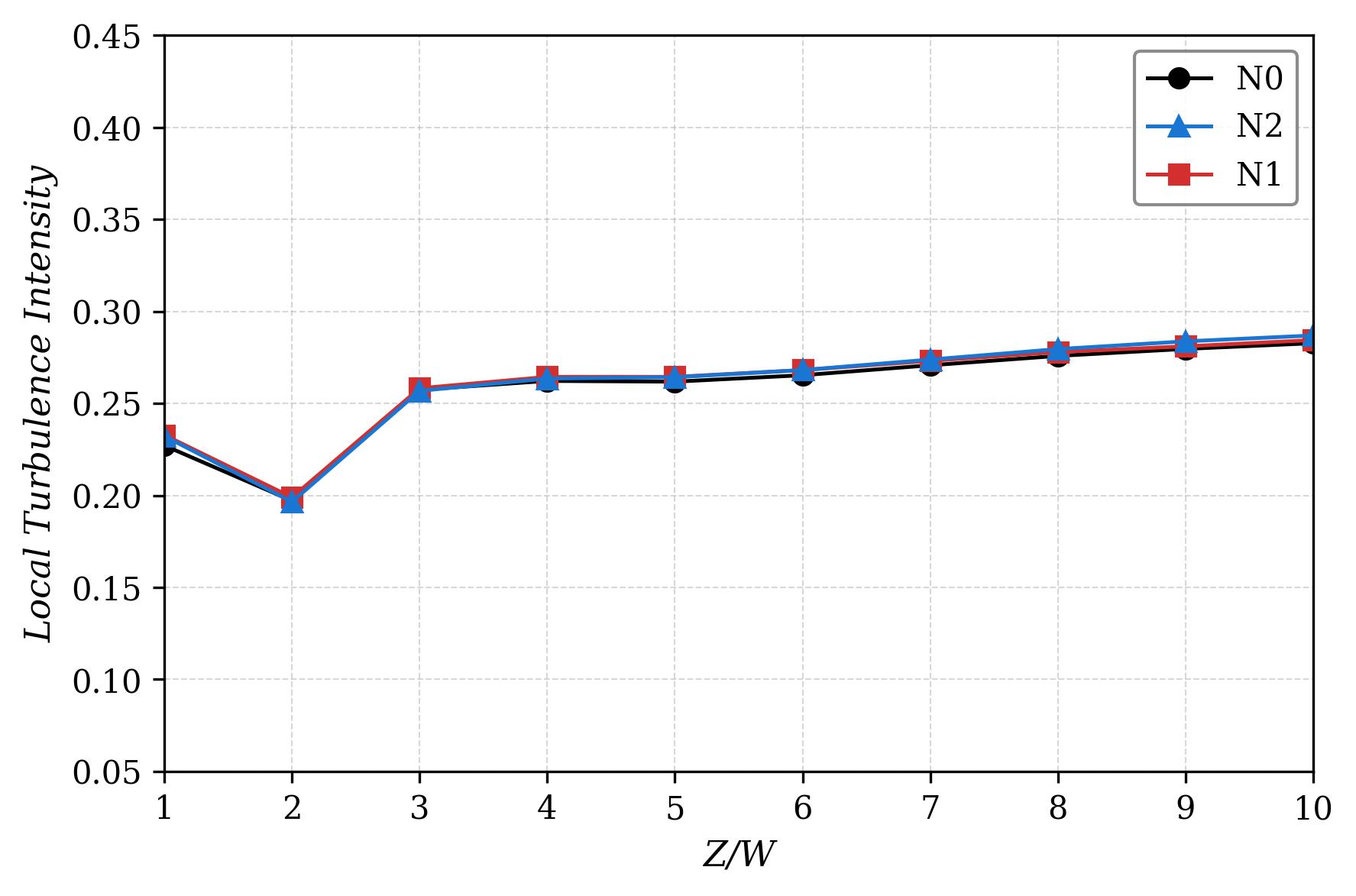}
			\vspace{5pt} \\ \small (c) Turbulence Intensity at s1p1
		\end{minipage}
		\hfill
		\begin{minipage}[b]{0.48\textwidth}
			\centering
			\includegraphics[width=\textwidth]{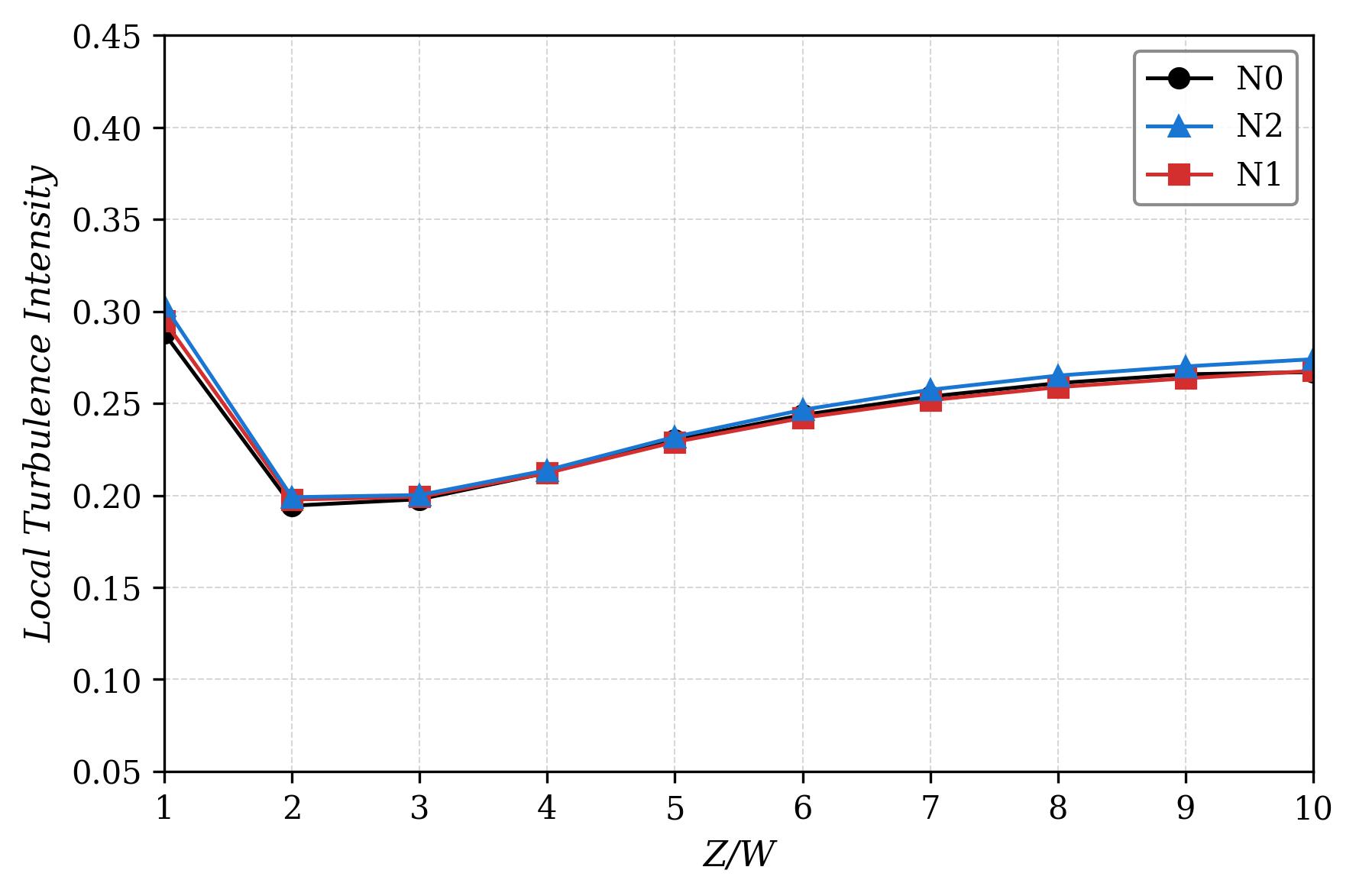}
			\vspace{5pt} \\ \small (d) Turbulence Intensity at s2p1
		\end{minipage}
		
		\caption{Influence of scaled fan rotational speeds ($N_1$, $N_0$, $N_2$) on downstream velocity and turbulence intensity profiles.}
		\label{fig:rpm_parametric_study}
	\end{figure*}

	As shown in Figure~\ref{fig:rpm_parametric_study}, increasing the rotational speed leads to a proportional increase in axial velocity levels throughout the domain. The streamwise decay rates, however, remain nearly identical across all cases, indicating that the downstream development is governed primarily by jet interaction and geometric spreading rather than the absolute momentum level.
	
	In contrast, the turbulence intensity (TI) profiles are essentially unchanged for all three configurations. Both the near-field levels and their downstream evolution collapse onto a single curve, independent of jet velocity. This behavior is consistent with the formulation of the fan boundary condition that was used, which imposes a pressure jump (momentum source) without modifying turbulence quantities (e.g., $k$, $\varepsilon$ or $\omega$). Consequently, the TI field is passively convected and evolves identically in all cases. The finite turbulence level is prescribed at the upstream pressure inlets, setting the baseline field. Downstream TI is then governed mainly by shear-layer production between adjacent jets, which is insensitive to uniform velocity scaling. As a result, TI profiles collapse across cases and fall within the experimental range, but not due to explicit modeling of fan-generated turbulence.
	
	\subsection{Sensitivity to Inlet Turbulence}
	
\begin{figure*}[htbp]
		\centering
		\begin{minipage}[b]{0.48\textwidth}
			\centering
			\includegraphics[width=\textwidth]{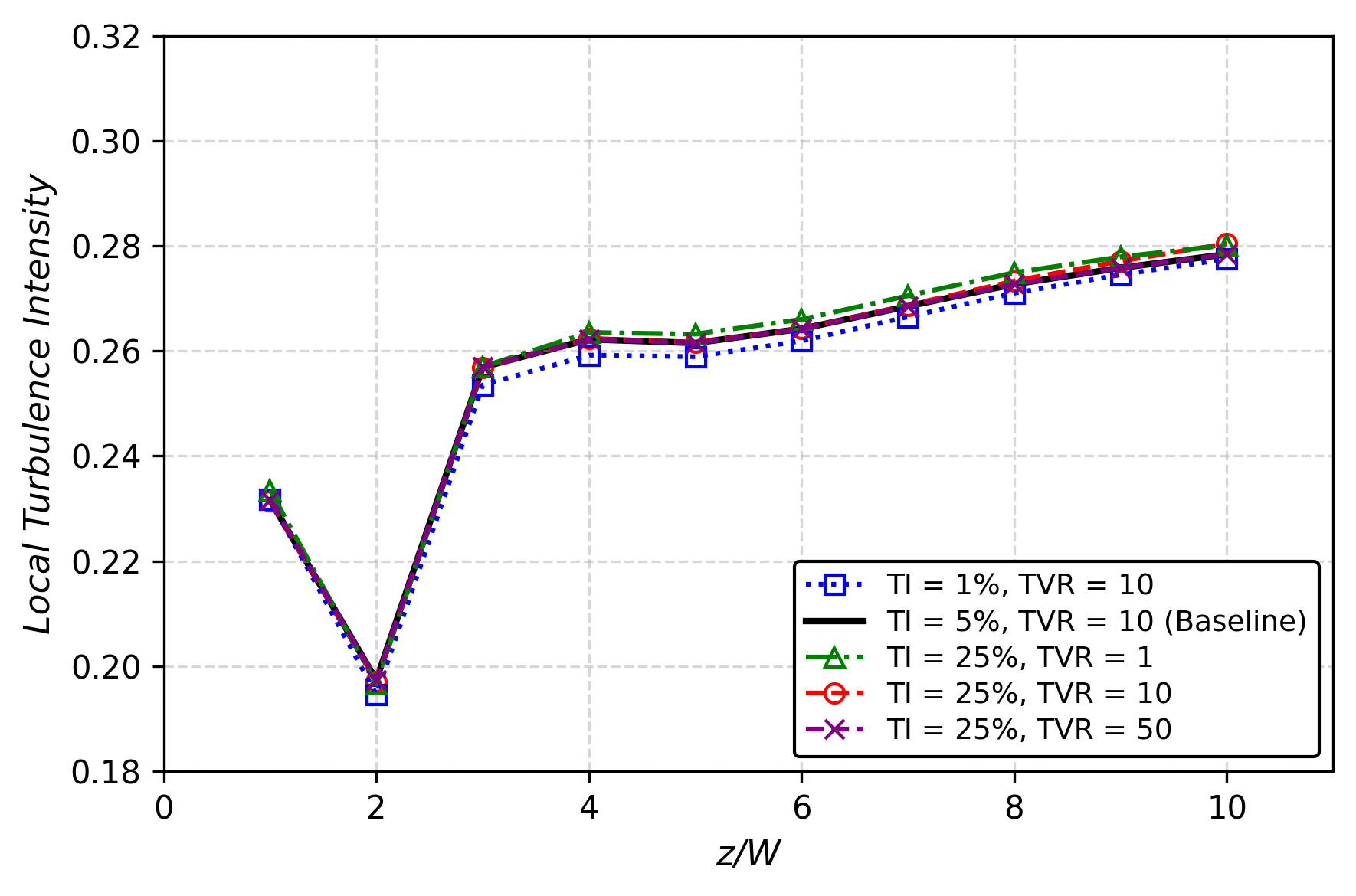}
			\vspace{5pt} \\ \small (a) Turbulence Intensity at s1p1
		\end{minipage}
		\hfill
		\begin{minipage}[b]{0.48\textwidth}
			\centering
			\includegraphics[width=\textwidth]{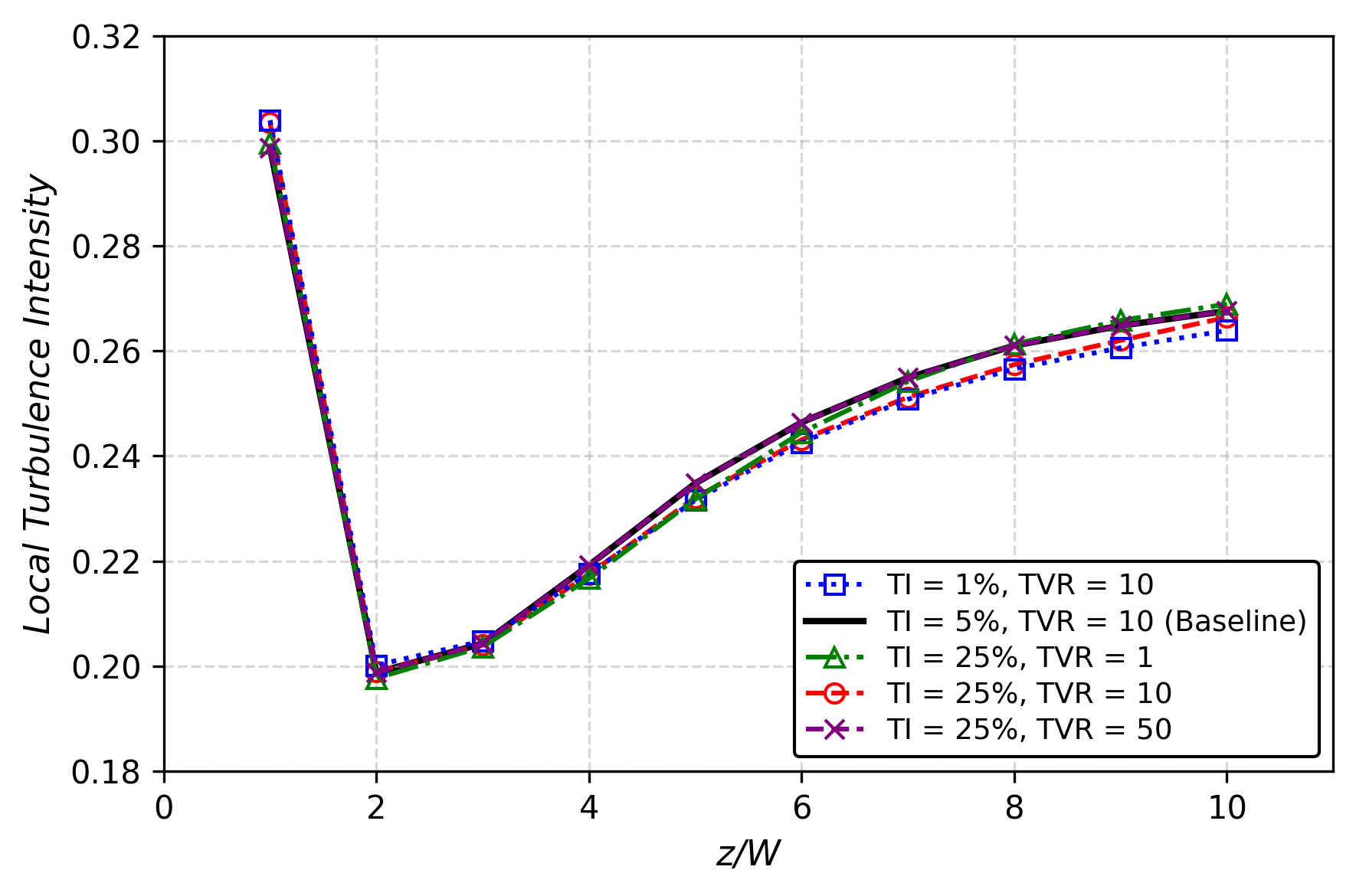}
			\vspace{5pt} \\ \small (b) Turbulence Intensity at s2p1
		\end{minipage}
		
		\caption{Influence of inlet turbulence specifications on the downstream turbulence intensity evolution in the core and peripheral regions.}
		\label{fig:ti_comparison_side_by_side}
	\end{figure*}

	Figure~\ref{fig:ti_comparison_side_by_side} presents the downstream evolution of turbulence intensity for different inlet turbulence specifications in both the core and peripheral regions. Despite substantial variations in the prescribed inlet conditions, including turbulence intensity values ranging from $1\%$ to $25\%$ and turbulent viscosity ratios between $1$ and $50$, the resulting turbulence intensity distributions remain similar across all cases. Minor differences are observed in the immediate near-field; however, these deviations rapidly diminish with downstream distance, leading to nearly identical profiles beyond $z/W \approx 4$.
	
	This behavior indicates that the FAWG-generated flow is largely insensitive to the inlet turbulence state imposed at the pressure boundary. When considered together with the earlier observation that variations in fan boundary condition parameters do not significantly affect turbulence levels, it becomes evident that the dominant mechanisms governing turbulence generation are internal to the flow field. In particular, shear layer development at jet boundaries, jet--jet interactions, and subsequent mixing processes act as primary sources of turbulence production. 
	
	\subsection{Flat Plate in FAWG Flow}
	
	To illustrate the aerodynamic implications of FAWG-generated inflow conditions, a canonical flat plate configuration is considered as a demonstrative application. The objective is to examine how the non-uniform, turbulent flow produced by the fan array influences the response of a simple lifting surface.
	
	The plate geometry is based on the low-Reynolds-number experimental study of Torres and Mueller \cite{torres2004}. The model is a low-aspect ratio ($AR=1.0$) square plate with a chord length of 300 mm, a thickness of 5.88 mm, and a 5:1 elliptical leading edge, followed by a blunt trailing edge. A single case with an angle of attack $\alpha=2\degree$ is considered. 
	
	The flow conditions correspond to a Reynolds number of approximately $Re \approx 1.1 \times 10^5$, consistent with the mean velocity levels generated by the FAWG. At this Reynolds number, transitional effects are expected to play a significant role. To account for this behavior, the $k$--$\omega$ SST model with the $\gamma$-transition formulation is activated in the code.
	
	To isolate the effect of the inflow conditions, two configurations are considered. In the first, the plate is placed within the FAWG-generated flow field. In the second, the same plate is simulated in a uniform freestream domain under equivalent Reynolds number conditions. Same problem domain is used, with the FAWG zone removed and the upstream boundary set as prescribed velocity boundary.
	
	The aerodynamic response is evaluated in terms of lift and drag coefficients, and further examined through comparisons of surface flow features, including velocity and shear distributions. By contrasting the FAWG and uniform inflow cases, this example highlights the influence of jet interaction and elevated turbulence levels on the predicted aerodynamic behavior.
	
	Under FAWG-generated inflow, both lift and drag increase significantly despite similar mean velocity levels. The lift coefficient increases by approximately 108\%, while the drag coefficient shows a larger rise of about 380\% (Table \ref{tab:flat_plate_results}). This indicates that the aerodynamic response is strongly influenced by inflow non-uniformity rather than the mean flow magnitude alone.
	
	The observed increase in aerodynamic forces is primarily associated with the interaction between discrete jet structures and the plate. Localized jet impingement regions, especially near the leading edge, produce strong spatial variations in velocity and pressure, leading to non-uniform surface loading. Compared to the uniform inflow case, the FAWG flow induces more pronounced gradients in both wall shear stress and surface pressure.
	
	This behavior is reflected in the surface distributions of skin friction coefficient ($C_f = 2(\tau_w)/\rho U_\infty^2$) and pressure coefficient ($C_p = 2(p-p_\infty)/\rho U_\infty^2$). The FAWG case exhibits localized peaks in both quantities, corresponding to high-momentum jet regions and enhanced boundary layer mixing. In contrast, the baseline case shows smoother and more uniform distributions.
	
	Overall, the results demonstrate that FAWG-generated inflow produces fundamentally different aerodynamic loading characteristics compared to uniform freestream conditions, with important implications for aerodynamic performance evaluation under realistic turbulent environments.
	
	\begin{table}[htbp]
		\centering
		\caption{Aerodynamic force comparison for the 5:1 elliptical flat plate at $\alpha = 2^\circ$.}
		\label{tab:flat_plate_results}
		\begin{ruledtabular}
			\begin{tabular}{lcccc}
				\textbf{Parameter} & \textbf{Ref.~\cite{torres2004}} &  \textbf{Freestream} & \textbf{FAWG Inflow}  & \textbf{\% Change} \\
				\midrule
				$C_L$ & 0.06 & 0.058  & 0.121  & +108\% \\
				$C_D$ & 0.015 & 0.015  & 0.072  & +380\% \\
			\end{tabular}
		\end{ruledtabular}
	\end{table}   
	
	\begin{figure*}[htbp]
		\centering
		\setlength{\tabcolsep}{4pt}
		\begin{tabular}{ccc}
			\begin{minipage}{0.32\textwidth}
				\centering
				\includegraphics[width=0.95\textwidth]{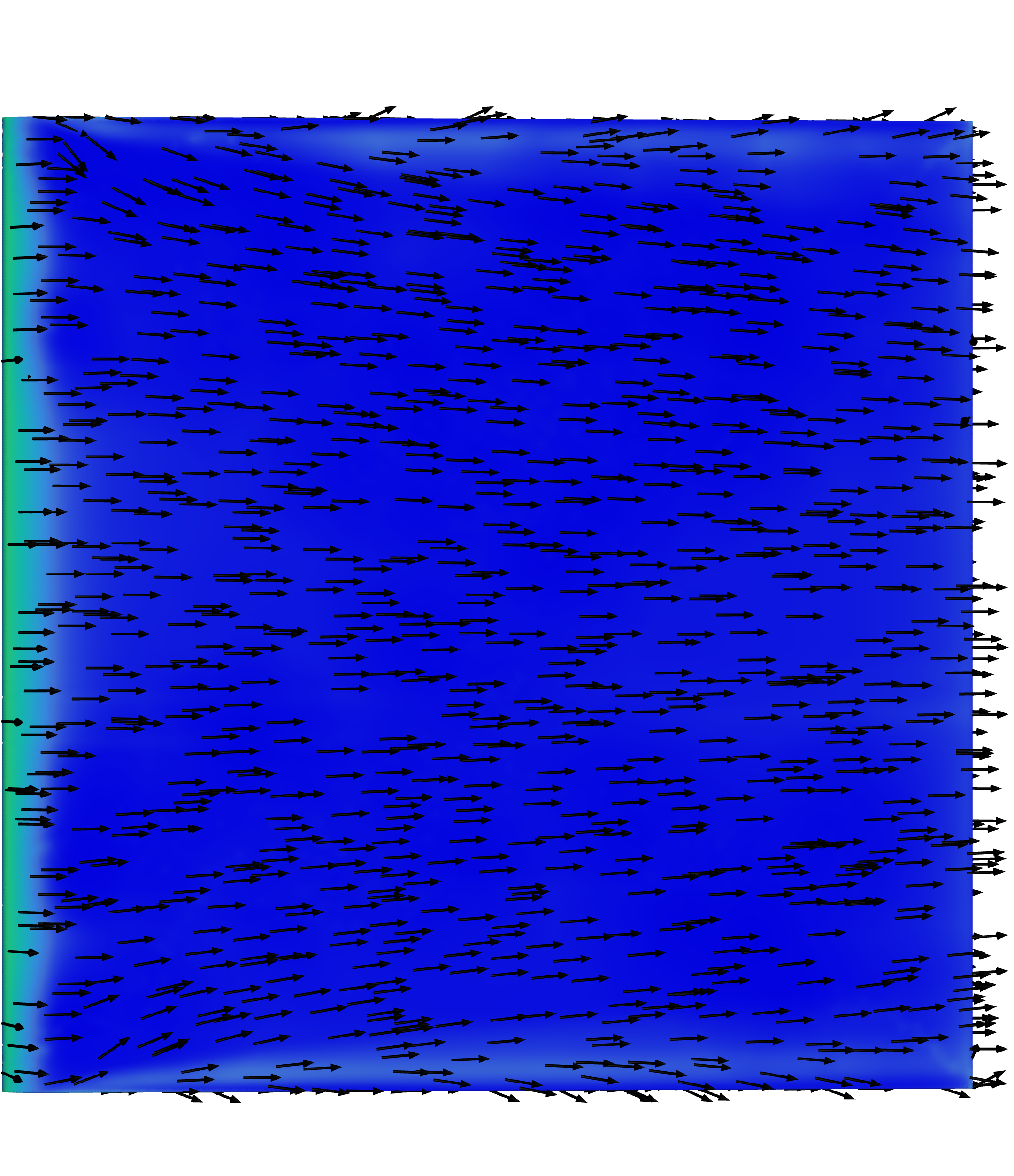}\\
				\small (a) Baseline suction side (top)
			\end{minipage} 
			&
			\multirow{2}{*}{
				\raisebox{-0.5\height}{
					\includegraphics[width=0.10\textwidth]{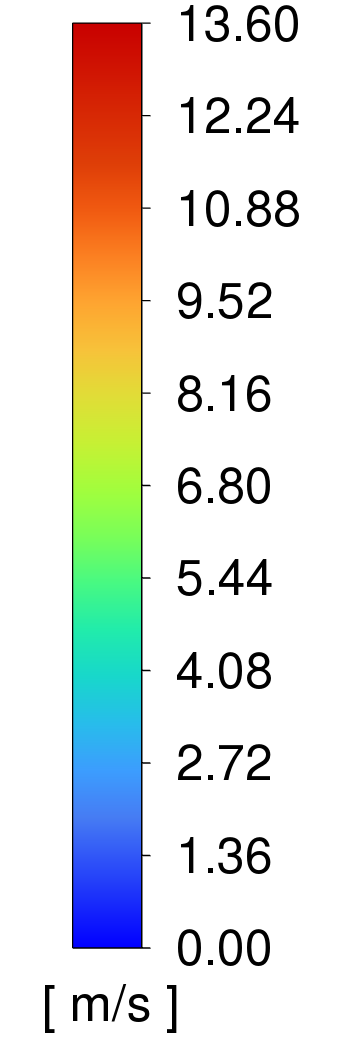}
				}
			} 
			&
			\begin{minipage}{0.32\textwidth}
				\centering
				\includegraphics[width=0.95\textwidth]{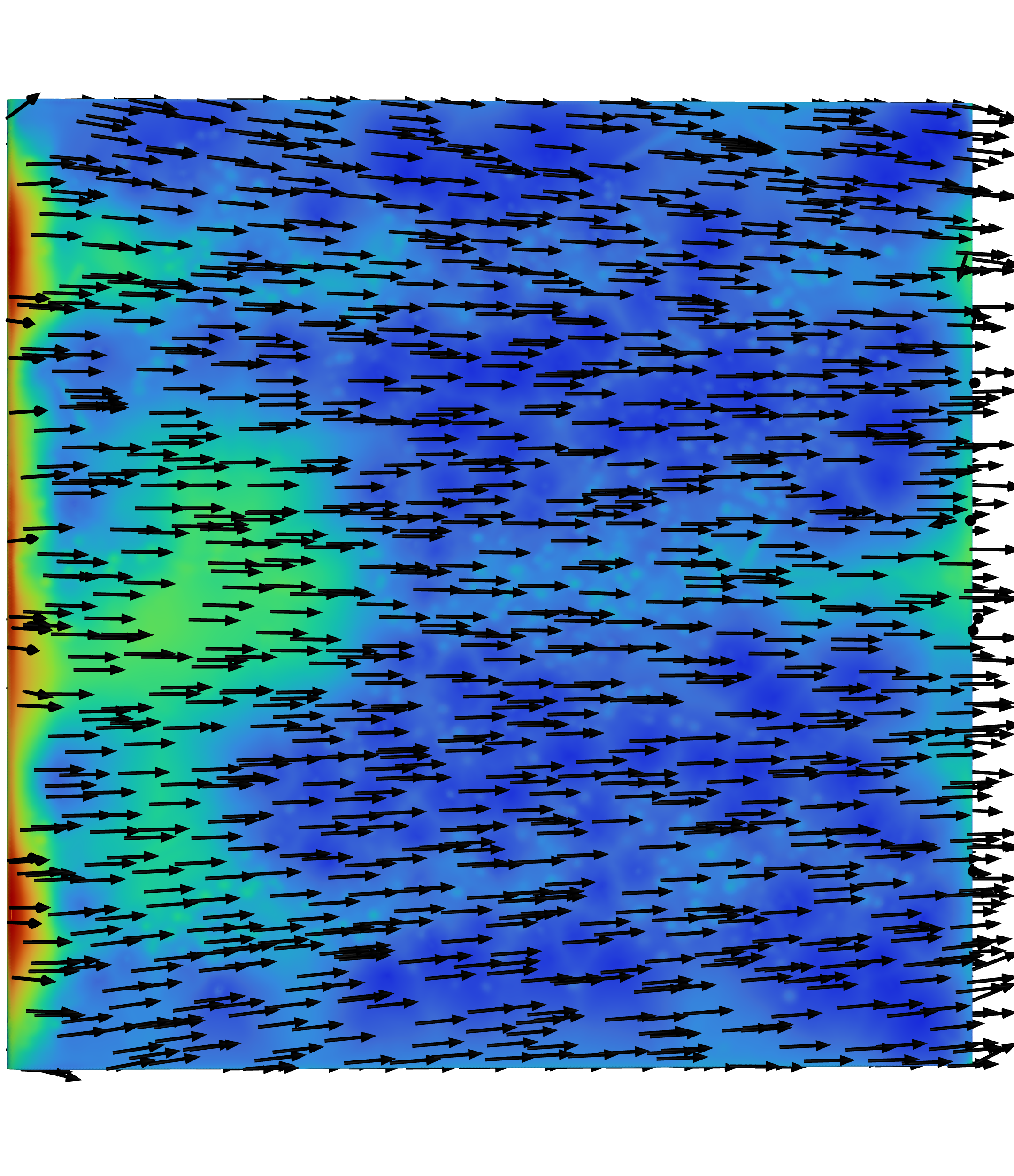}\\
				\small (b) FAWG suction side (top)
			\end{minipage} \\[20pt]
			
			\begin{minipage}{0.32\textwidth}
				\centering
				\includegraphics[width=0.95\textwidth]{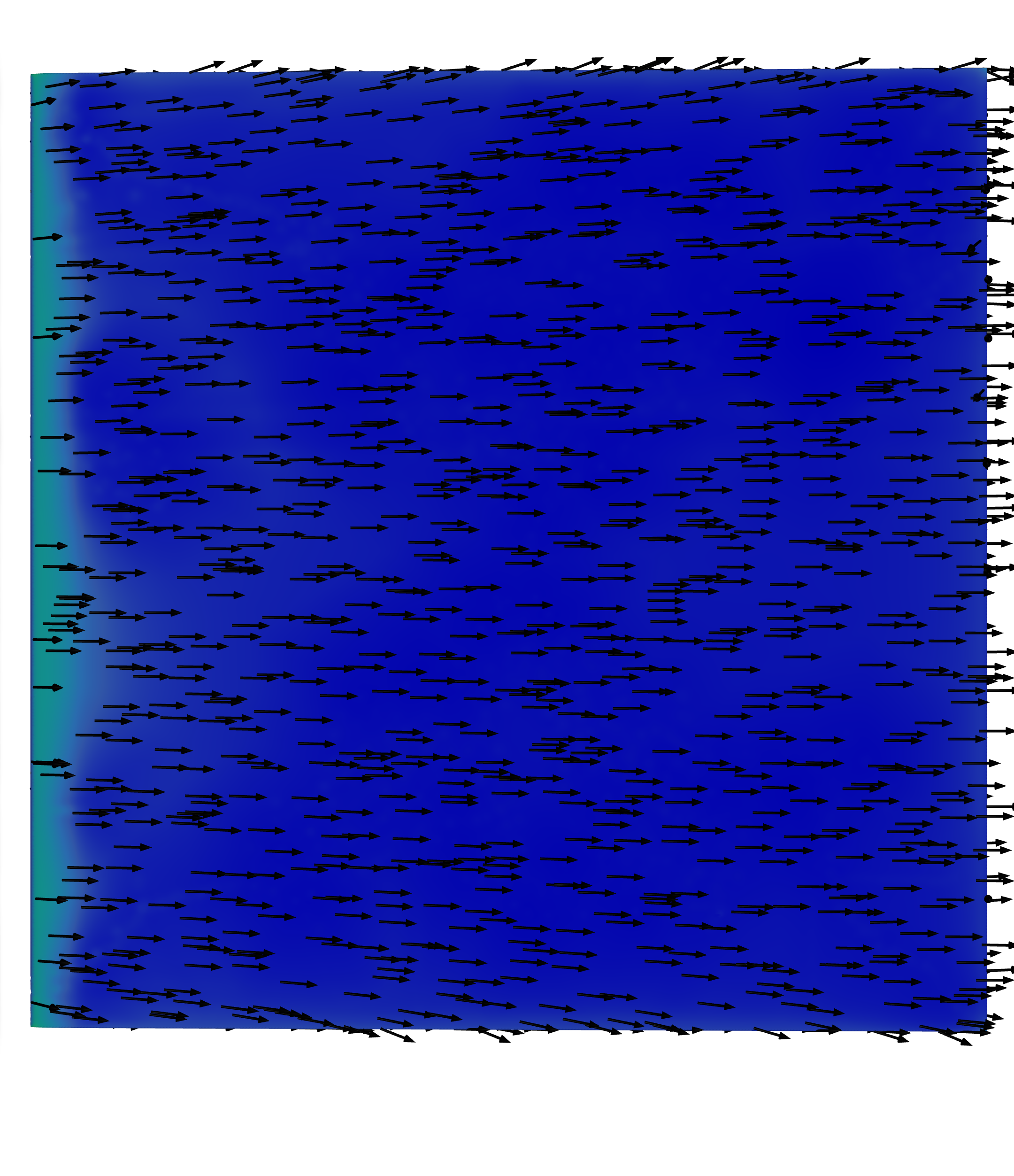}\\
				\small (c) Baseline pressure side (bottom)
			\end{minipage} 
			&
			&
			\begin{minipage}{0.32\textwidth}
				\centering
				\includegraphics[width=0.95\textwidth]{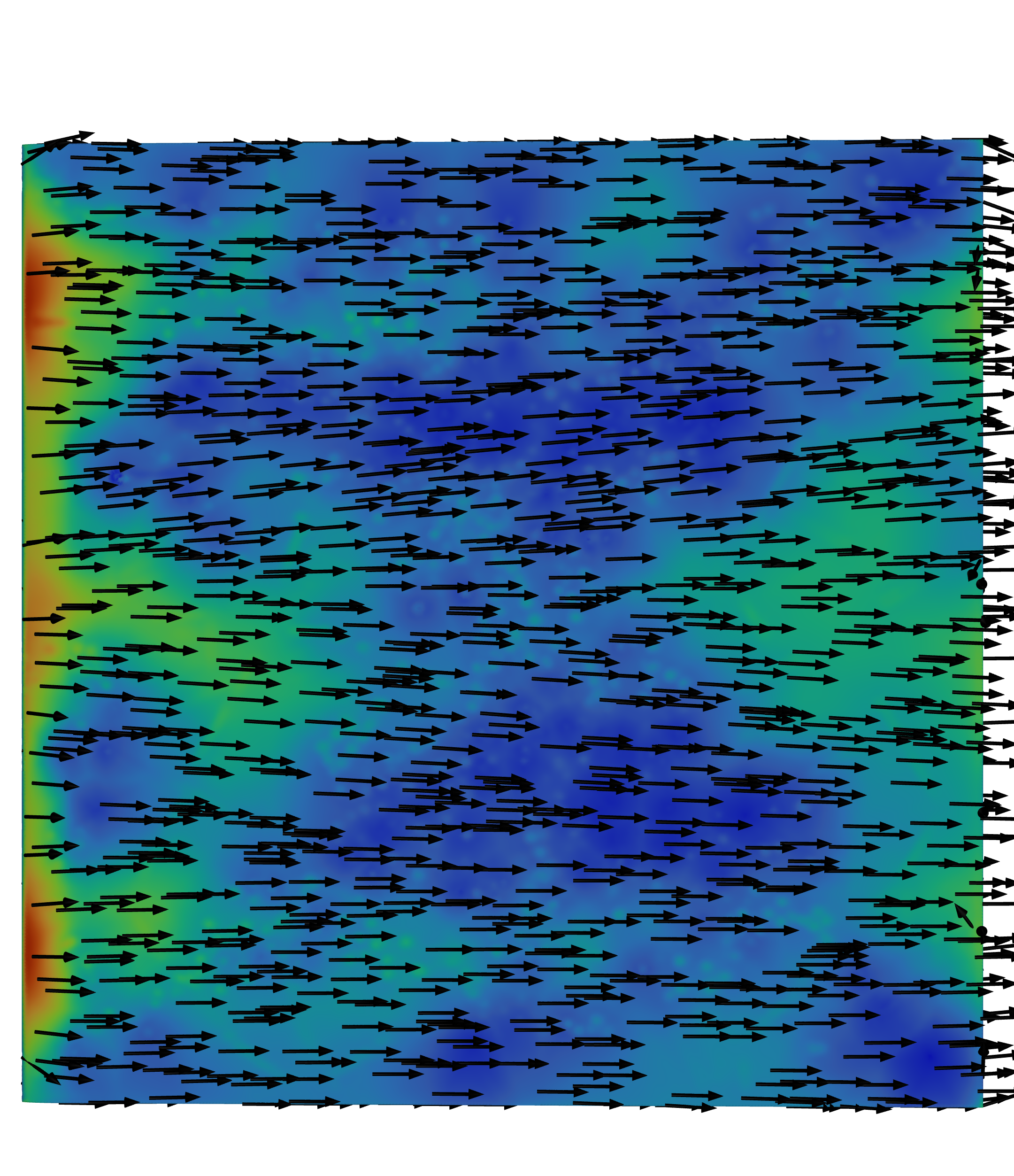}\\
				\small (d) FAWG pressure side (bottom)
			\end{minipage} \\
		\end{tabular}
		\caption{Surface velocity contours for baseline and FAWG configurations. Flow direction is from left to right.}
		\label{fig:velocity_comparison}
	\end{figure*}
	
	\begin{figure*}[htbp]
		\centering
		\setlength{\tabcolsep}{4pt}
		\begin{tabular}{ccc}
			\begin{minipage}{0.32\textwidth}
				\centering
				\includegraphics[width=0.95\textwidth]{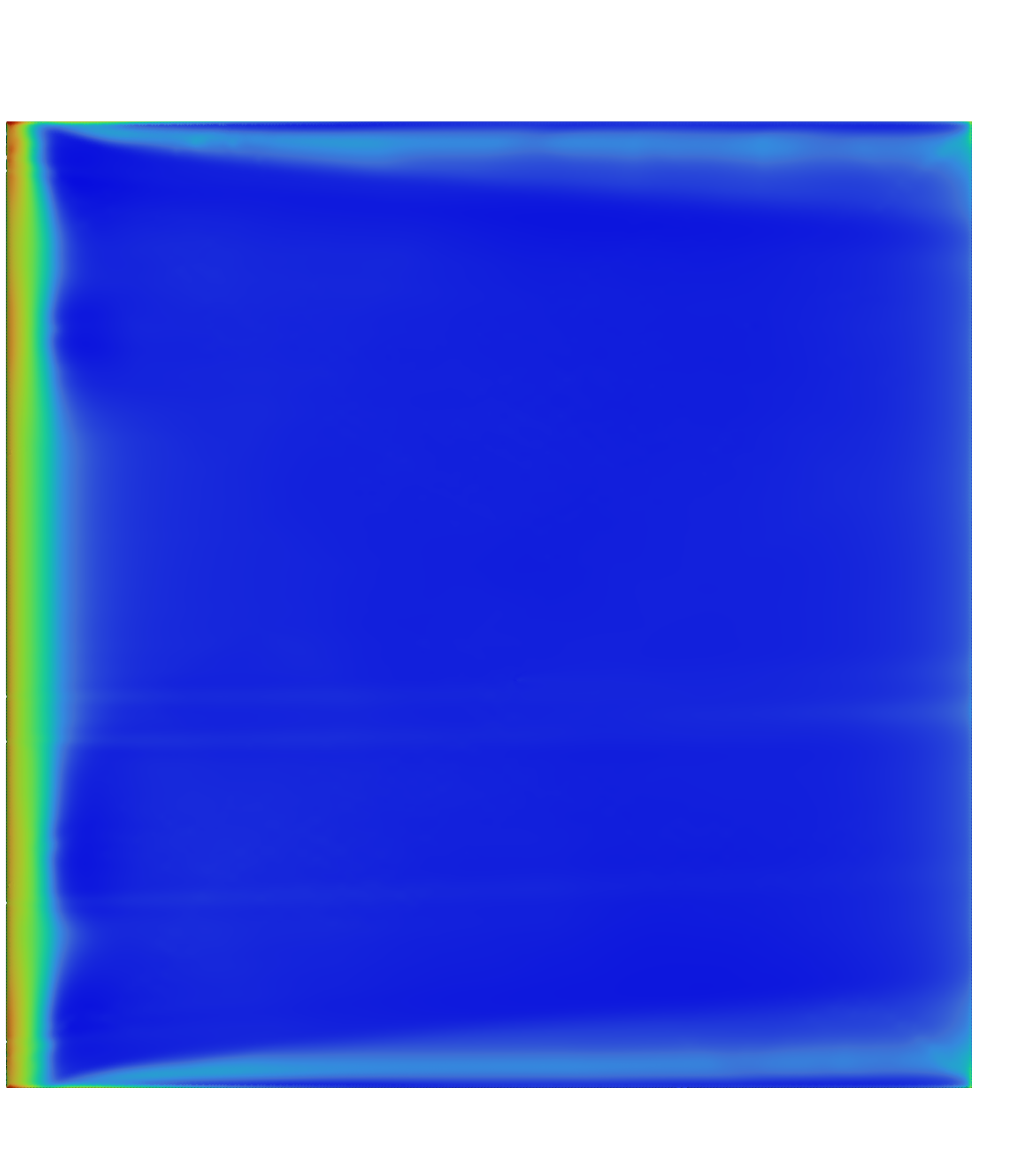}\\
				\small (a) Baseline suction side (top)
			\end{minipage} 
			&
			\multirow{2}{*}{
				\raisebox{-0.5\height}{
					\includegraphics[width=0.10\textwidth]{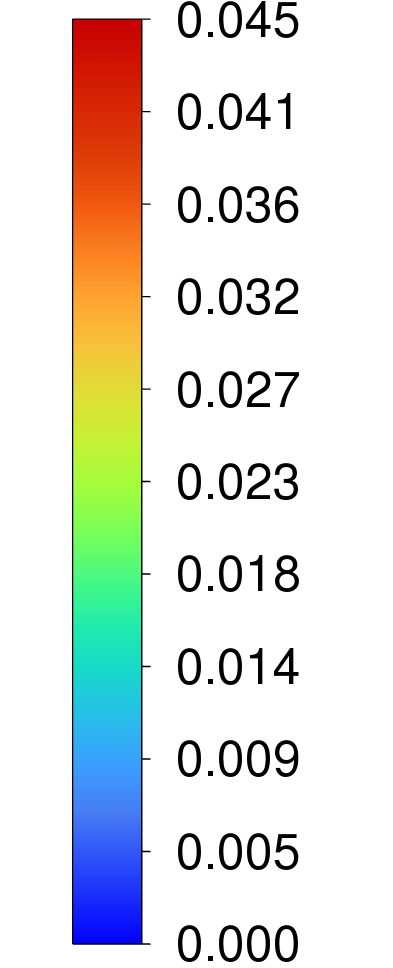}
				}
			} 
			&
			\begin{minipage}{0.32\textwidth}
				\centering
				\includegraphics[width=0.95\textwidth]{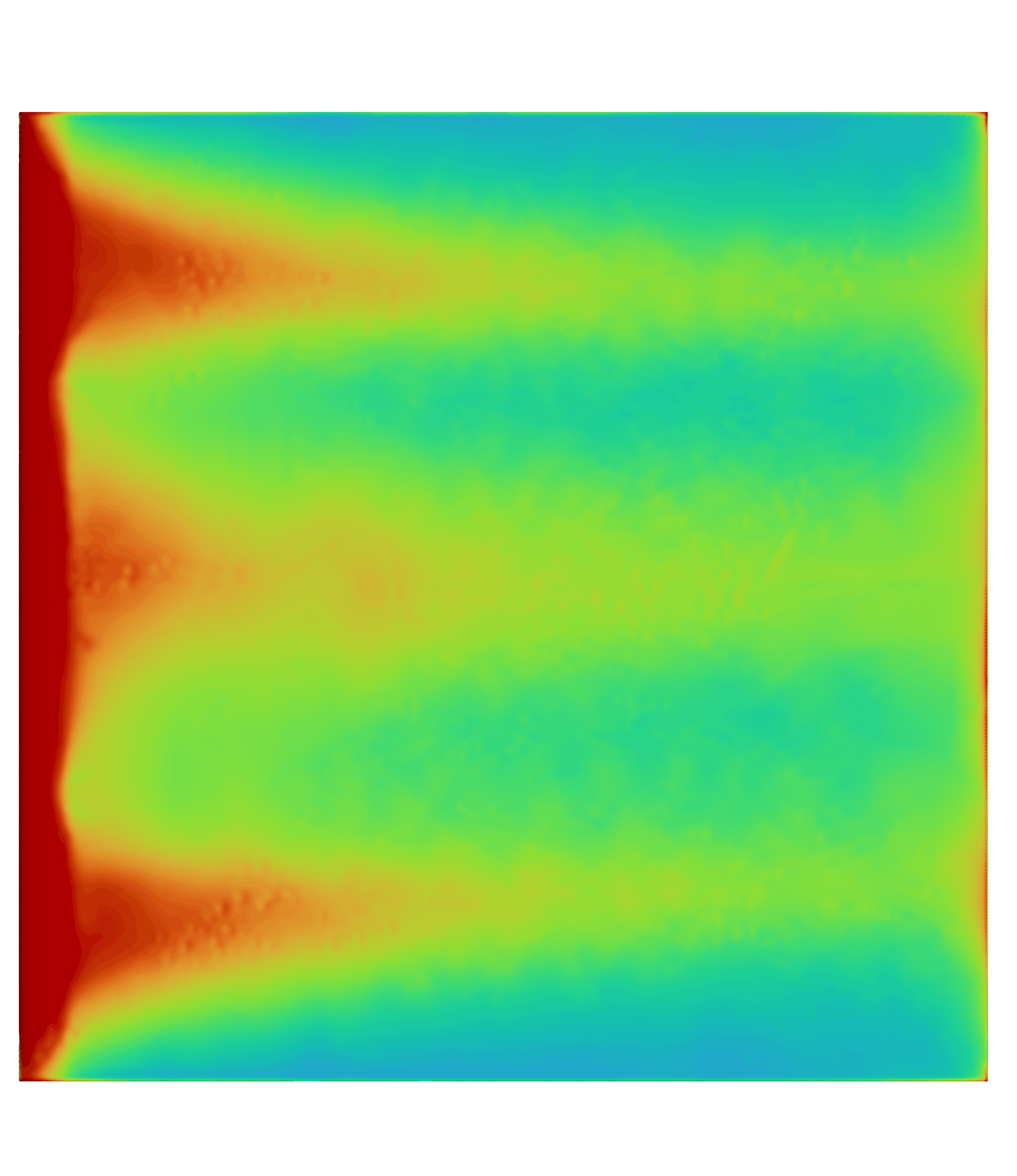}\\
				\small (b) FAWG suction side (top)
			\end{minipage} \\[20pt]
			
			\begin{minipage}{0.32\textwidth}
				\centering
				\includegraphics[width=0.95\textwidth]{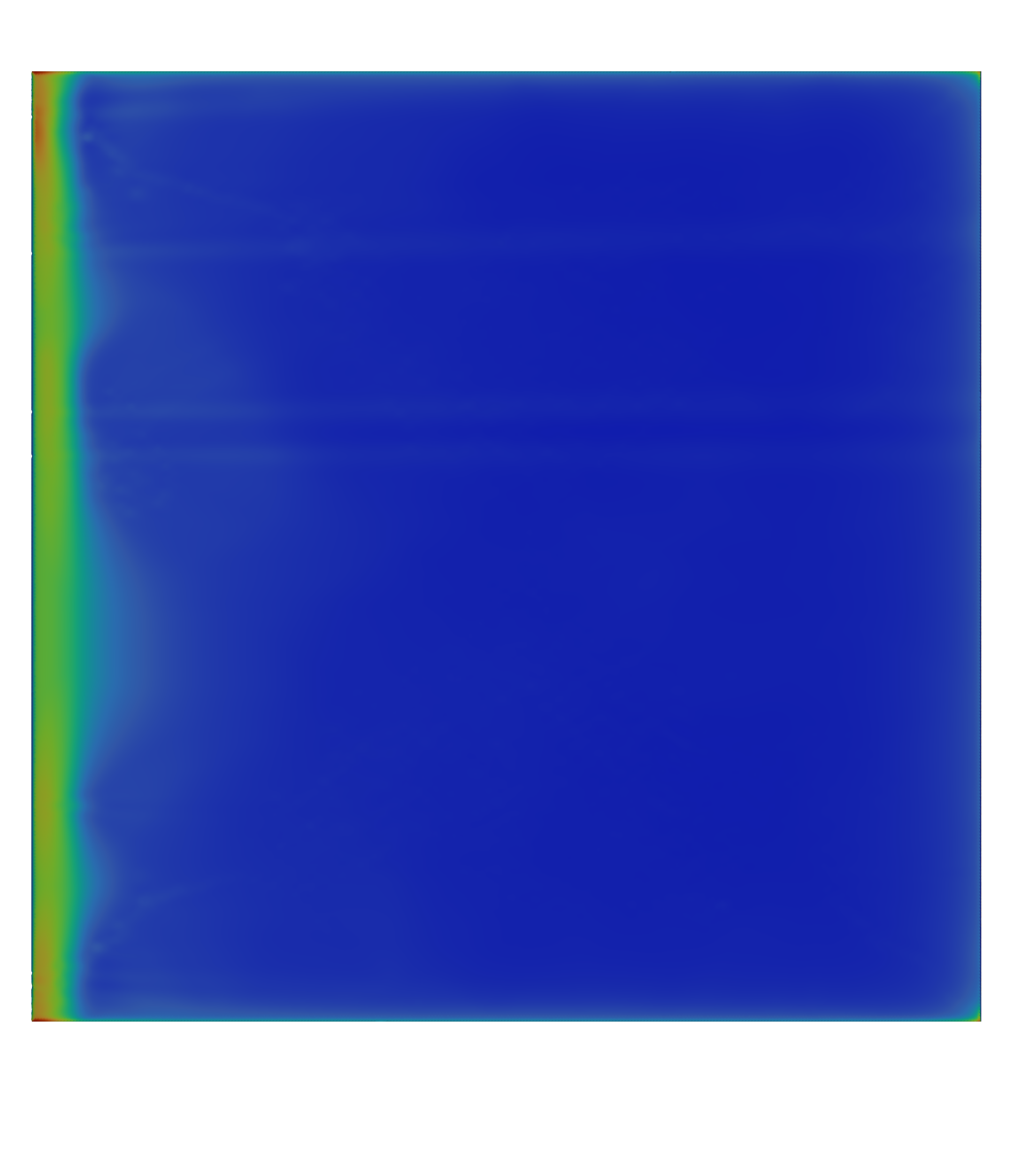}\\
				\small (c) Baseline pressure side (bottom)
			\end{minipage} 
			&
			&
			\begin{minipage}{0.32\textwidth}
				\centering
				\includegraphics[width=0.95\textwidth]{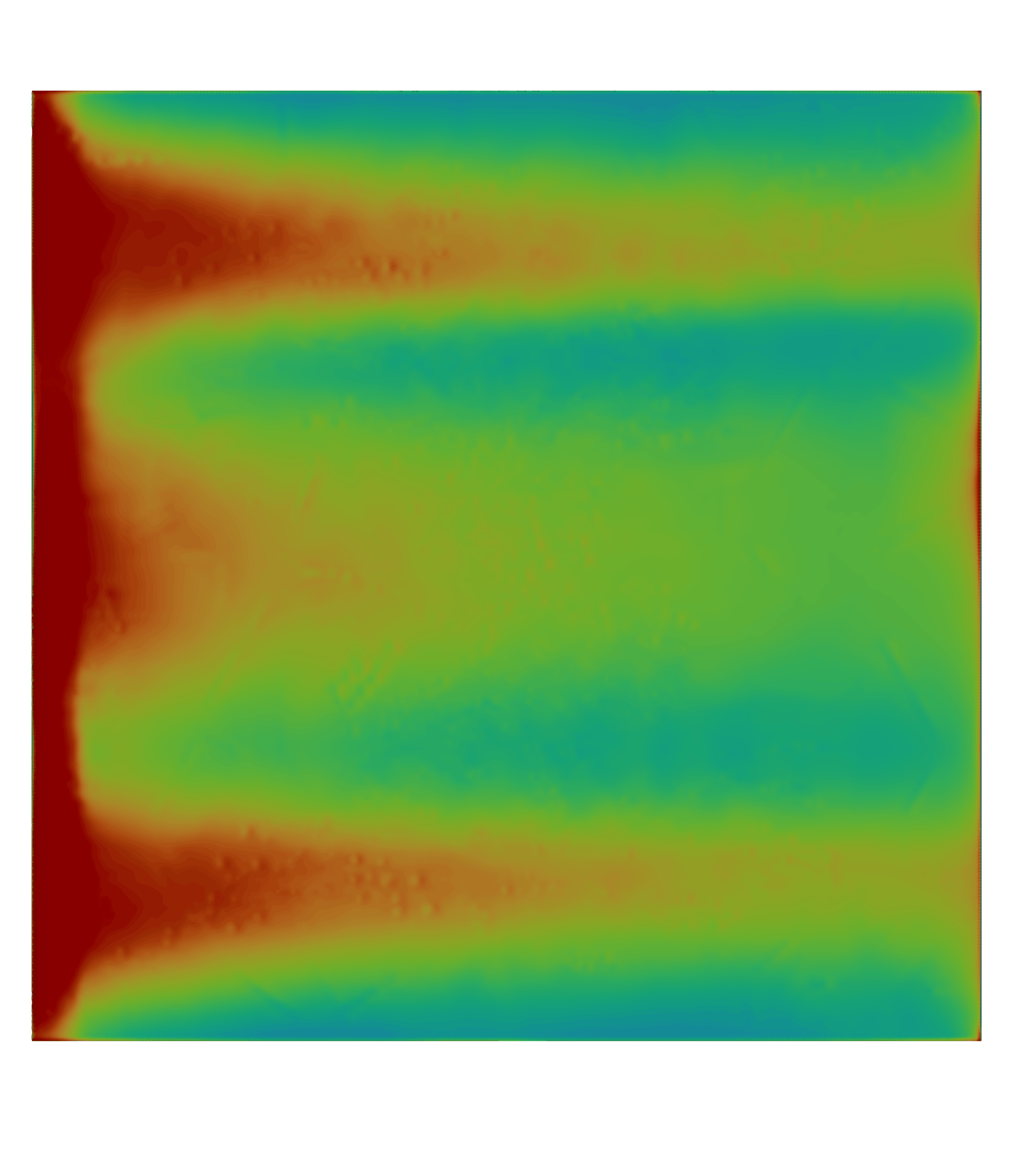}\\
				\small (d) FAWG pressure side (bottom)
			\end{minipage} \\
		\end{tabular}
		\caption{Surface skin friction coefficient ($C_f$) contours for baseline and FAWG configurations. All subplots share the same color scale to ensure direct comparison of wall shear stress distributions. Flow direction is from left to right.}
		\label{fig:cf_comparison}
	\end{figure*}
	
	\begin{figure*}[htbp]
		\centering
		\setlength{\tabcolsep}{4pt}
		\begin{tabular}{ccc}
			\begin{minipage}{0.32\textwidth}
				\centering
				\includegraphics[width=0.95\textwidth]{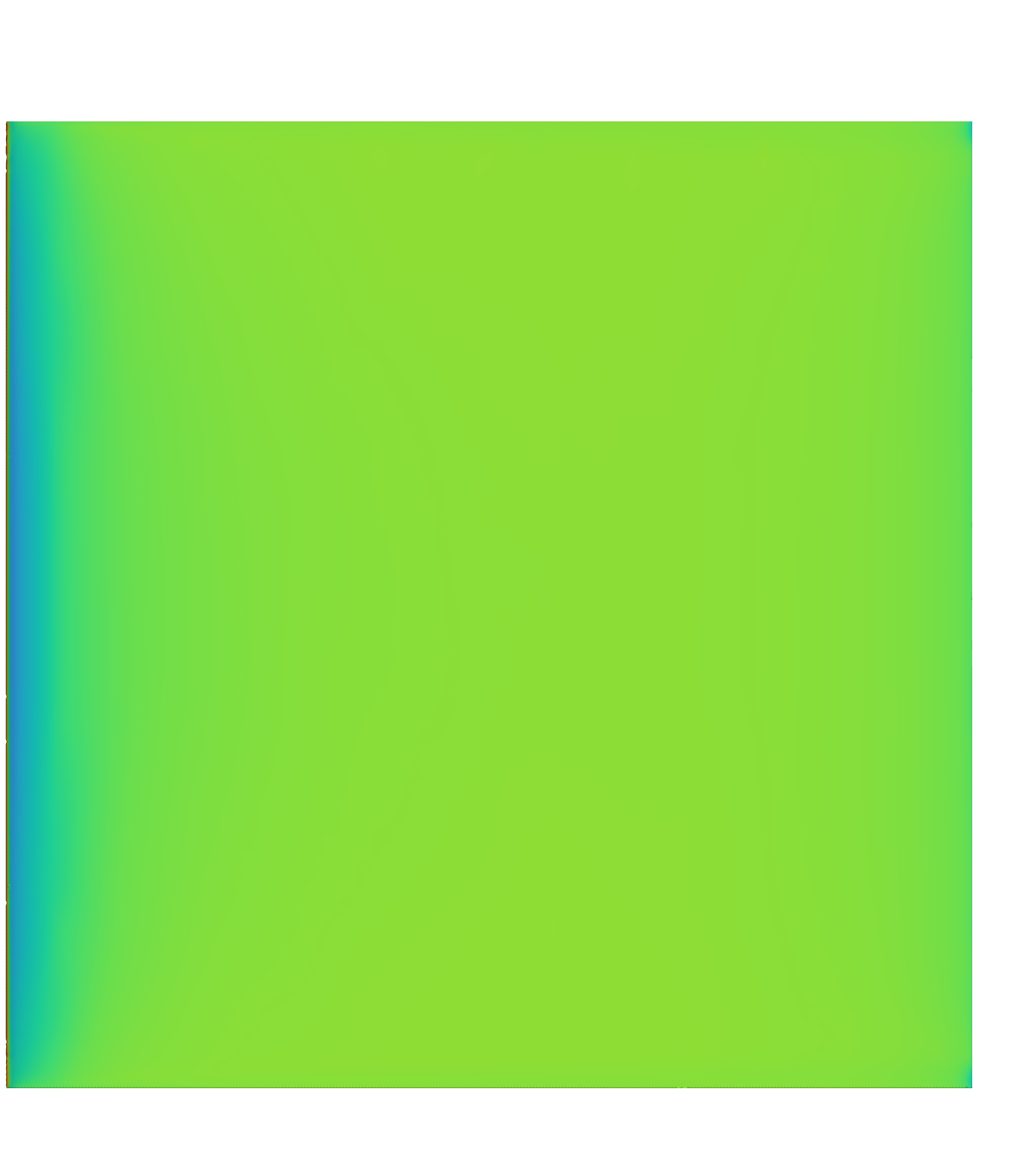}\\
				\small (a) Baseline suction side (top)
			\end{minipage} 
			&
			\multirow{2}{*}{
				\raisebox{-0.5\height}{
					\includegraphics[width=0.10\textwidth]{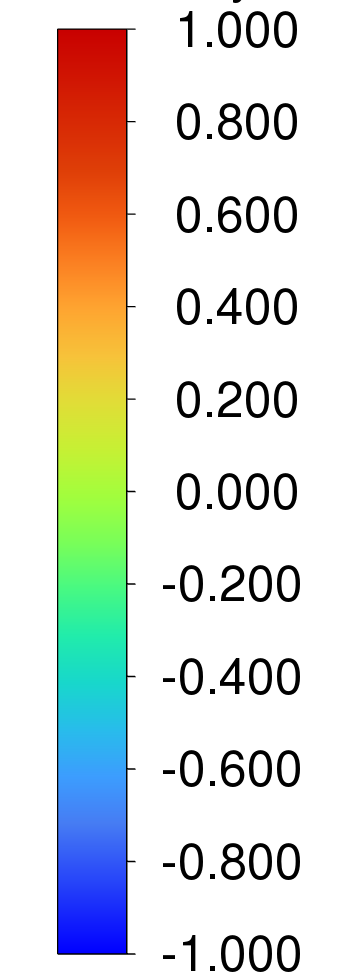}
				}
			} 
			&
			\begin{minipage}{0.32\textwidth}
				\centering
				\includegraphics[width=0.95\textwidth]{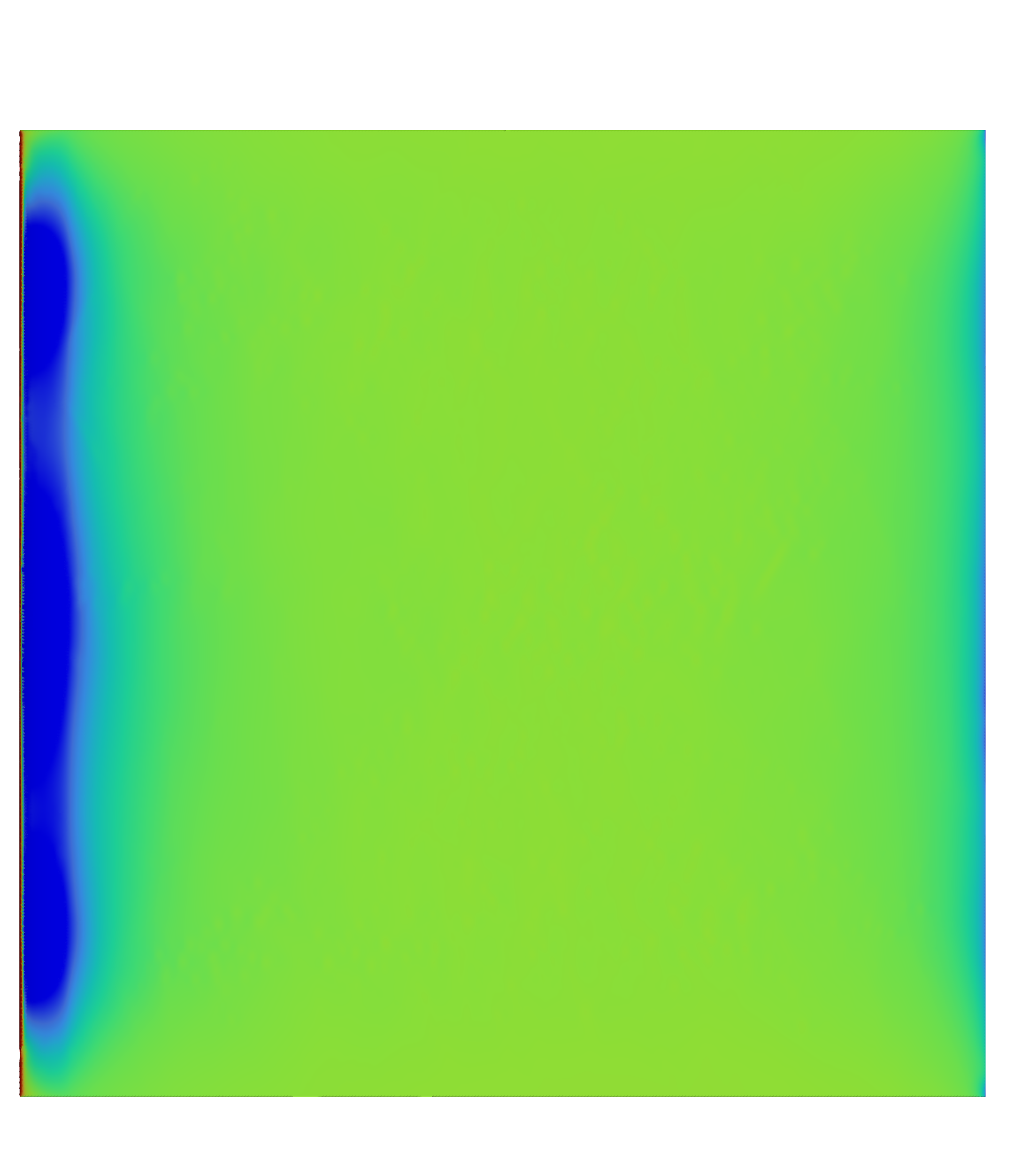}\\
				\small (b) FAWG suction side (top)
			\end{minipage} \\[20pt]
			
			\begin{minipage}{0.32\textwidth}
				\centering
				\includegraphics[width=0.95\textwidth]{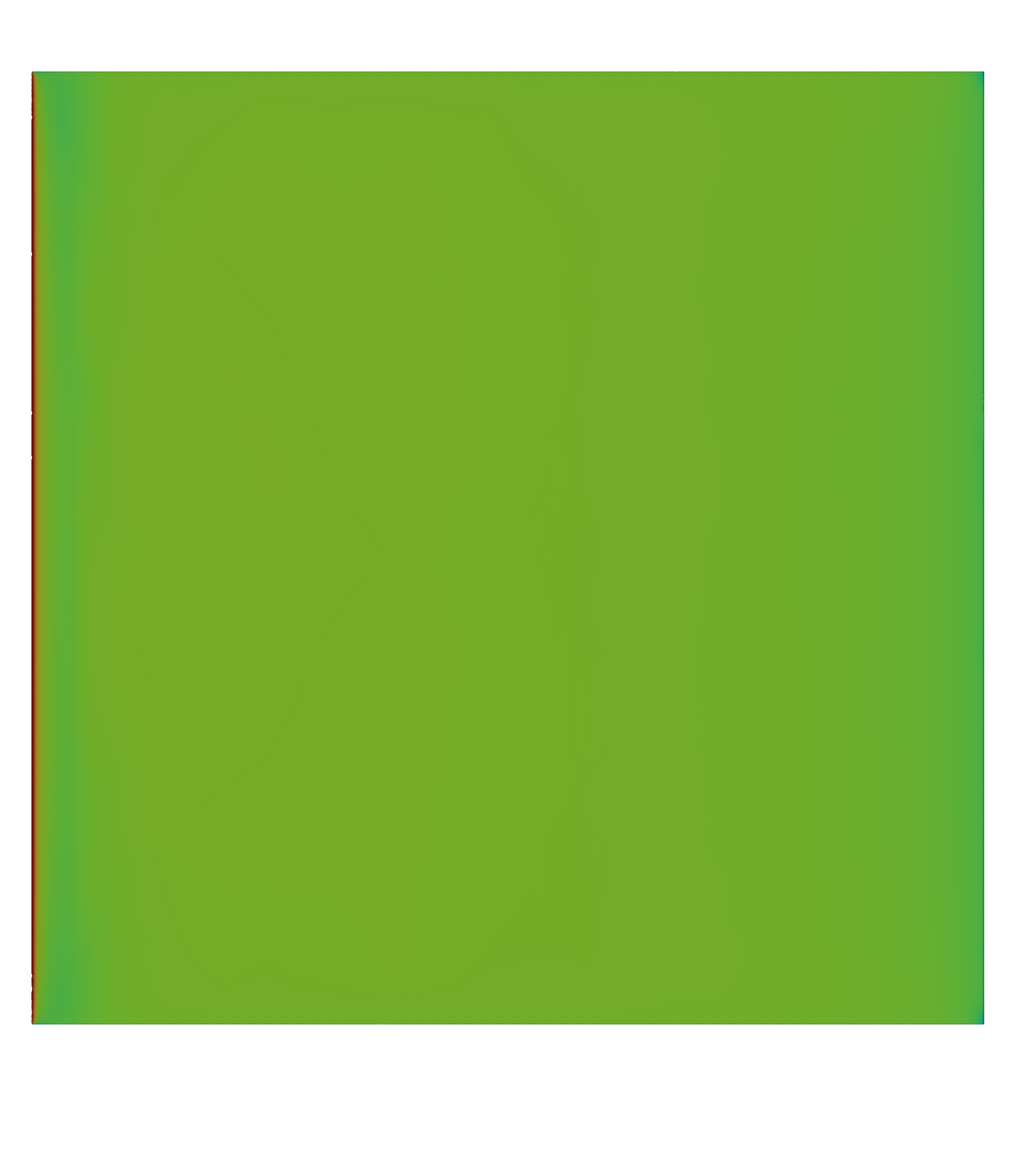}\\
				\small (c) Baseline pressure side (bottom)
			\end{minipage} 
			&
			&
			\begin{minipage}{0.32\textwidth}
				\centering
				\includegraphics[width=0.95\textwidth]{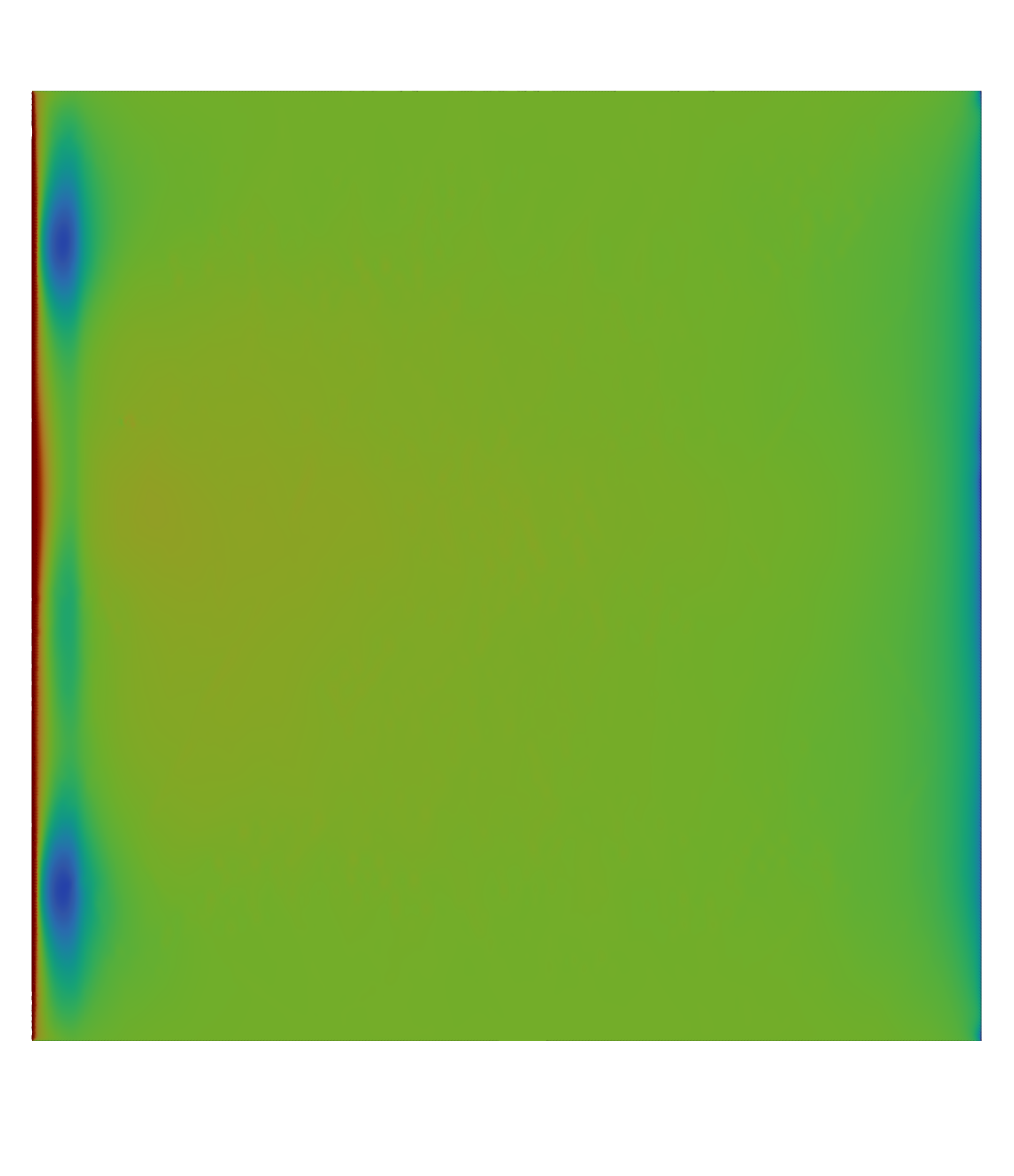}\\
				\small (d) FAWG pressure side (bottom)
			\end{minipage} \\
		\end{tabular}
		\caption{Surface pressure coefficient ($C_p$) contours for baseline and FAWG configurations. Flow direction is from left to right.}
		\label{fig:cp_comparison}
	\end{figure*}
	
	\section{Conclusions}
	
	In this study, a computational framework for modeling fan array wind generators (FAWG) was developed using a pressure jump boundary condition to represent individual fans. This approach provides a practical alternative to fully resolved blade simulations, which are computationally prohibitive for large fan arrays.
	
	The methodology was first assessed against the experimental configurations reported in \cite{arc2025WS} for two distinct fan activation patterns. The RANS simulations showed good qualitative agreement with the measured velocity and TI distributions in front of the FAWG. While the overall magnitudes were captured consistently, discrepancies in spatial trends were observed. These differences are attributed to the inherent limitations of the pressure jump representation of fans, as well as the known constraints of steady RANS modeling in complex, highly non-uniform flows.
	
	Following validation, the influence of our modeling choices was systematically investigated. The placement of the fan boundary condition was examined by comparing a bare configuration with a configuration embedded within a hollow flow domain that accounts for hub and shroud walls. The two approaches produced noticeably different jet structures, however, both captured a jet deflection into the central core.
	
	The effect of fan operating conditions was then explored by varying the jet velocity, corresponding to different fan speeds. While the velocity field scaled as expected, the turbulence intensity field remained largely unchanged. Similarly, variations in inlet turbulence specifications, including turbulence intensity and turbulent viscosity ratio, had negligible impact on the downstream TI distribution. These findings indicate that the turbulence characteristics of the FAWG-generated flow are governed primarily by internal mechanisms, such as shear layer development and jet--jet interactions, rather than by inlet conditions or operating speed. There are no other sources to artificially modify the modeled TI in the current simulation setup.
	
	To demonstrate the aerodynamic implications of FAWG-generated inflow, a low aspect ratio flat plate was simulated under both uniform freestream and FAWG conditions. The freestream simulations showed good agreement with the experimental results of Torres and Mueller~\cite{torres2004}, supporting the validity of the numerical setup. Under FAWG inflow, however, significant increases in both lift and drag were observed. These changes were associated with localized jet impingement regions, particularly near the leading edge, leading to highly non-uniform surface loading. The magnitude of these effects is linked to the ratio of jet diameter to the characteristic length scale of the body, suggesting that configurations with larger arrays and larger aerodynamic bodies may exhibit more distributed and less concentrated interactions.
	
	Overall, the results highlight that FAWG-generated flows introduce fundamentally different inflow characteristics compared to conventional uniform freestream conditions, with important implications for aerodynamic performance assessment. For future work, the present framework can be extended using scale-resolving unsteady simulations to capture transient effects associated with duty-cycle modulation of the fans with prescribed swirl.
	
	\section*{Data Availability Statement}
	The data that support the findings of this study are available from the corresponding author upon reasonable request.
	\\
	\\
	\\
	\\\\\\\\\\\\\\\\\\\\\\\\\\\\\\\\\\\\
	\bibliography{references}

@article{liu2025fanarray,
	author  = {Liu, Yutong and Noack, Bernd R. and Hu, Gang and Chen, Jialong and Gao, Nan and Raps, Franz},
	title   = {Aerodynamic characterization of a wind generator with 40 × 40 individually controllable fans},
	journal = {Physics of Fluids},
	year    = {2025},
}

@article{arc2025WS,
	author  = {Cao, Gengshou and Shaqarin, Tamir and Jiang, Zhutao and Liu, Yutong and Li, Yiqing and Gao, Nan and He, Xiaozhou and Noack, Bernd R.},
	title   = {Turbulence enhancement of a fan array wind generator using geometric texturing and optimization-based control},
	journal = {arXiv preprint arXiv:2512.14215},
	year    = {2025},
}

@article{li2024aerodynamic,
	title={Aerodynamic characterization of a fan-array wind generator},
	author={Li, Songqi and Liu, Yutong and Jiang, Zhutao and Hu, Gang and Noack, Bernd R and Raps, Franz},
	journal={AIAA Journal},
	volume={62},
	number={1},
	pages={291--301},
	year={2024},
	publisher={American Institute of Aeronautics and Astronautics}
}

@article{longobardi2024aerodynamic,
	title={Aerodynamic modeling of a delta-wing UAV for model-based navigation},
	author={Longobardi, Paolo and Skaloud, Jan},
	journal={CEAS Aeronautical Journal},
	volume={15},
	pages={283--301},
	year={2024},
	publisher={Springer},
}

@inproceedings{noca2021flow,
	title={Flow Profiling in a WindShaper for Testing Free-Flying Drones in Adverse Winds},
	author={Noca, Flavio and Bujard, Thomas and Visvaratnam, Guillaume and Catry, Guillaume and Bosson, Nicolas},
	booktitle={AIAA AVIATION 2021 FORUM},
	year={2021},
}

@article{wang2024coarse,
	title={Coarse-graining characterization of the room flow circulations due to a fan-array wind generator},
	author={Wang, X. and Cornejo Maceda, G. Y. and Liu, Y. and Hu, G. and Gao, N. and Raps, F. and Noack, B. R.},
	journal={Physics of Fluids},
	volume={36},
	year={2024},
	publisher={AIP Publishing},
}

@article{menter1994two,
	title={Two-equation eddy-viscosity turbulence models for engineering applications},
	author={Menter, Florian R.},
	journal={AIAA Journal},
	volume={32},
	number={8},
	pages={1598--1605},
	year={1994},
	publisher={American Institute of Aeronautics and Astronautics}
}

@manual{ansys_fluent_fan,
	title        = {Ansys Fluent User's Guide, Section 7.2.3: Fan Model},
	author       = {{Ansys Inc.}},
	organization = {ENEA AFS Project},
	year         = {2026},
	note         = {Available at: \url{https://www.afs.enea.it/project/neptunius/docs/fluent/html/ug/node254.htm}},
	url          = {https://www.afs.enea.it/project/neptunius/docs/fluent/html/ug/node254.htm}
}

@book{roache1998verification,
	title={Verification and Validation in Computational Science and Engineering},
	author={Roache, Patrick J.},
	year={1998},
	publisher={Hermosa Publishers},
	address={Albuquerque, New Mexico}
}

@article{celik2008procedure,
	title={Procedure for estimation and reporting of uncertainty due to discretization in CFD applications},
	author={Celik, Ismail B. and Ghia, Urmila and Roache, Patrick J. and Freitas, Christopher J.},
	journal={Journal of Fluids Engineering},
	volume={130},
	number={7},
	year={2008},
	publisher={American Society of Mechanical Engineers},
	doi={10.1115/1.2960953}
}

@techreport{torres2004,
	author      = {Gabriel E. Torres and Thomas J. Mueller},
	title       = {Low Reynolds Number Aerodynamics of Low Aspect Ratio, Thin/Flat/Cambered-Plate Wings},
	institution = {Department of Aerospace and Mechanical Engineering, University of Notre Dame},
	year        = {2004},
	month       = {June},
	type        = {Final Report},
	number      = {ADA397533},
	address     = {Notre Dame, Indiana}
}

@article{RajasakeraBabu2025,
	author  = {K. B. Rajasakera Babu and Gang Hu and Bernd R. Noack and Kenny C. S. Kwok},
	title   = {From active grids to fan-array wind generators: A review of turbulence generation, control, and artificial intelligence integration in wind tunnels},
	journal = {Physics of Fluids},
	volume  = {37},
	number  = {8},
	pages   = {081304},
	year    = {2025},
	doi     = {10.1063/5.0279910}
}

@article{Velsmann2021,
	author  = {Marcel Velsmann and Christopher Dougherty and Jason Rabinovitch and Amelia Quon and Morteza Gharib},
	title   = {Low-density multi-fan wind tunnel design and testing for the Ingenuity Mars Helicopter},
	journal = {Experiments in Fluids},
	volume  = {62},
	number  = {9},
	pages   = {193},
	year    = {2021},
	doi     = {10.1007/s00348-021-03278-5},
	publisher = {Springer}
}
	
\end{document}